\documentclass[twocolumn]{aastex701}

\newcommand{\kms}{km s$^{-1}$}
\newcommand{\feh}{[Fe/H]}

\newcommand{\rinvlmc}{$R_{\rm inv, LMC}$}
\newcommand{\rinvlmckpc}{3.2}
\newcommand{\rdisklmc}{$R_{\rm disk, LMC}$}
\newcommand{\mgfe}{[Mg/Fe]}

\newcommand{\yIa}{$y^{\rm Ia}_{\rm Fe}$}
\newcommand{\yCC}{$y^{\rm CC}_{\rm Fe}$}
\newcommand{\yCCMg}{$y^{\rm CC}_{\rm Mg}$}
\newcommand{\MgFeCC}{[Mg/Fe]$_{\rm CC}$}
\newcommand{\outflow}{$\eta$}
\newcommand{\sfe}{$\tau_\star$}
\newcommand{\Fb}{$F_{\rm b}$}
\newcommand{\taub}{$\tau_b$}
\newcommand{\sigmab}{$\sigma_b$}

\newcommand{\mhrgb}{[M/H]$_{\rm RGB}$}
\newcommand{\taurgb}{$\tau_{\rm RGB}$}

\newcommand{\dmhrgb}{$\Delta$[M/H]$_{\rm RGB}$}

\newcommand{\Dmu}{$D_\mu$}

\newcommand{\lmccenterHIra}{78.77}
\newcommand{\lmccenterHIdec}{$-69.01$}
\newcommand{\smccenterphotra}{13.19}
\newcommand{\smccenterphotdec}{$-72.83$}
\newcommand{\thetalmc}{149.23}
\newcommand{\incllmc}{25.86}
\newcommand{\lmccenterphotra}{82.25}
\newcommand{\lmccenterphotdec}{$-69.5$}

\newcommand{\nfieldsscyllatot}{96}
\newcommand{\nfieldscohenlmc}{111}
\newcommand{\nfieldscohensmc}{83}
\newcommand{\nfieldscohensmcarchival}{33}

\newcommand{\medfeherr}{0.015}
\newcommand{\medmgerr}{0.022}
\newcommand{\nsampsmc}{1546}
\newcommand{\nsamplmc}{4793}

\newcommand{\agergblmc}{2.8$^{+4.3}_{-1.7}$}
\newcommand{\agergbsmc}{2.8$^{+2.8}_{-1.7}$}

\newcommand{\agelmc}{$4.5^{+4.3}_{-3.1}$}
\newcommand{\agesmc}{$3.6^{+3.5}_{-2.3}$}

\newcommand{\agergblmcabund}{$7.1^{+5.5}_{-0.8}$ Gyr}
\newcommand{\agergbsmcwingabund}{$11.7^{+0.9}_{-6.6}$ Gyr}

\newcommand{\meanfehwingnear}{$-0.95$}
\newcommand{\meanfehwingfar}{$-0.98$}
\newcommand{\meanmgfewingnear}{$-0.06$}
\newcommand{\meanmgfewingfar}{$-0.02$}

\newcommand{\factoryIa}{$1.6 \pm 0.01$}

\newcommand{\factorfb}{1.8 $\pm$ 0.1}
\newcommand{\sigmafb}{6}
\newcommand{\sigmataub}{1.5$\sigma$}
\newcommand{\factortaub}{1.06 $\pm$ 0.05}
\newcommand{\sigmasigmab}{1$\sigma$}
\newcommand{\factorsigmab}{1.2 $\pm$ 0.2}

\newcommand{\sigmasfewingbody}{2.3$\sigma$} 
\newcommand{\factorsfewingbody}{2.1 $\pm$ 0.3}
\newcommand{\sigmafbwingbody}{1.8$\sigma$}
\newcommand{\factorfbwingbody}{1.6 $\pm$ 0.3}
\newcommand{\sigmataubwingbody}{3.7$\sigma$}
\newcommand{\factortaubwingbody}{1.4 $\pm$ 0.1}

\newcommand{\sigmasecondarysmceta}{6.3$\sigma$}
\newcommand{\sigmasecondarysmcsfe}{13.2$\sigma$}

\usepackage{threeparttable,hyperref}
\usepackage{amsmath,physics,mathrsfs,amssymb}

\received{22 December 2025}
\revised{16 Mar 2026}
\accepted{19 Mar 2026}

\shorttitle{Scylla at APOGEE: Starbursts \& Chemical Evolution of the MCs}
\shortauthors{Escala et al.}

\graphicspath{{./}{figures/}}

\begin{document}

\title{Scylla at APOGEE: The Impact of Starbursts on the Chemical Evolution of the Magellanic Clouds}

\author[0000-0002-9933-9551, gname=Ivanna, sname=Escala]{Ivanna Escala}
\affiliation{Space Telescope Science Institute, 3700 San Martin Drive, Baltimore, MD 21218, USA}
\email{iescala@stsci.edu}

\author[0000-0001-5538-2614,gname=Kristen, sname=McQuinn]{Kristen B.~W. McQuinn}
\affiliation{Space Telescope Science Institute, 3700 San Martin Drive, Baltimore, MD 21218, USA}
\affiliation{Department of Physics and Astronomy, Rutgers the State University of New Jersey, 136 Frelinghuysen Rd., Piscataway, NJ, 08854, USA}
\email{kmcquinn@stsci.edu}

\author[0000-0001-5388-0994, gname=Sten, sname=Hasselquist]{Sten Hasselquist}
\affiliation{Space Telescope Science Institute, 3700 San Martin Drive, Baltimore, MD 21218, USA}
\email{shasselquist@stsci.edu}

\author[0000-0002-2970-7435, gname=Roger, sname=Cohen]{Roger E.~Cohen}
\affiliation{Department of Physics and Astronomy, Rutgers the State University of New Jersey, 136 Frelinghuysen Rd., Piscataway, NJ, 08854, USA}
\email{rc1273@physics.rutgers.edu}

\author[0000-0002-6534-8783, gname=James, sname=Johnson]{James W.~Johnson}
\affiliation{The Observatories of the Carnegie Institute for Science, 813 Santa Barbara Street, Pasadena, CA 91101, USA}
\email{jjohnson10@carnegiescience.edu}

\author[0000-0003-2969-2445, gname=Christian, sname=Hayes]{Christian R.~Hayes}
\affiliation{Space Telescope Science Institute, 3700 San Martin Drive, Baltimore, MD 21218, USA}
\email{chayes@stsci.edu}

\author[0009-0005-0339-015X, gname=Clare, sname=Burhenne]{Clare Burhenne}
\affiliation{Department of Physics and Astronomy, Rutgers the State University of New Jersey, 136 Frelinghuysen Rd., Piscataway, NJ, 08854, USA}
\email{cdb201@physics.rutgers.edu}

\author[0000-0002-4863-8842, gname=Alexander, sname=Ji]{Alexander P.~Ji}
\affiliation{Department of Astronomy \& Astrophysics, University of Chicago, 5640 South Ellis Avenue, Chicago, IL 60637, USA; Kavli Institute for Cosmological Physics, University of Chicago, Chicago, IL 60637, USA}
\affiliation{Joint Institute for Nuclear Astrophysics—Center for Evolution of the Elements (JINA), East Lansing, MI 48824, USA}
\email{alexji@uchicago.edu}

\author[0000-0003-0588-7360]{Christina W. Lindberg}
\email{christina.lindberg@live.com}
\affiliation{Center for Astrophysics, Harvard \& Smithsonian, 60 Garden Street, Cambridge, MA 02138, USA}
\affiliation{Space Telescope Science Institute, 
3700 San Martin Drive, 
Baltimore, MD 21218, USA}

\author[0000-0002-9912-6046, gname=petia, sname=yanchulova merica-jones]{Petia Yanchulova Merica-Jones}
\affiliation{Institute of Astronomy, Bulgarian Academy of Sciences, 72, Tsarigradsko Chaussee Blvd, 1784, Sofia, Bulgaria}
\affiliation{Space Telescope Science Institute, 3700 San Martin Drive, Baltimore, MD 21218, USA}
\affiliation{University of Sofia, Faculty of Physics, 5 James Bourchier Blvd., 1164 Sofia, Bulgaria}
\email{pyanchulova@astro.bas.bg}

\author[0000-0003-1680-1884]{Yumi Choi}
\affiliation{NSF National Optical-Infrared Astronomy Research Laboratory, 950 North Cherry Avenue, Tucson, AZ 85719, USA}
\email{yumi.choi@noirlab.edu}

\author[0000-0001-8416-4093, gname=andrew, sname=dolphin]{Andrew E. Dolphin}
\affiliation{Raytheon Technologies, 1151 East Hermans Road, Tucson, AZ 85756, USA}
\affiliation{ Steward Observatory, University of Arizona, 933 North Cherry Avenue, Tucson, AZ 85719, USA}
\email{adolphin@rtx.com}

\author[0000-0002-7502-0597]{Benjamin F. Williams}
\affiliation{Department of Astronomy, University of Washington, Box 351580, Seattle, WA 98195, USA}
\email{benw1@uw.edu}

\author[0000-0002-7743-8129]{Claire E.~Murray}
\affiliation{Space Telescope Science Institute, 3700 San Martin Drive, Baltimore, MD 21218, USA}
\affiliation{The William H. Miller III Department of Physics \& Astronomy, Bloomberg Center for Physics and Astronomy, Johns Hopkins
University, 3400 N. Charles Street, Baltimore, MD 21218, USA}
\email{cmurray@stsci.edu}


\correspondingauthor{Ivanna Escala}

\begin{abstract}

Owing to their proximity to the Milky Way, the Large and Small Magellanic Clouds (L/SMC) 
uniquely probe the evolution of low-mass galaxies  
undergoing mutual interactions. In this work, 
we investigate the connection between the star formation histories (SFHs) of the L/SMC measured from 
HST imaging in the Scylla survey 
and APOGEE chemical abundances. 
We model the chemical evolution of the L/SMC in the [Mg/Fe]--[Fe/H] plane within a robust statistical framework to predict chemical abundance signatures resulting directly from 
starbursts in Scylla SFHs. 
Both the L/SMC rapidly enrich to high metallicity ([Fe/H] $\gtrsim$ $-1$) within 3 Gyr, 
followed by slower chemical evolution regulated by sequential starbursts, where the SMC may require higher Fe yields from Type Ia supernovae than the LMC.
We also
model 
the [Mg/Fe]--[Fe/H] plane to infer starburst properties 
across distinct spatial regions in the L/SMC. 
We identify dominant starbursts in the L/SMC with broadly similar timing, though the SMC may host an earlier burst, and 
larger 
burst strength in the LMC. 
The global starburst properties are nearly uniform across the LMC disk, whereas the dominant SMC population experiences a stronger and later-onset burst 
in its eastern wing compared to the main body. We also find evidence for a  chemically distinct secondary population in the SMC that preferentially traces the foreground and may originate from the LMC.
We discuss the implications of these results for the evolutionary history of the L/SMC and for starbursts in interacting low-mass galaxy pairs.

\end{abstract}



\section{Introduction} \label{sec:intro}

The evolution of low-mass galaxies ($\log (M_\ast/ M_\odot) \lesssim 9$) fundamentally drives the hierarchical assembly of galaxies across different mass scales (e.g., \citealt{BullockJohnston2005,Cooper2010,Wetzel2015}) and cosmic time given their status as the most common systems at all redshifts (e.g., \citealt{Karachentsev2013,Grazian2015}). In our cosmological paradigm, interactions and mergers between dwarf galaxies are not only theoretically expected, but should occur more frequently within a given volume compared to massive galaxies (e.g., \citealt{Deason2014}), in the context of both relatively isolated dwarf galaxy associations \citep{Stierwalt2017,Besla2018,Paudel2024} and dwarf galaxy satellite systems 
\citep{MartinezDelgado2012,Rich2012,Paudel2017,Annibali2020,Sacchi2024}.
Due to their low masses and thus shallow gravitational potential wells, dwarf galaxies should experience significant disturbances as a consequence of dwarf--dwarf interactions. Indeed, observations of interacting pairs of dwarf galaxies in the local Universe have demonstrated that these systems are preferentially star-forming compared to their isolated counterparts (\citealt{Stierwalt2015,Privon2017,Sun2020,KadoFong2020,Subramanian2024,Huang2025,Chauhan2025}; c.\@f.\ \citealt{Paudel2018,KadoFong2024}) without theoretically requiring a direct merger between dwarf galaxies \citep{Martin2021,WilliamsonHugo2021}.

Owing to their proximity (50 and 62 kpc, respectively; \citealt{deGrijs2014,deGrijsBono2015}), the Large and Small Magellanic Clouds (L/SMC) provide an unparalleled opportunity to study the impact of dwarf--dwarf interactions on the evolution of low-mass galaxies 
in exquisite detail. The Magellanic Clouds (MCs) have likely been mutually interacting for the last $\sim$6--7 Gyr (\citealt{Besla2012,DiazBekki2012}), experiencing first infall into the Milky Way halo as a binary pair (\citealt{Besla2007,Besla2012,Kallivayalil2013}; but see also \citealt{Vasiliev2024} for a second passage scenario, with the first $\gtrsim$5 Gyr ago at a large pericenter $\gtrsim$100 kpc, \added{although this scenario is inconsistent with the observed properties of the LMC corona; \citealt{Lucchini2025}}). In the first infall scenario, the interaction likely involves multiple passages between the MCs \citep{Patel2020} as well as a direct collision with an impact parameter of $\lesssim$5 kpc in the LMC disk in the last $\lesssim$200 Myr \citep{Zivick2018,Zivick2019,Choi2022,Rathore2025,Rathore2025smc}. 

Therefore any structural and dynamical disturbances dating before the LMC--SMC system crossed the Milky Way's (MW's) virial radius $\sim$1--2 Gyr ago \citep{Patel2020} likely result from 
dwarf--dwarf interactions, as opposed to tidal or ram pressure stripping by the MW (e.g., \citealt{Lucchini2020,WilliamsonHugo2021}). These features include the Magellanic Bridge (e.g., \citealt{Nidever2013,Noel2013}), Magellanic Stream \citep{Putman1998,Nidever2008,Chandra2023}, warping of the LMC disk \citep{Olsen2002,Choi2018,Oscar2025,Oden2025,Garver2026}, and extended stellar tidal debris in the LMC--SMC system \citep{Cullinane2022,Cullinane2023,Grady2021,Massana2024}. In addition, the lower mass SMC shows signs of significant perturbations in both its stellar and gaseous components (e.g., \citealt{DeLeo2020,Zivick2021,Niederhofer2021,Petia2021,Murray2024smc,Rathore2025smc}), with position-dependent line-of-sight distance variations ($\gtrsim$5 kpc) apparent throughout the main body of the galaxy (e.g., \citealt{Subramanian2012,Tatton2021,ElYoussoufi2021}). The eastern wing of the SMC (which is distinct from the gas-dominated inter-Cloud structure that constitutes the Magellanic Bridge) is a particularly salient feature that demonstrates a well-known distance bimodality driven by a population of stars $\sim$10 kpc in front of the SMC (e.g., \citealt{Nidever2013,Subramanian2017,ElYoussoufi2021,Omkumar2021}) that may originate from tidal material stripped from the inner SMC by recent interactions with the LMC \citep{Almeida2024,Garver2026}.

As a probable consequence of the interaction between the LMC and SMC, studies of the stellar mass assembly history of the system have revealed recent ($\lesssim$5 Gyr ago) global epochs of enhanced star formation \citep{HarrisZaritsky2004,HarrisZaritsky2009,Rubele2012,Rubele2015,Rubele2018,Cignoni2013,Weisz2013,Meschin2014,Monteagudo2018,Ruiz-Lara2020,Massana2022,Burhenne2025}. In particular, \citet{Massana2022} reported synchronous starbursts occurring between the LMC and SMC within the last $\sim$3.5 Gyr based on photometry of resolved stellar populations from the 
ground-based Survey of the MAgellanic Stellar History (SMASH; \citealt{Nidever2017}), where they interpreted the correlated star formation histories (SFHs) as evidence of repeated dynamical interactions between the MCs. 
Similarly, a systematic analysis of starbursts in the MCs based on data from the pure-parallel HST-based Scylla survey \citep{Murray2024,Cohen2024a,Cohen2024b} found evidence for 
a single coincident event $\sim$3.2 Gyr ago that occurred globally within the LMC and locally in the SMC (within its main body, but absent from some regions in the eastern SMC; \citealt{Burhenne2025}, hereafter B25). By comparing to orbital models for the MW-LMC-SMC interaction by \citet{Patel2020}, B25 concluded that, although the mutual interactions were capable of triggering starbursts in the MCs \added{(c.\@f.\@ \citealt{WilliamsonHugo2021})}, the measured starbursts did not clearly match dynamical events in the models such as pericentric passages between the LMC and SMC or crossing the virial radius of the MW. Moreover, if the measured starbursts shared the same dynamical trigger, the SMC responded earlier and with a greater enhancement in star formation by exhibiting a global starburst $\sim$2 Gyr before the $\sim$3 Gyr ago event in the LMC.

Regardless of their precise timing, the recent global starbursts in the MCs have impacted their chemical evolution. Early analyses based on chemical abundance measurements of relatively small samples of individual red giant branch (RGB) stars \citep{Pompeia2008,Lapenna2012,vanderSwaelmen2013,Mucciarelli2014} showed a slower efficiency of star formation in the MCs compared to the MW, but limited evidence for chemical signatures associated with starburst events (see \citealt{BekkiTsujimoto2012}). However, the thousands of measurements provided by the SDSS-III/IV Apache Point Observatory Galactic Evolution Experiment (APOGEE; \citealt{Majewski2017}) revealed a more complete picture of the chemical abundance patterns of the MCs. Both the LMC and SMC are characterized by an initially steep decline in the abundance of $\alpha$-elements (O, Ne, Mg, Si, S, Ar, Ca, Ti), which empirically track contributions from core-collapse supernovae (CCSNe), at low metallicity ([Fe/H]). This decline in $\alpha$-enhancement is due to the delayed onset of 
Type Ia supernovae (SNe Ia), which produce Fe at the exclusion of $\alpha$-elements, coupled with low star formation efficiency (\citealt{Nidever2020}, hereafter N20). The chemical abundance patterns of the MCs subsequently demonstrate an increase in [$\alpha$/Fe] of $\sim$0.1 dex between $-1.5 \lesssim {\rm [Fe/H]} \lesssim -0.5$, where these chemical signatures are reproducible with galactic chemical evolution models that include a major starburst event in each dwarf galaxy (N20; \citealt{Hasselquist2021}). During a starburst, enhanced star formation (SF) and therefore CCSNe production enriches the interstellar medium (ISM) with $\alpha$-elements, followed by a decline in [$\alpha$/Fe] as SNe Ia from stars formed at the burst onset inject Fe, but not $\alpha$-elements, into the ISM \citep{JohnsonWeinberg2020}.

In detail, \citet{Hasselquist2021}, hereafter H21, modeled the chemical evolution of the LMC and SMC as being characterized by a single dominant starburst in each galaxy triggered by an increase in star formation efficiency (SFE). 
From fitting median chemical abundance tracks in the [$\alpha$/Fe]--[Fe/H] plane, their models suggested that the LMC experienced stronger outflows, a higher SFE, and a stronger and later (by $\sim$3--4 Gyr) major starburst event compared to the SMC. Although the starbursts were not temporally coincident as implied by later works deriving SFHs from photometry (e.g., \citealt{Massana2022}; B25), 
H21 concluded that the epochs of increased SF in the MCs likely resulted from their mutual interactions, demonstrating that chemical evolution models offer a viable pathway for exploring the impact of starbursts on low-mass galaxy evolution.

Here, we further constrain the chemical evolution of the MCs in the context of their major starburst events by incorporating the first direct constraints from SFHs measured from photometry combined with $\gtrsim$6000 individual chemical abundance measurements of RGB stars from APOGEE (N20; \citealt{Povick2024}). In particular, we leverage the SFH precision provided by deep, space-based imaging from the Scylla survey based on 194 HST fields throughout both the LMC and SMC (\citealt{Cohen2024a,Cohen2024b}; hereafter C24a,b). We incorporate numerous improvements into the chemical evolution modeling procedure, such as corrections for selection biases in APOGEE in order to simultaneously predict abundance distribution functions alongside median chemical tracks in the [$\alpha$/Fe] versus [Fe/H] plane, while employing a flexible modeling framework designed to implement starbursts (\citealt{JohnsonWeinberg2020}, hereafter JW20). In addition, we take advantage of the extended coverage of both the Scylla and APOGEE surveys to explore spatial variations in the modeled starbursts across the LMC disk and in regions of the SMC such as its chemically distinct main body and wing (see \citealt{Almeida2024}). In combination with this work, the standardized framework for burst identification applied to Scylla SFHs by B25 enables straightforward comparisons between the spatial and temporal distributions of observed versus modeled starbursts in the MCs for the first time.

This paper is organized as follows. In Section~\ref{sec:data}, we present the data used in this work, including HST imaging analyzed as part of the Scylla survey (Section~\ref{sec:scylla}) and APOGEE chemical abundance measurements (Section~\ref{sec:apogee}). We also define spatial subregions in the MCs (Section~\ref{sec:region}). In Section~\ref{sec:results}, we provide a detailed overview of Scylla SFHs (Section~\ref{sec:csfh}) and APOGEE chemical abundance distributions (Section~\ref{sec:abund}) for the spatial subregions. We describe our chemical evolution modeling methodology in Section~\ref{sec:gce}. Specifically, we constrain global models for the chemical evolution of the MCs by Scylla SFHs in Section~\ref{sec:gce_scylla} and model the spatial subregions 
to predict starburst characteristics in Section~\ref{sec:gce_sfedriven}. In Section~\ref{sec:discuss}, we conclude by summarizing our findings and their implications for the evolution of the MCs (Section~\ref{sec:evo}) and the propagation of starbursts in interacting pairs of low-mass galaxies (Section~\ref{sec:burst}).

\section{Data} \label{sec:data}

\begin{figure*}
    \centering
    \includegraphics[width=\linewidth]{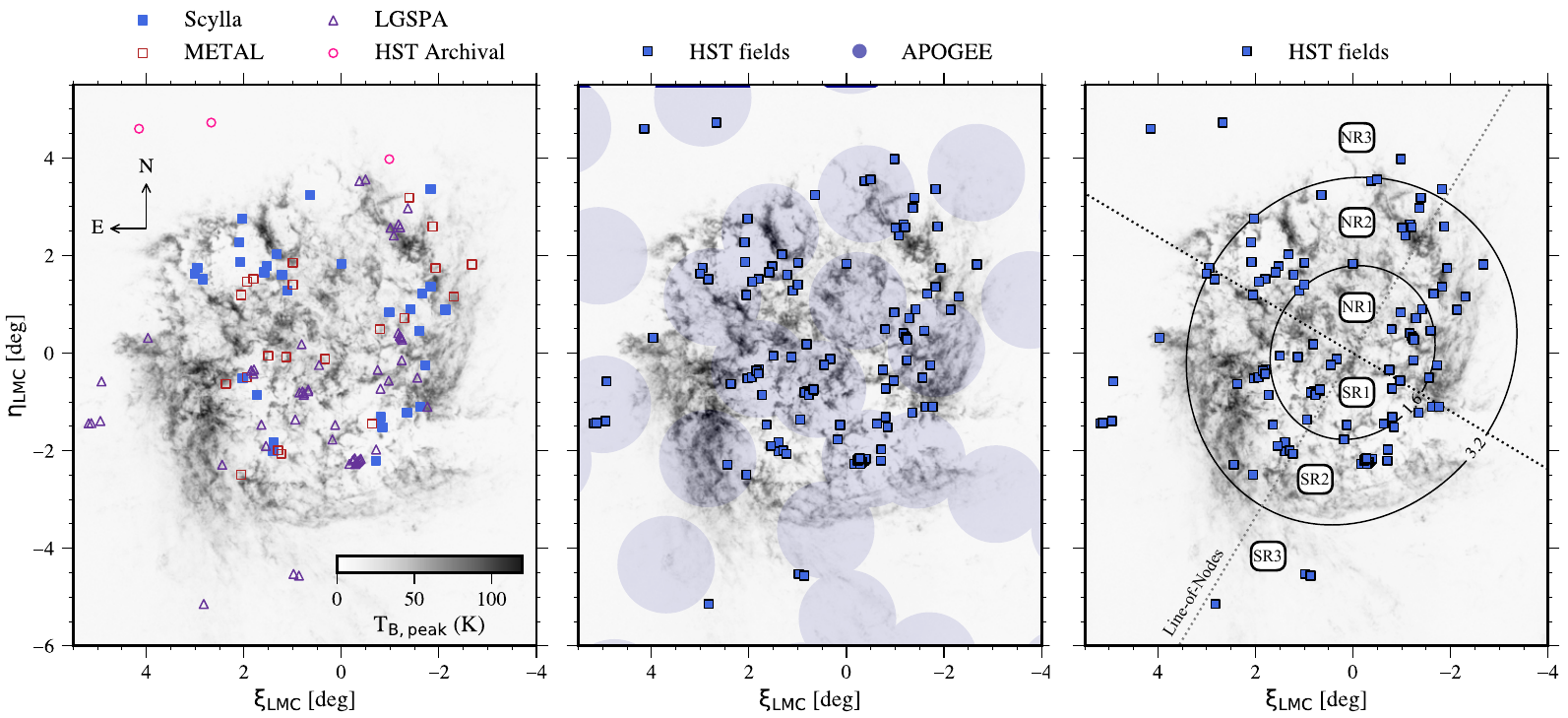}
    \caption{\textbf{HST and APOGEE field locations in the LMC.} Maps of the peak brightness temperature ($T_{B, {\rm peak}}$) of 21 cm emission along the line-of-sight to the LMC measured by the ATCA \citep{Kim1999}, overlaid with the locations of HST and APOGEE fields (Section~\ref{sec:data}). The maps are in tangent-plane coordinates, assuming an LMC center of ($\alpha, \delta$) = (\lmccenterHIra$^\circ$,\lmccenterHIdec$^\circ$), the dynamical center of the HI disk \citep{LuksRohlfs1992,Kim1999}. (Left panel) HST field sample from C24a, which includes WFC3 imaging from Scylla (blue squares; \citealt{Murray2024}) and METALS (red open squares; \citealt{RomanDuval2019}), WFPC2 imaging from the LGSPA (purple open triangles; \citealt{Holtzman2006}), and other archival HST ACS and WFC3 imaging (pink open circles). (Middle panel) HST field sample (black-outlined blue squares) compared to the footprint of 2$^\circ$ diameter APOGEE fields targeting the LMC (blue shaded circles; N20; \citealt{Povick2024}). (Right panel) HST fields compared to spatial regions defined in Section~\ref{sec:region_lmc} based on de-projected in-plane radial distances and location in the northern versus southern halves of the LMC disk. Black contours delineating radial zones correspond to $R_{\rm inv, LMC}$/2 = 1.6 kpc and $R_{\rm inv, LMC}$ = 3.2 kpc, where $R_{\rm inv, LMC}$ is the inversion radius (C24a). We also show the adopted line-of-nodes PA, $\theta \sim 149$ deg E of N \citep{Choi2018} used to define the in-plane LMC coordinate system and its orthogonal line of maximum line-of-sight depth.
    }
    \label{fig:lmc}
\end{figure*}

\begin{figure*}
    \centering
    \includegraphics[width=0.8\linewidth]{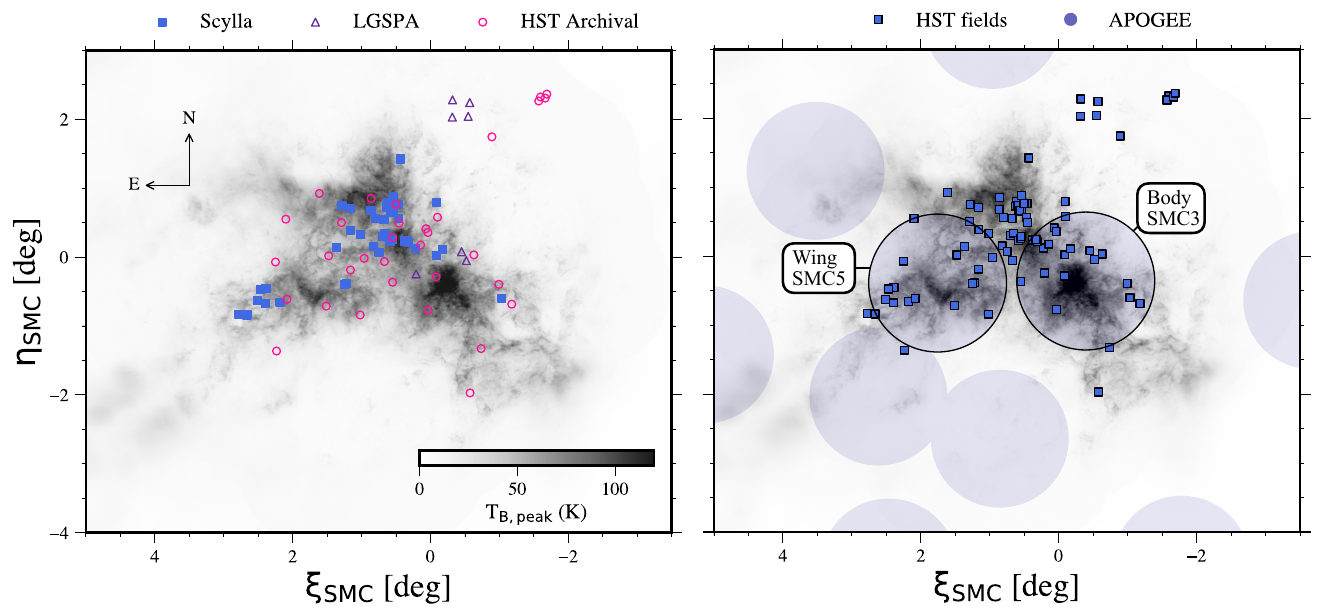}
    \caption{\textbf{HST and APOGEE field locations in the SMC.} Similar to Figure~\ref{fig:lmc}, except for the SMC (HI maps from the GASKAP survey; \citealt{Pingel2022}). We assumed an optical center for the SMC of ($\alpha, \delta$) = (\smccenterphotra$^\circ$,\smccenterphotdec$^\circ$) \citep{Crowl2001,Subramanian2012}. (Left panel) HST field sample from C24b, which includes WFC3 imaging from Scylla (blue squares; \citealt{Murray2024}), archival WFPC2 LGSPA imaging \citep{Holtzman2006}, and other archival HST ACS and WFC3 imaging (pink open circles). (Right panel) HST field sample (black-outlined blue squares) compared to APOGEE fields targeting the SMC (blue shaded circles; \citealt{Nidever2020,Povick2024}), where \added{the APOGEE fields SMC3 and SMC5} that overlap \added{with the HST fields} designated as the SMC ``body'' and ``wing'' \added{respectively} are highlighted (Section~\ref{sec:region_smc}).
    }
    \label{fig:smc}
\end{figure*}

In this work, we combined an expanded set of HST pointings built on the Scylla survey (Section~\ref{sec:scylla}) with chemical abundance measurements for thousands of 
RGB stars from APOGEE (Section~\ref{sec:apogee}). We show the spatial distributions 
of the HST and APOGEE fields in the MCs, overlaid on maps of the peak brightness temperature ($T_{B, \rm peak}$) of 21 cm HI emission in each galaxy \citep{Kim1999,Pingel2022}, in Figures~\ref{fig:lmc} and~\ref{fig:smc}.

Throughout this work, we assume an LMC center of ($\alpha, \delta$) = (\lmccenterHIra$^\circ$, \lmccenterHIdec$^\circ$), which corresponds to the dynamical center of its HI disk \citep{LuksRohlfs1992,Kim1999}, as well as its stellar kinematical center \citep{vanderMarel2014}. For the SMC, we adopt its optical center of ($\alpha, \delta$) = (\smccenterphotra$^\circ$, \smccenterphotdec$^\circ$) \citep{Crowl2001, Subramanian2012}. For reference, the right panels of Figures~\ref{fig:lmc} and~\ref{fig:smc} also show the spatial regions used in our analysis 
to compare SFHs inferred from the chemical abundance distributions to those measured from color-magnitude diagrams (CMDs) in the MCs (Section~\ref{sec:sfedriven_lmc},~\ref{sec:sfedriven_smc}). 
We motivate the choice of these regions based on SFH characteristics, and further discuss the adopted coordinate systems in the MCs, in Section~\ref{sec:region}.

\subsection{Scylla} \label{sec:scylla}

The Scylla survey is a pure-parallel complement to the Ultraviolet Legacy Library of Young Stars as Essential Standards (ULLYSES; \citealt{RomanDuval2020}) spectroscopic survey of massive stars. Scylla obtained multi-band HST Wide Field Camera (WFC3) imaging of \nfieldsscyllatot\ fields in the LMC and SMC for up to seven filters between ultraviolet (F225W) and near-infrared (F160W) wavelengths, producing homogeneous F475W and F814W optical photometry across all pointings \citep{Murray2024}. In this work, we used \nfieldscohenlmc\ and \nfieldscohensmc\ HST fields across the LMC and SMC, respectively, based on an expanded Scylla sample combined with archival imaging from C24a,b.
The majority ($\gtrsim$85\%) of fields, 
including \textit{all} Scylla fields, are 80\% complete to at least 1 mag faintward of the oldest main sequence turn-off (oMSTO; C24a,b).
We summarize the relevant fields here, but refer the reader to C24a and C24b for further detail on HST imaging in the LMC and SMC, respectively, and \citet{Murray2024} for additional information on the Scylla survey science goals and observing strategy.

In the LMC, C24a supplemented the Scylla fields with additional parallel HST WFC3 (F475W, F814W) imaging from the Metal Evolution Transport and Abundance in the Large Magellanic Cloud program (METAL; \citealt{RomanDuval2019}). 
C24a also included 
archival HST (F555W, F814W) imaging with Wide Field Planetary Camera 2 (WFPC2) of sufficient photometric depth (faintward of the oMSTO) from the Local Group Stellar Photometry Archive (LGSPA; \citealt{Holtzman2006}, with a subset analyzed by \citealt{Weisz2013}). 
The sample also includes archival fields imaged with two or more optical and/or near-infrared filters of sufficient depth in the outer LMC disk ($R_{\rm LMC} \gtrsim 5$ kpc; C24a). In the SMC, C24b 
included \nfieldscohensmcarchival\ archival HST fields of sufficient depth in optical and/or near-infrared filters, incorporating WFPC2 fields provided in the LGSPA.

The data reduction procedure and point spread function (PSF) photometry for Scylla imaging are described in detail by \citet{Murray2024}, with similar methods applied to the supplemental archival HST imaging for the LMC and SMC by C24a,b. In summary, 
\texttt{DOLPHOT}\footnote{\url{http://americano.dolphinsim.com/dolphot/}} \citep{Dolphin2000,Dolphin2016} was used to preprocess the reduced HST WFC3 images and subsequently model PSFs of individual sources, customized to each filter, in each image. \texttt{HSTphot} \citep{Dolphin2000} was used to produce the archival stellar catalogs based on WFPC2 imaging.
The photometric quality flags set by \citet{Williams2014,Williams2021} 
were adopted to produce the final vetted stellar catalogs for all fields in the Vega photometric system. 

\subsubsection{Measuring Star Formation Histories}
\label{sec:sfh}

C24a,b 
modeled the observed 
CMDs constructed from the per-field stellar catalogs using linear combinations of single stellar populations 
generated by \texttt{MATCH} \citep{Dolphin2002} assuming PARSEC stellar evolutionary models \citep{Bressan2012} and a \citet{Kroupa2001} initial mass function (IMF).\footnote{Scylla SFHs were also measured using MIST \citep{Dotter2016,Choi2016} and BaSTI18 \citep{Hidalgo2018} models to check the dependence of the results on the assumed isochrones (see Appendix C of C24a,b for further discussion).}
\texttt{MATCH} also predicts age-metallicity relations (AMRs) from SFH fitting, where C24a,b demonstrated that the AMRs for the LMC and SMC are in good agreement with independent age-metallicity determinations informed by resolved stellar spectroscopy \citep{Carrera2008lmc,Carrera2008smc,Povick2024} and from star clusters \citep{Perren2017,Maia2019,Dias2021,Dias2022,Oliveira2023}.
During 
SFH fitting, 
\texttt{MATCH} was allowed to explore a grid in distance modulus ($m-M$)$_0$, foreground extinction $A_V$, and internal differential extinction $\delta A_V$ 
(c.\@f.\ \citealt{Lewis2015,Lazzarini2022}).\footnote{
C24a successfully recover per-field distances and extinctions in the LMC compared to independent observational constraints \citep{Choi2018,Skowron2021}, except for fields with high internal dust attenuation excluded from the sample presented in Section~\ref{sec:scylla}. 
The impact of the SMC's complex distance distribution on its SFH recovery is negligible within the total uncertainties (see Appendix B of C24b).}
\texttt{MATCH} 
convolves the model CMDs with observational uncertainties and photometric incompleteness and bias, which are determined as a function of color, magnitude, and spatial location from artificial star tests, with the observed CMDs by computing a Poisson maximum likelihood statistic. 
Statistical uncertainties on the best-fit SFHs are computed with a hybrid Monte Carlo approach \citep{Dolphin2013}, whereas systematic uncertainties were calculated by refitting the observed CMD after perturbing the effective temperature and bolometric luminosity in 50 Monte Carlo trials \citep{Dolphin2012}. 
We used \texttt{MATCH} to combine best-fit SFHs 
measured from individual HST fields
according to the spatial groupings defined in Section~\ref{sec:region}, thereby enforcing proper statistical handling of the random and systematic uncertainties in the combination process.

\subsubsection{Generating Fake Color-Magnitude Diagrams}
\label{sec:fake}

\begin{figure}
    \centering
    \includegraphics[width=\linewidth]{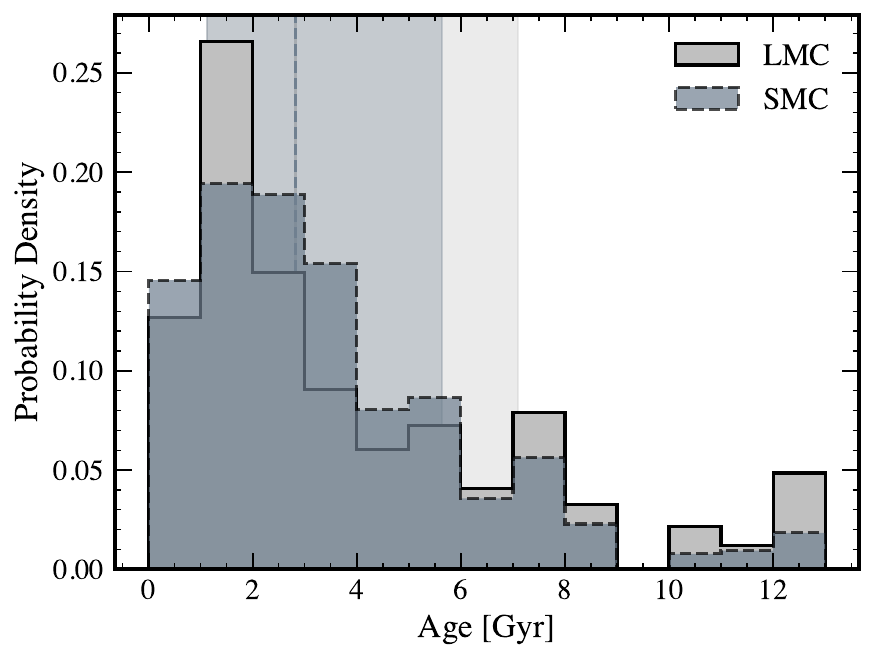}
    \caption{\textbf{Age distributions predicted for APOGEE RGB populations in the L/SMC from Scylla SFHs.} The ages are estimated using \texttt{MATCH} to generate synthetic JHK photometry from global Scylla SFHs for the L/SMC (Section~\ref{sec:fake}) and applying the APOGEE RGB selection from Section~\ref{sec:apogee}. Histograms assume bin sizes of 1 Gyr. The solid (dashed) vertical line and shaded vertical band represent the median and 1$\sigma$ ranges for the estimated age distribution of the LMC (SMC). Ages younger than 1 Gyr are present in the histograms due to minor contamination in the RGB selection region from red super giants. The median ages and 1$\sigma$ ranges for the LMC (\taurgb\ = \agergblmc) and SMC (\taurgb\ = \agergbsmc) are consistent with each other.
    }
    \label{fig:fake}
\end{figure}

To estimate the age distribution of RGB stars in the APOGEE sample (Section~\ref{sec:apogee}), we additionally used \texttt{MATCH} to generate synthetic JHK photometry from PARSEC stellar evolutionary models and input best-fit Scylla SFHs measured globally for the L/SMC (C24a,b). For these simulations, we assumed fixed distances to both the L/SMC of 49.9 kpc and 62 kpc \citep{deGrijs2014,deGrijsBono2015} respectively. 
We assumed fixed values for the input foreground and internal differential extinction of $A_V$ = 0.25 mag and $\delta A_V$ = 0.29 mag for the LMC (C24a), and  $A_V$ = 0.1 mag and $\delta A_V$ = 0.1 mag for the SMC \citep{SchlaflyFinkbeiner2011,Skowron2021}. 
By applying the RGB selection from Section~\ref{sec:apogee} to the synthetic ($J-H,K$) CMDs, we calculated the predicted median stellar age \added{and 1$\sigma$ age range} of APOGEE RGB stars ($\tau_{\rm RGB}$) from the 16$^{\rm th}$, 50$^{\rm th}$, and $84^{\rm th}$ percentiles of the simulated age distributions (\added{Figure~\ref{fig:fake}}).
For the LMC (SMC), we estimated $\tau_{\rm RGB}$ = \agergblmc\ (\agergbsmc) Gyr. 
In comparison, the predicted median stellar age \added{and 1$\sigma$ age range} of \textit{all} stellar populations in the LMC (SMC) is \agelmc\ (\agesmc) Gyr. \added{Within both the LMC and SMC, there are minor differences in the predicted median stellar ages between spatial subregions (Section~\ref{sec:region}) that follow the age trends probed by lifetime SFH metrics (Appendix~\ref{sec:csfh}, Table~\ref{tab:sfh_abund}), but these differences are not significant when considering the 1$\sigma$ age ranges for each spatial subregion.
}

\added{Figure~\ref{fig:fake} also shows that the simulated age distributions contain stars younger than 1 Gyr that fall outside the 1$\sigma$ age ranges for the L/SMC due to minor contamination at bluer colors in the APOGEE RGB selection region (Section~\ref{sec:apogee}) from faint, metal-rich red super giant stars (see N20 and H21). However, we chose to select RGB stars over a broad color range to avoid biasing the metallicity distributions of the L/SMC against blue and metal-poor RGB stars, and for simplicity in characterizing the APOGEE selection function (Section~\ref{sec:apogee}).
}

\subsection{APOGEE} \label{sec:apogee}

We used spectroscopic data from the second phase of 
APOGEE \citep{Majewski2017}, which includes data from the Southern Hemisphere  \citep{Blanton2017} as part of the seventeenth data release of SDSS-IV (DR17; \citealt{sdss4dr17}).  The targeting method and observing plan for APOGEE-2S is described by \citet{Zasowski2017} and \citet{Santana2021}.
The APOGEE-2S near-infrared fiber-fed spectrograph \citep{Wilson2019} operates from the 2.5 m du Pont Telescope at Las Camapanas Observatory, producing R$\sim$22,5000 spectra for up to 300 targets in the H-band (1.51--1.70 $\mu$m) over a 2 deg diameter field-of-view. In particular, we use data from the APOGEE MC survey 
(N20, with a DR17 expansion of the survey 
by \citealt{Povick2024}), which provides broad radial and azimuthal coverage out to $\sim10^\circ$ in the LMC and $\sim5^\circ$ in the SMC (Figures~\ref{fig:lmc} and~\ref{fig:smc}).

The stellar spectra were reduced with the APOGEE data reduction pipeline (\citealt{Nidever2015}; Holtzmann et al., in preparation) and radial velocities were measured using \texttt{DOPPLER} \citep{doppler}. Stellar parameters and chemical abundances were derived using the APOGEE Stellar Parameter and Chemical Abundance (ASPCAP; \citealt{Holtzman2015,GarciaPerez2016}) pipeline, which compares continuum-normalized observed spectra against \texttt{SYNSPEC} synthetic spectra \citep{synspec} using the least-squares minimization package \texttt{FERRE} \citep{AllendePrieto2006}. From the best-fit synthetic spectrum, ASPCAP determines global stellar parameters and bulk chemical abundances (e.g., $T_{\rm eff}$, $\log g$, [M/H], [$\alpha$/M]) prior to fitting spectral ``windows'' sensitive to variations in individual elements to determine their abundances (e.g., O, Mg, Si, S, Ca, Ti, Fe). The bulk [$\alpha$/M] ratio from APOGEE includes the aforementioned $\alpha$-elements, 
which are incorporated into the $\alpha$-abundance of the MARCS model atmospheres \citep{Gustafsson2008,Jonsson2020} used to generate synthetic spectra.
We adopt Mg as a representative $\alpha$-element, 
where Mg empirically tracks contributions from CCSNe with negligible contributions to its yields 
from SNe Ia (e.g., \citealt{Weinberg2019}).

We selected RGB stars from the APOGEE MC survey using similar criteria to N20.  We sourced data from 
deep APOGEE fields targeting the LMC and SMC, including fields targeting globular clusters along the line-of-sight to the Clouds (47Tuc and NGC362) and an HII region in the LMC (30Dor). We identified RGB candidates from their Two Micron All Sky Survey (2MASS)  photometry, where $0.55 < J - K_s < 1.3$ and $16 < H < 12.35$ ($16 < H < 12.9$) for the LMC (SMC). We refined our RGB sample by restricting it to giant stars with $T_{\rm eff} < 5200\ K$ and $\log g < 3.4$, as defined using ASPCAP stellar parameters. Furthermore, we required that all RGB stars had reliable ASPCAP stellar parameters, following the updated DR17 quality flags for the APOGEE MC survey from \citet{Povick2024}: \texttt{TEFF\_BAD}, \texttt{LOGG\_BAD}, \texttt{VMICRO\_BAD}, \texttt{M\_H\_BAD}, \texttt{ALPHA\_M\_BAD}, \texttt{C\_M\_BAD}, \texttt{N\_M\_BAD}, and \texttt{NO\_ASPCAP\_RESULT}. 
We adopted a signal-to-noise ratio threshold of S/N $>$ 40, 
which is sufficient for accurate, albeit less precise, detailed chemical abundance measurements while 
maximizing the sample size. The median chemical abundance uncertainties are $\delta$([Fe/H]) = \medfeherr\ and 
$\delta$([Mg/Fe]) = \medmgerr\ 
for the RGB sample. \added{The APOGEE abundance uncertainties from \texttt{FERRE} are underestimated, thus the reported DR17 abundance uncertainties (adopted in this work) are empirically derived from repeat observations, although metal-poor stars in APOGEE still suffer from  systematic errors \citep{Mead2024}.
}

We classified RGB stars as MC members using their radial velocities and \textit{Gaia} eDR3 proper motions \citep{gaiaedr3}, which are included in the DR17 catalog. Following N20, we adopted the $\pm$2.5$\sigma$ thresholds on the radial velocity distributions, corresponding to 160--348 \kms\ (71--220 \kms) for the LMC (SMC), taking into account radial velocity errors. We also applied proper motion cuts based on the ellipses adopted by N20: $\langle \mu_{\rm RA} \rangle = 1.82$ mas yr$^{-1}$, $\langle \mu_{\rm DEC} \rangle = 0.26$ mas yr$^{-1}$, a = 1.5 mas yr$^{-1}$, b = 0.8 mas yr$^{-1}$, $\theta$ = 100$^\circ$ for the LMC, and $\langle \mu_{\rm RA} \rangle = 0.78$ mas yr$^{-1}$, $\langle \mu_{\rm DEC} \rangle = -1.21$ mas yr$^{-1}$, a = 0.9 mas yr$^{-1}$, b = 0.6 mas yr$^{-1}$, $\theta$ = 0$^\circ$ for the SMC. The final sample consists of \nsamplmc\ (\nsampsmc) RGB member stars with reliable APOGEE chemical abundances in the LMC (SMC).

In addition, we calculated an approximate APOGEE selection function from the ratio of the number of all observed targets with S/N $>$ 40 and the number of all candidate objects in the 2MASS source catalog, both within the adopted ($J-K_s, H$) RGB selection box, for each spectroscopic field (e.g., \citealt{Bovy2014,Mackereth2020}). The selection function therefore takes the form of a 2D piecewise constant function of CMD position for a given APOGEE field. We account for the selection function in the chemical evolution modeling procedure described in Section~\ref{sec:gce}, \added{where the net effect of the selection function is to increase the likelihood that a metal-poor star in the less dense, more complete stellar outskirts of the MCs is observed relative to a metal-rich star in the denser, less complete central regions of the galaxies by a factor of $\sim$3--5. However, the impact of the APOGEE selection function on the underlying metallicity distribution functions (MDFs) is expected to be minor (see Appendix~A of \citealt{Hayden2015} for a discussion in the case of the MW).
}

\subsection{Defining Spatial Subregions} \label{sec:region}

\begin{figure*}
    \centering
    \includegraphics[width=0.8\linewidth]{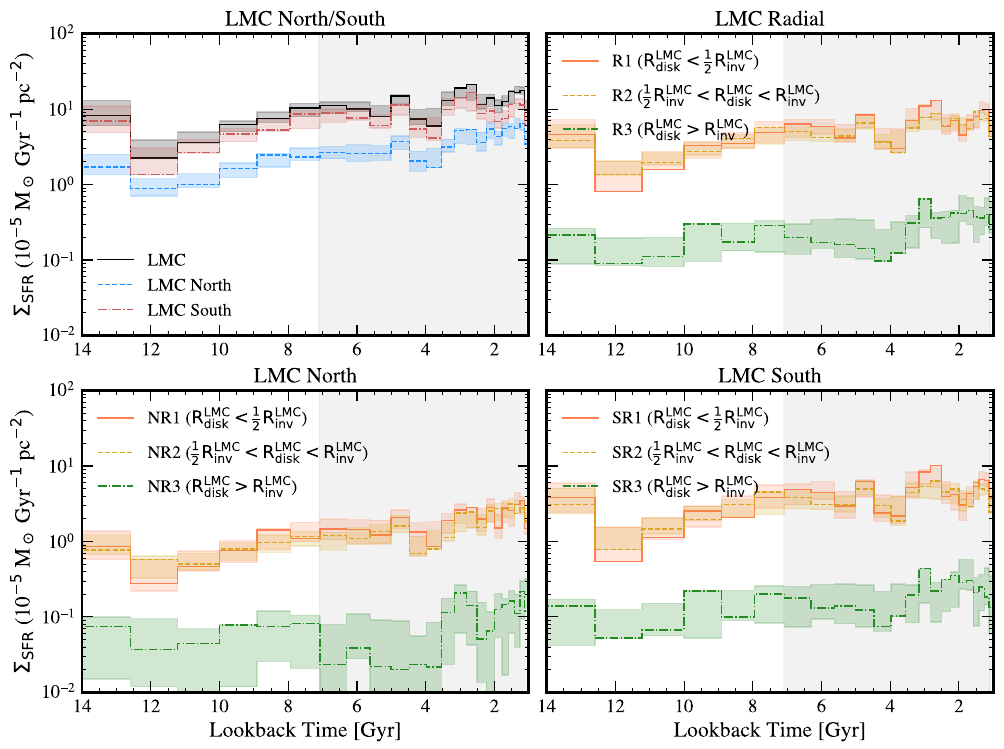}
    \caption{\textbf{Scylla area-normalized SFRs in the LMC.} Best-fit area-normalized SFRs as a function of lookback time derived from HST CMDs using PARSEC stellar evolutionary models (Section~\ref{sec:sfh}). Shaded colored regions represent 1$\sigma$ uncertainties, including both systematic and random uncertainty contributions.  The shaded grey vertical region in each panel represents the 68\% \added{percentile range} on the age distribution of APOGEE RGB stars estimated from simulations with the global LMC SFH 
    as input (\taurgb\ = \agergblmc\ Gyr; Section~\ref{sec:fake}). We show lookback times $\gtrsim$1 Gyr, where $\sim$1 Gyr corresponds to the age of the youngest RGB stars. (Top panels) We show SFRs along north/south (left) and radial (right) spatial divisions in the LMC (defined by the radius at which the LMC's age-radius relation inverts, \rinvlmc\ = \rinvlmckpc\ kpc; C24a). 
    The bottom panels show further radial subdivisions of the northern (left) and southern (right) halves of the LMC disk. The LMC experienced a global burst of star formation $\sim$3.2 Gyr ago (B25), which coincides with \taurgb. The peak SFR enhancement during the burst is fairly uniform across the LMC 
    (Section~\ref{sec:csfh}).
    }
    \label{fig:sfr_lmc}
\end{figure*}

We divided the LMC and SMC into subregions based on known spatial variations in their SFH characteristics (Section~\ref{sec:intro}).  These spatial subregions provide additional tests, alongside considering each galaxy as a whole, of whether 
differences in starburst properties are captured in the chemical abundance distributions of 
RGB stars (Section~\ref{sec:gce_sfedriven}). We detail our subregion definitions, which are shown in the right panels of Figures~\ref{fig:lmc} and~\ref{fig:smc}, in Sections~\ref{sec:region_lmc} and~\ref{sec:region_smc} for the LMC and SMC respectively. 

\subsubsection{LMC Subregions} \label{sec:region_lmc}

Recent radial age gradient measurements from Scylla, enabled by the spatial resolution and SFH precision of the survey, have demonstrated that the LMC has an inside-out gradient in its inner disk, prior to inverting beyond a radius of \rinvlmc\ $\approx$ \rinvlmckpc\ kpc (C24a).
We thus separated the LMC into radial subregions defined relative to its measured inversion radius \rinvlmc\ and azimuthal spatial regions defined relative to the 
tangent-plane projection of the LMC on the sky.
To calculate the in-plane de-projected radial distance from the center of the LMC (\rdisklmc), we assumed an LMC center based on HI kinematics (Section~\ref{sec:data}). In contrast to assuming a photometric center for the LMC offset from its kinematical center (($\alpha, \delta$) = (\lmccenterphotra$^\circ$, \lmccenterphotdec$^\circ$); \citealt{vanderMarel2001}), adopting a dynamical center enables direct comparison between radial stellar and gas properties, and minimizes scatter in radial trends with HI column density and lifetime SFH metrics (C24a). 
We assumed $\theta =$ \thetalmc\ deg E of N for the line-of-nodes position angle (PA) of the LMC (Figure~\ref{fig:lmc}) and $i$ = \incllmc$^\circ$ for the inclination of the LMC disk plane \citep{Choi2018}, and a line-of-sight distance to the \textit{optical} center of the LMC of 49.9 kpc \citep{deGrijs2014}. 

Although the inside-out radial age gradient in the inner LMC disk (\rdisklmc\ $<$ \rinvlmc) is statistically significant, particularly for more recent lookback times ($\lesssim$2.5 Gyr), it is not formally well-described by a linear relationship due to instrinsic field-to-field SFH variations (C24a). Thus, we divided the inner LMC disk into two spatial regions delineated by \rinvlmc/2 $\approx$ 1.6 kpc, with the intention of reflecting differences in mass assembly history between the innermost (\rdisklmc\ $<$ \rinvlmc/2, hereafter R1) and intermediate radii (\rinvlmc/2 $<$ \rdisklmc\ $<$ \rinvlmc, hereafter R2) in the LMC disk, in addition to a single region for the outer LMC disk defined by \rdisklmc\ $>$ \rinvlmc\ (hereafter R3). Furthermore, C24a found that HST fields in the northern LMC disk ($-90 <$ PA $< 90$ deg E of N) exhibited a stronger linear correlation between age and radius, or less intrinsic SFH variation, across all lifetime SFH metrics compared to the southern LMC disk.

We thus defined northern (N) and southern (S) spatial regions in the LMC
using a physically motivated coordinate system in the LMC disk plane, where the disk is divided by the line of maximum line-of-sight depth in the LMC orthogonal to the line-of-nodes PA computed by \citet{Choi2018} (Figure~\ref{fig:lmc}). For context, NR2 contains the LMC's single interaction-induced spiral arm \citep{Besla2016,Pardy2016,Ruiz-Lara2020},
SR1 contains the titled dynamically offset stellar bar \citep{Choi2018,Rathore2025obs,Rathore2025}, and SR3 contains the southeastern (SE) disk (C24a).

We also placed restrictions on PA for the spatial selection of RGB stars from APOGEE to ensure that they cover a similar area compared to the HST fields. Although the HST fields span nearly the full PA range for the inner LMC disk (\rdisklmc\ $<$ \rinvlmc; right panel of Figure~\ref{fig:lmc}), HST fields in the outer disk (\rdisklmc\ $>$ \rinvlmc)  have a sparser spatial distribution in azimuth, particularly when compared to the APOGEE fields (middle panel of Figure~\ref{fig:lmc}). For the outer LMC disk, we thus limited our RGB sample to the conjunction of angular regions defined by PA $>$ 50 or PA $< -150$ deg E of N, and PA $> -65$ and PA $<$ 40 deg E of N, in addition to the maximum in-plane radial extent of the HST fields (\rdisklmc\ $<$ 7 kpc).

\subsubsection{SMC Subregions} \label{sec:region_smc}

\begin{figure}
    \centering
    \includegraphics[width=\columnwidth]{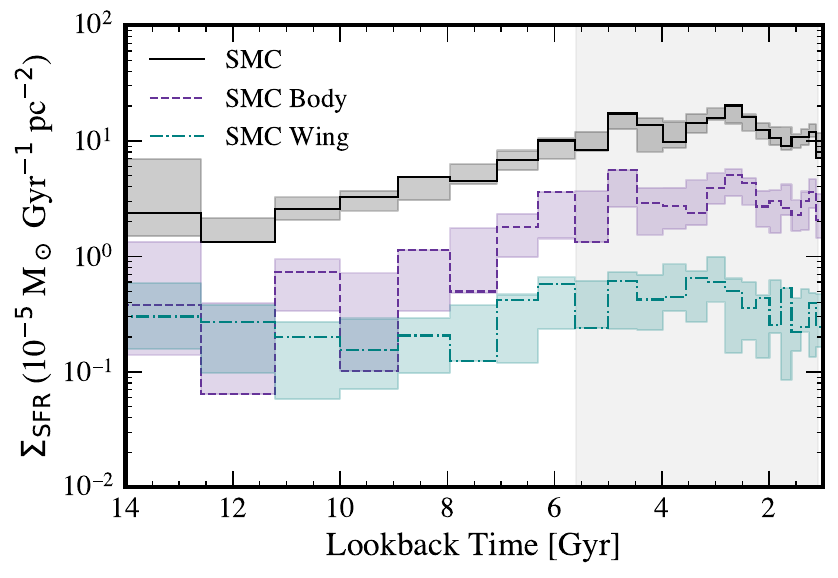}
    \caption{\textbf{Scylla area-normalized SFRs in the SMC.} Similar to Figure~\ref{fig:sfr_lmc}, except for the SMC. The 68\% \added{percentile range} on the stellar age distribution of APOGEE RGB stars in the SMC inferred from the global SFH (shaded grey vertical region) is \taurgb\ = \agergbsmc\ Gyr (Section~\ref{sec:fake}). The main body of the SMC shows a starburst beginning $\sim$5 Gyr ago, with a localized burst occurring $\sim$3 Gyr ago (B25), 
    \added{both of which} coincide with \taurgb\ in the SMC. In contrast, the SMC wing shows a short $\sim$4 Gyr ago burst followed by near-constant SF over the last $\sim$3 Gyr. Each burst shows an SFR enhancement of a factor of $\sim$3 (Section~\ref{sec:csfh}).
    }
    \label{fig:sfr_smc}
\end{figure}

Taking into account the relatively limited spatial sampling of the APOGEE fields in the SMC compared to the LMC (right panel of Figure~\ref{fig:smc}), we adopted a simple spatial division in the SMC corresponding to the overlap of the HST and APOGEE fields. We designate the spatial footprint of the APOGEE fields SMC5 and SMC3, respectively, as representing the ``wing'' and ``body'' of the SMC.
Although SMC5, our wing field, samples its inner population, the relative contribution of the foreground and inner SMC populations exists on a continuum in the direction of the wing, and remains present in SMC5 (\citealt{Almeida2024}; see also Figure~\ref{fig:abund_smc}).


\section{Star Formation Histories and Chemical Abundance Distributions} \label{sec:results}

Here, we provide an overview of the starburst characteristics of the MCs as measured from Scylla (C24a,b; B25)
and their implications for the associated chemical abundance distributions of each spatial subregion (Section~\ref{sec:csfh}). 
For completeness, we present cumulative SFHs 
for the MCs, including lifetime SFH and photometric metallicity metrics, in Appendix~\ref{sec:appendix}.
We summarize the characteristics of the APOGEE chemical abundance distributions for each spatial subregion prior to evaluating whether the chemical abundance trends align with expectations from the Scylla SFHs in Section~\ref{sec:abund}. We further analyze the relationship between SFHs and chemical abundances in the MCs in the context of galactic chemical evolution (GCE) modeling in Section~\ref{sec:gce}.

\subsection{Scylla Starbursts}
\label{sec:csfh}

\begin{figure*}
    \centering
    \includegraphics[width=\linewidth]{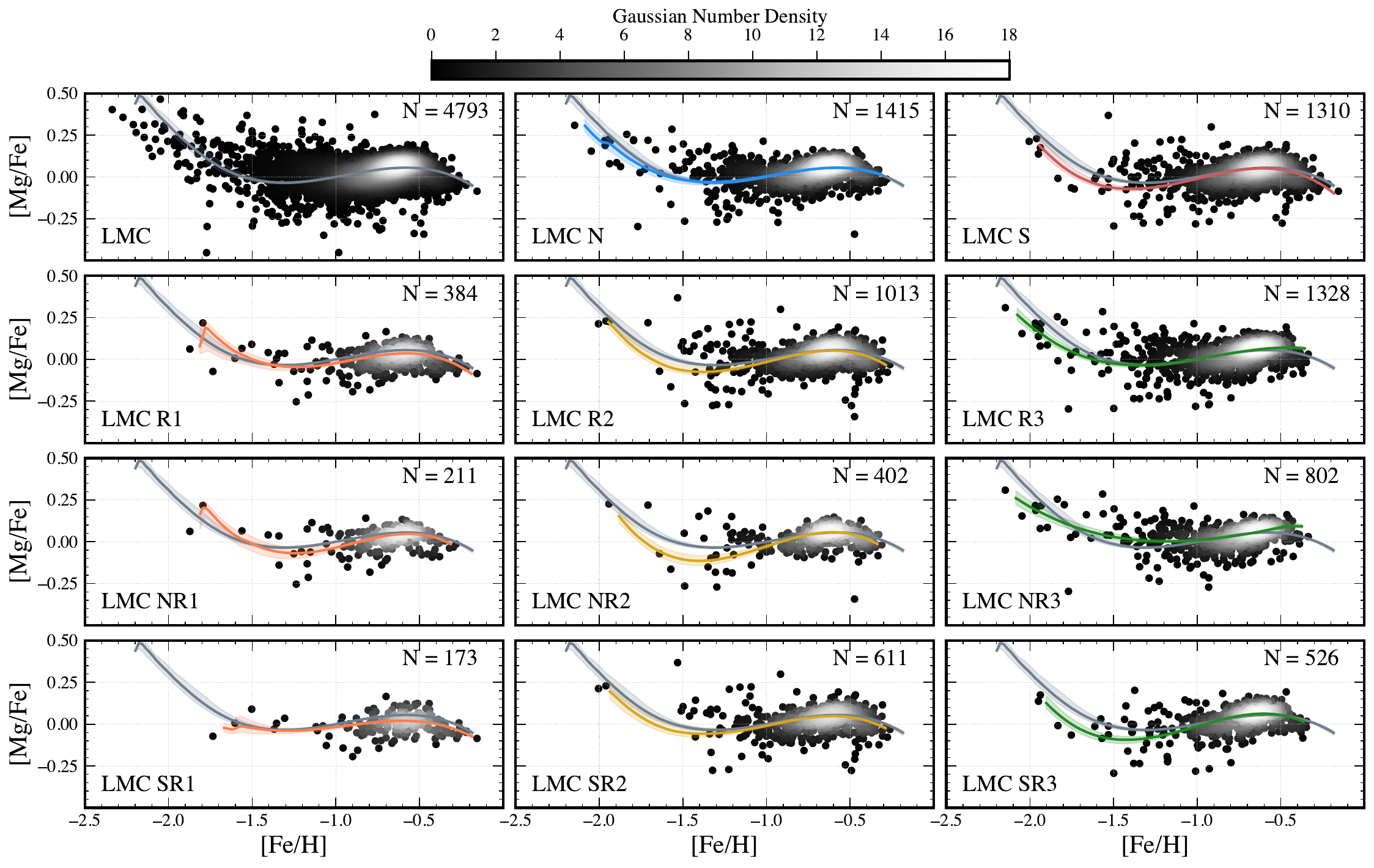}
    \caption{\textbf{[Mg/Fe] vs.\ [Fe/H] chemical abundance trends in the LMC.}  We show 2D chemical abundance distributions for all LMC stars, as well as for north/south and radial subregions in the LMC 
    (Section~\ref{sec:region_lmc}). For ease of visualization, each RGB star is color-coded by an estimate of its neighboring density in chemical abundance space using a Gaussian kernel. 
    We show BSpline fits to the chemical abundance distributions (Section~\ref{sec:abund}) of the LMC (solid \added{grey} lines) and its spatial subregions (\added{colored} solid lines) and their uncertainties (shaded regions) in each panel to estimate 2D chemical abundance trends, and to visually compare subregions. The increase in [Mg/Fe] at [Fe/H] $\sim-1.3$
    across all subregions is driven by a starburst in the LMC (N20; H21), with a secondary decline in [Mg/Fe] at [Fe/H] $\sim-0.6$ seen in all subregions except the outer LMC (R3, or \rdisklmc\ $>$ \rinvlmc). This secondary decline is more prominent in the southern LMC (SR3). The 
    trendlines show more variation when divided into radial zones in each half of the disk (bottom two rows; see discussion in Section~\ref{sec:abund}).
    }
    \label{fig:abund_lmc}
\end{figure*}

Figures~\ref{fig:sfr_lmc} and~\ref{fig:sfr_smc} show the area-normalized SFRs measured from Scylla ($\Sigma_{\rm SFR}$) for the LMC, SMC, and their spatial subregions, which we constructed by statistically combining the best-fit SFHs for each HST field in a given subsample using \texttt{MATCH}, assuming PARSEC stellar evolutionary models (Section~\ref{sec:sfh}).
For reference, Figures~\ref{fig:sfr_lmc} and~\ref{fig:sfr_smc} also show 1$\sigma$ \added{ranges} corresponding to the age distribution estimated for APOGEE RGB stars based on simulated stellar populations using the global SFH for each galaxy, which span $\sim$4.5--1 Gyr ago for both the LMC and SMC (Section~\ref{sec:fake}). These age ranges therefore set a base assumption for SF 
properties 
most pertinent to APOGEE RGB stars.


\textit{LMC:} Figure~\ref{fig:sfr_lmc} shows that 
the predicted median stellar age of APOGEE RGB stars traces the global starburst in the LMC occurring $\sim$3.2 Gyr ago, in which the SFR increased by a factor of 
$\gtrsim$2 relative to the mean lifetime SFR of the galaxy (B25). The duration ($\sim$1-2 Gyr) and strength ($\times$2-3 above the mean) of the $\sim$3 Gyr starburst is fairly uniform across all LMC subregions. 
The relative intensity of the starburst at intermediate radii (R2) appears to be the weakest ($\times$2 versus $\times$3 in other radial regions). 
In addition, the relative burst intensity may be stronger in the outer northern LMC (NR3), as well as the inner southern LMC (SR1), which contains the LMC bar, and the outer southern LMC (SR3). Indeed, B25 found that the burst intensity is strongest in the SE disk (followed by the northwestern portion of the LMC; see also \citealt{Massana2022}).


Thus, 
signatures of the global starburst should be encoded in the chemical abundance distributions of the RGB stars. 
Moreover, the APOGEE chemical abundance patterns may show increases in [$\alpha$/Fe] as a function of [Fe/H] that depend on spatial location. These increases should be correlated with localized differences burst strength (e.g., N20, JW20, H21), but 
may occur at fixed metallicity (i.e., burst timing).



\begin{figure*}
    \centering    \includegraphics[width=\textwidth]{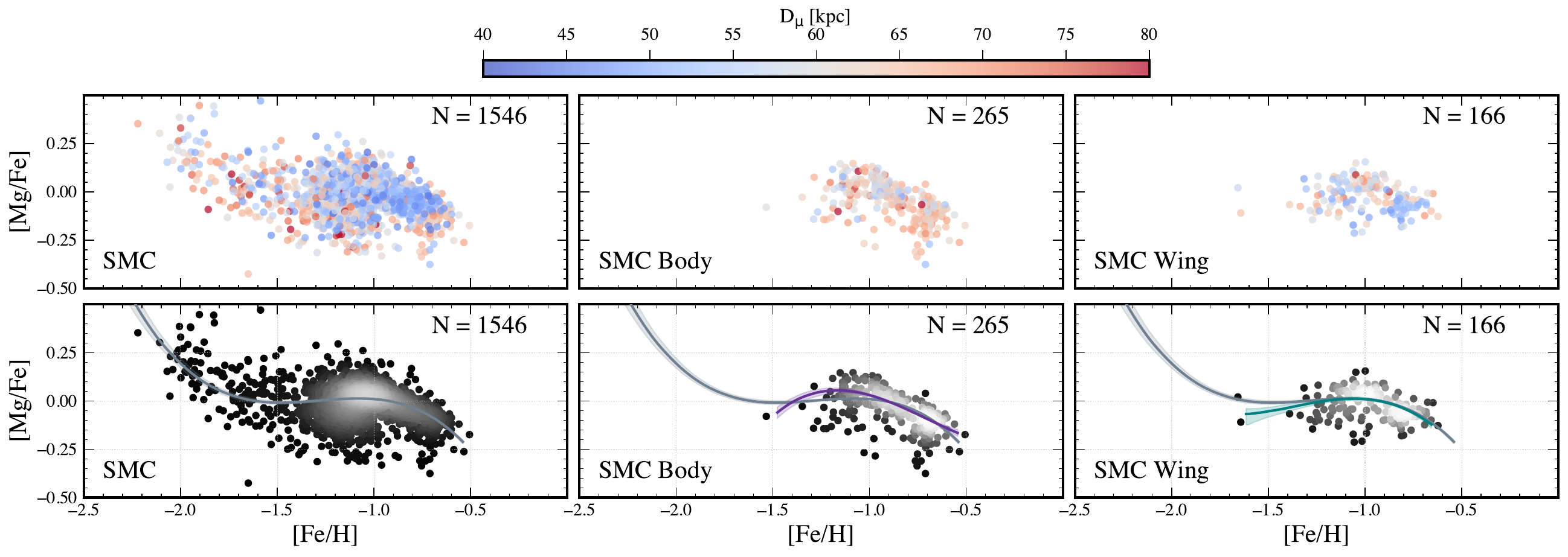}
    \caption{\textbf{[Mg/Fe] vs.\ [Fe/H] chemical abundance trends in the SMC.} Similar to Figure~\ref{fig:abund_lmc} (normalized to a number density of 12), except for all SMC stars (left panels), the SMC body (middle panels), and SMC wing (right panels; Section~\ref{sec:region_smc}). (Top) Chemical abundance distributions color-coded by a Gaia EDR3 parallax-based relative distance metric ($D_\mu$) to each RGB star (Section~\ref{sec:abund}; \citealt{Almeida2024}), demonstrating the well-known distance bimodality 
    along the line-of-sight to the wing (e.g., \citealt{Nidever2013}). The foreground population in the wing is metal-rich and $\alpha$-poor compared to the background population. (Bottom)
    Compared to the LMC, the SMC is characterized by weaker earlier star formation efficiency, and weaker signatures of an increase in [Mg/Fe] associated with a starburst (N20; H21), which peaks at [Fe/H] $\sim -1$ before declining. The SMC body has a declining pattern compared to the wing, and a more pronounced metal-rich and $\alpha$-poor population.
    }
    \label{fig:abund_smc}
\end{figure*}

\textit{SMC}: Figure~\ref{fig:sfr_smc} shows that the bulk of APOGEE RGB stars in the main body of the SMC, which dominates the global SMC SFR, should have 
formed in a starburst $\sim$3 Gyr ago 
in which the SFR approximately 
tripled (B25). An earlier global SF enhancement of similar strength occurred across all SMC subregions, including the wing, $\sim$5 Gyr ago (B25). 
Following this global starburst, 
our SMC wing region shows a short enhancement of the SFR $\sim$4 Gyr ago prior to remaining approximately
constant over the last $\sim$3 Gyr \added{within the SFH uncertainties}.
Thus, in contrast to the main body, a strong burst signature is not naively expected in the chemical abundance distribution of the wing. However, the presence of multiple stellar populations along the line-of-sight to the wing complicates interpretations of its SFH (e.g., C24b; \citealt{Almeida2024}), where a $\sim$3 Gyr ago localized starburst is known to have occurred in at least some regions of the SMC wing (B25) encapsulated by the subregion in this work.

\subsection{APOGEE Chemical Abundance Distributions}
\label{sec:abund}

In Figures~\ref{fig:abund_lmc} and~\ref{fig:abund_smc}, we show the [Mg/Fe]--[Fe/H] abundance plane for spatial subregions in the LMC and SMC (Section~\ref{sec:region}). To highlight the most populated location in the 2D chemical abundance distributions, we estimated the neighboring point density for each RGB star using a standard Gaussian kernel (determined by Scott's rule). For ease of comparison between subregions, we also quantified the trend between [Mg/Fe] and [Fe/H] by fitting a third-order BSpline to the 2D distribution, iteratively performing 2$\sigma$ clipping until the fit converged to a constant number of data points. We repeated this procedure for 1000 iterations in which we perturbed the [Fe/H] and [Mg/Fe] measurements by their Gaussian uncertainties (Section~\ref{sec:apogee}) to derive the median trendlines and 1$\sigma$ credible intervals from the 16$^{\rm th}$, 50$^{\rm th}$, and 84$^{\rm th}$ percentiles of the resulting distributions.

Given the well-known distance bimodality in the eastern 
SMC (e.g., \citealt{Nidever2013,Subramanian2017}), we also calculated a relative distance metric \Dmu\ 
based on Gaia EDR3 parallaxes ($\mu$) following \citealt{Almeida2024}, where $D_\mu = v_{\rm tan} / (4.74 \mu$) and $v_{\rm tan} = 398$ km s$^{-1}$ is the adopted systemic SMC tangential velocity. This distance proxy 
provides a consistent method for identifying foreground and background populations in Figure~\ref{fig:abund_smc} delineated by $D_\mu = 52$ kpc, 
where $D_\mu = 60$ kpc corresponds to the SMC center \citep{Almeida2024}.

In addition, we computed basic moments of the chemical abundance distributions for each subregion in each galaxy by performing 1000 bootstrap resamplings in which we perturbed each data point assuming Gaussian measurement uncertainties (Appendix~\ref{sec:appendix}, Table~\ref{tab:sfh_abund}). The errors on each statistic, which are $\leq$0.01 due to the large number of data points and the precise chemical abundance measurements, are calculated from the 16$^{\rm th}$ and 84$^{\rm th}$ percentiles of the resampled distribution. 

 
\textit{LMC}: As previously found globally for the LMC (N20; H21), an increase in \mgfe\ at \feh\ $\sim-1.3$, followed by secondary decline in [Mg/Fe] at \feh\ $\sim-0.6$ is uniformly apparent across the majority of subregions as approximated from the \mgfe\ vs.\ \feh\ trendlines.\footnote{There are minor differences in the metallicity at which the increase in \mgfe\ begins with respect to spatial location in the LMC. 
However, we do not over-interpret these apparent differences because the BSpline fits provide only an estimate of this \feh\ value given the small number of stars at low metallicity. 
}  
When separating the LMC disk into northern and southern halves, a more pronounced decline in [Mg/Fe] at high metallicity is apparent in the south. For radial regions in the LMC, the secondary decline in [Mg/Fe] is absent in the outer LMC (R3),
likely owing to the lack of metal-rich stars at larger LMC-centric radii  (e.g., \citealt{Choudhury2021,Grady2021,Povick2023,Massana2024}).
Figure~\ref{fig:abund_lmc} also demonstrates that the enhancement in [Mg/Fe] as traced by the trendlines is similarly uniform across all subregions ($\lesssim$0.005 dex).

More variation in the [Mg/Fe] versus [Fe/H] trendlines with respect to spatial location in the LMC becomes evident when separately dividing the northern and southern halves of the LMC disk into radial zones. 
The NR2 subregion containing the LMC Arm 
shows the clearest increase, and subsequent decrease, in [Mg/Fe] with increasing [Fe/H]. Similar trends, but with comparatively flatter increases in [Mg/Fe], are present throughout the inner LMC (\rdisklmc\ $<$ \rinvlmc), with the caveat of fewer stars populating the metal-poor tail of the NR1 and SR1 subregions. In addition, the more pronounced secondary decline in [Mg/Fe] at high [Fe/H] in the southern LMC compared to the north
may be driven by differences in its stellar outskirts (SR3 versus\ NR3).

While subtle spatial variations exist, the lack of a significant change in the \mgfe\ enhancement with spatial location therefore provides strong evidence for a single global burst of star formation
(\citealt{Weisz2013,Rubele2012,Massana2022}; B25) occurring within the predicted 1$\sigma$ age range of the RGB stars (\agergblmc\ Gyr). 
This also implies that any differences in burst strength, or peak SFR, inferred from Scylla SFHs in LMC subregions (Section~\ref{sec:csfh}) may  not be sufficiently large to have a detectable impact on the chemical abundance distributions.
We further assess whether the chemical abundance distributions, in combination with modeling, can provide robust constraints on the timing, strength, and duration of bursts in the LMC with spatial location in agreement with observed SFHs in Section~\ref{sec:gce}.


\textit{SMC}: Figure~\ref{fig:abund_smc} shows that the chemical abundance distributions for the SMC subregions show more substantial differences than those in the LMC. The foreground population in the SMC wing is metal-rich and $\alpha$-poor compared to the background population, where we computed $\langle$\feh$\rangle$ = \meanfehwingnear\ (\meanfehwingfar) and $\langle$\mgfe$\rangle$ = \meanmgfewingnear\ (\meanmgfewingfar) for the foreground (background) population when delineated by \Dmu\ = 52 kpc \citep{Almeida2024}.
The SMC body shows a declining chemical abundance pattern, 
with a less clear signature corresponding to the $\sim$3 Gyr burst in this region (Section~\ref{sec:csfh}) excepting a tentative increase in [Mg/Fe] peaking at \feh\ $\sim$ $-1.2$ (H21), with the caveat that the metal-poor tail of the SMC MDF may be poorly sampled owing to our restricted spatial selection over the Scylla footprint. In contrast, the wing shows a flatter chemical abundance pattern for \feh\ $\lesssim-1$ before declining at higher \feh, likely owing to the two chemically distinct stellar populations present in this subregion. Another explanation for the flatter chemical abundance pattern in the wing could be the absence of a clear $\sim$3 Gyr starburst in the RGB age range (Figure~\ref{fig:sfr_smc}; Section~\ref{sec:csfh}), which is present in the SMC body. 
We evaluate whether the chemical abundance trends in the SMC body and wing can be attributed to the presence or absence of a starburst by modeling their chemical evolution in Section~\ref{sec:gce}.

\section{Galactic Chemical Evolution Modeling}
\label{sec:gce}

\begin{figure}
    \centering
    \includegraphics[width=\linewidth]{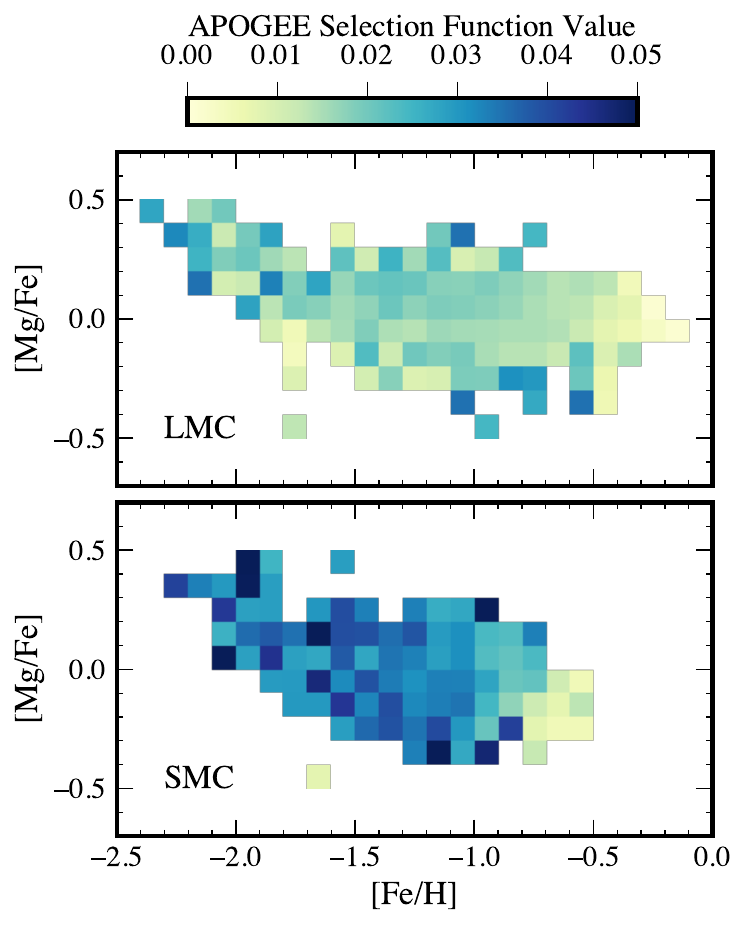}
    \caption{\textbf{Map of APOGEE selection function values in the [Mg/Fe]--[Fe/H] plane for the L/SMC}. We computed the selection function from the ratio of observed to candidate candidate targets for each APOGEE field (Section~\ref{sec:apogee}), from which we constructed a 2D map of the average value of the selection function in 0.1 dex bins in chemical abundance space (Section~\ref{sec:gce}) to more accurately reproduce the observed stellar density predicted by the GCE models. The net effect of weighting by the APOGEE selection function (Equation~\ref{eq:loglike}) is to increase the likelihood of observing a metal-poor star. The inferred GCE model parameters are minimally affected by the selection function weighting for the global SFH-constrained models (Section~\ref{sec:gce_scylla}). GCE models assuming SFE-driven starbursts (Section~\ref{sec:gce_sfedriven}) show more significant impacts if the selection function is not accounted for in metal-rich spatial subregions (i.\@e., the inner L/SMC) that are less observationally complete in APOGEE.
    }
    \label{fig:sfunc}
\end{figure}

\begin{table*}
\centering
\begin{threeparttable}
    \caption{Priors on Free Parameters and Values of Fixed Parameters in Fiducial Galactic Chemical Evolution Models}
    \begin{tabular*}{0.7\textwidth}{ccc}
    \hline\hline
    Parameter & Description & Prior/Value \\ \hline
    \multicolumn{3}{c}{Free Parameters}\\ \hline
    \multicolumn{3}{c}{\textit{All Models}}\\ 
    $\eta$  &  $=\dot{M}_{\rm out}/\dot{M}_\star$, mass-loading factor & $\mathcal{U}$(0, 15) \\
    $\tau_\star$ & $=M_{\rm gas} / \dot{M}_\star$, star-formation efficiency timescale in Gyr & $\mathcal{U}$(0, 30) \\
    & \\
    \multicolumn{3}{c}{\textit{Scylla SFH Models (Section~\ref{sec:gce_scylla})}}\\ 
    \yCC\ & IMF-integrated CCSNe Fe yield & $\mathcal{U}$(3.4, 6.7) $\times$ 10$^{-4}$ \\
    \yIa\ & IMF-integrated SNe Ia Fe yield & $\mathcal{U}$(0, 20) $\times$ 10$^{-4}$ \\
    & \\
    \multicolumn{3}{c}{\textit{SFE-Driven Models (Section~\ref{sec:gce_sfedriven})}}\\
    \Fb\ & peak SFE factor of increase during starburst & $\mathcal{U}$(1, 15) \\
    \taub\ & simulation time at which peak SFE occurs in Gyr & $\mathcal{U}$(0, 5) \\
    \sigmab\ & scale factor for burst duration in Gyr & $\mathcal{U}$(0, 1) \\
    $dt$ & time step size for total star formation duration in Gyr & $\mathcal{U}$(2, 5) $\times$ 10$^{-3}$ \\ \hline
    \multicolumn{3}{c}{Fixed Parameters}\\ \hline
    \multicolumn{3}{c}{\textit{Scylla SFH Models (Section~\ref{sec:gce_scylla})}}\\ 
    $y^{\rm CC}_{\rm Mg}$ & IMF-integrated CCSNe Mg yield (absolute yield scale) & 6.56 $\times$ 10$^{-4}$ \\
    $y^{\rm Ia}_{\rm Mg}$ & IMF-integrated SNe Ia Mg yield & 0 \\
    & \\
    \multicolumn{3}{c}{\textit{SFE-Driven Models (Section~\ref{sec:gce_sfedriven})}}\\
    $y^{\rm CC}_{\rm Mg, def}$ & default \texttt{VICE} CCSNe Mg yield & 4.97 $\times$ 10$^{-4}$ \\
    $y^{\rm Ia}_{\rm Mg, def}$ & default SNe Ia Mg yield & 8.83 $\times$ 10$^{-6}$ \\
    $y^{\rm CC}_{\rm Fe, def}$ & default CCSNe Fe yield & 2.46 $\times$ 10$^{-4}$ \\
    $y^{\rm Ia}_{\rm Fe, def}$ & default SNe Ia Fe yield & 2.58 $\times$ 10$^{-3}$ \\
    $M_{\rm g,0}$ & initial gas mass in $M_\odot$ & 3 $\times$ 10$^{9}$ \\
    $M_{\rm in}$ & gas inflow mass scale in $M_\odot$ & $6 \times 10^{10}$ \\
    $\tau_{\rm in}$ & gas inflow timescale in Gyr & 2 \\
    \hline
    \end{tabular*}
    \label{tab:gce_priors}
    \begin{tablenotes}
    \item For Scylla SFH models, we adopt an absolute yield scale following W24 based on the mean Fe mass yield measured for CCSNe by \citet{Rodriguez2023}, solar abundances from \citet{Mag2022}, and assuming 
    [Mg/Fe]$_{\rm CC}$ = 0.45. For SFE-driven models, we adopt consistent values for the gas inflow history across the MCs following H21 to facilitate comparisons across galaxies, as well as with the literature.
    \end{tablenotes}
\end{threeparttable}
\end{table*}

In this section, we systematically model the chemical evolution of the MCs in a Bayesian framework following a two-pronged approach. We investigate (1) predictions for chemical abundance distributions when observationally constrained by CMD-based SFHs (Section~\ref{sec:gce_scylla}) and (2) inferred SFH, particularly starburst, properties when constrained by chemical abundance measurements (Section~\ref{sec:gce_sfedriven}). For the first time to our knowledge, we assess whether measured SFHs---specifically from the Scylla survey---can match observed chemical abundance distributions in the [$\alpha$/Fe]--[Fe/H] plane of the MCs given reasonable GCE modeling assumptions. We also explore the implications of the level of agreement between the predicted and observed chemical abundance tracks for the global evolutionary history of the MCs. In the second approach, we model the observed APOGEE chemical abundance distributions assuming an SFH dominated by a single SFE-driven starburst for consistency with previous work (N20; H21). We then evaluate whether observed chemical abundance distributions can provide robust constraints on the timing, strength, and duration starbursts for spatial subregions with distinct SFH characteristics in the MCs (Section~\ref{sec:region}).

\begin{figure*}
    \centering
    \includegraphics[width=\textwidth]{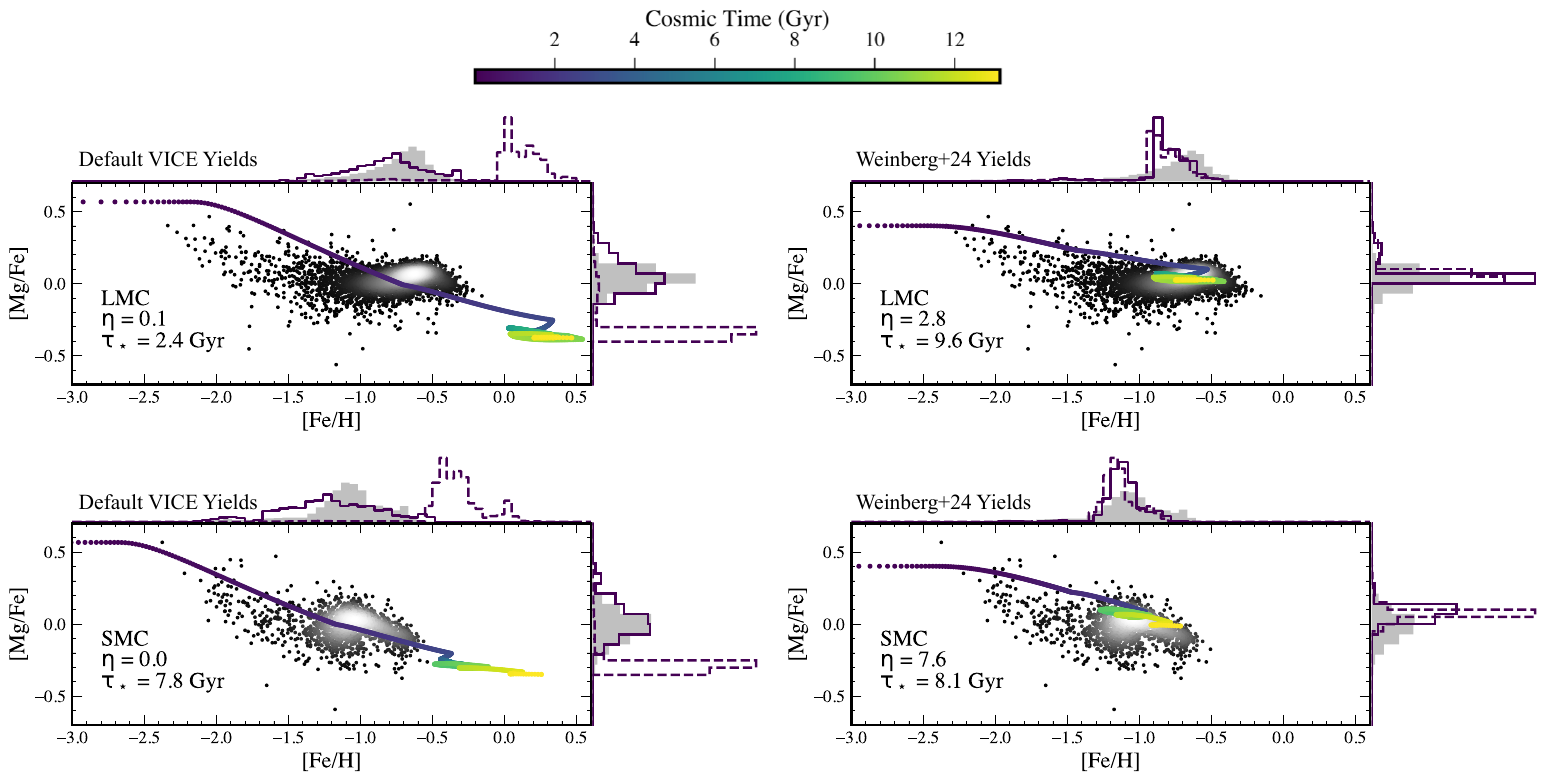}
    \caption{\textbf{Scylla SFH-constrained GCE models fitted to the MCs for different fixed yield sets.} Best-fit \texttt{VICE} 
    model tracks (colored points; Section~\ref{sec:gce_scylla}) in the [Mg/Fe] vs.\ [Fe/H] plane fitted to APOGEE chemical abundance distributions (similar to Figure~\ref{fig:abund_lmc}) based on the full sample of stars in the LMC (top panels) and SMC (bottom panels). Best-fit model parameters are calculated from the 16$^{\rm th}$, 50$^{\rm th}$, and 84$^{\rm th}$ percentiles of the marginalized posterior probability distributions (Table~\ref{tab:gce_pars}).  The metallicity and $\alpha$-element distribution functions for each galaxy (grey histograms), including the best-fit MDFs and ADFs from the GCE model (purple outlined histograms), are shown at the top and to the right of each panel. We show MDFs and ADFs constructed with (solid lines) and without (dashed lines) weighting by the APOGEE selection function, which is taken into account in the likelihood function (Equation~\ref{eq:loglike}).
    The input SFR for each galaxy is fixed to the global Scylla SFR 
    with a star formation duration of 13.1 Gyr (Section~\ref{sec:sfh}; Figures~\ref{fig:sfr_lmc} and~\ref{fig:sfr_smc}), where the model tracks are color-coded by 
    time 
    from the start of each simulation.
    For these models, we adopted the default yields in \texttt{VICE} (left panels) and the W24 yields assuming an [Mg/Fe] plateau of 0.45 dex (Section~\ref{sec:yields}). When assuming an absolute yield scale, $\eta$ is inversely related to \yIa\ (i.e., the larger \yIa\ yields in the \texttt{VICE} set produce models with weaker outflows) and \sfe\ is inversely related to \yCC (see \citealt{Johnson2023}).
    The \texttt{VICE} yields result in models that rapidly enrich to high metallicity and provide poor fits to the abundance distributions when varying only \outflow\ and \sfe\ (as opposed to having \yIa\ and \yCC\ as free parameters after setting the absolute yield scale; Figure~\ref{fig:photsfh}). The W24 yields produce more reasonable fits 
    particularly in the case of the LMC, but result in models that are not as able to match the lower $\alpha$-enhancement of the SMC. 
    }
    \label{fig:photsfh_fixedyield}
\end{figure*}

To model the chemical evolution of the LMC and SMC, we used the \texttt{Versatile Integrator for Chemical Evolution} (\texttt{VICE}; JW20)\footnote{GitHub: \url{https://github.com/giganano/VICE}; Science Documentation: \url{https://vice-astro.readthedocs.io/en/latest/science_documentation/index.html}}, a flexible Python package designed to implement starbursts via user-specified SFHs or gas accretion histories. In contrast to \texttt{flexCE} \citep{Andrews2017}, which has previously been used to model the chemical evolution of the MCs from APOGEE data (N20; H21),  \texttt{VICE} operates from user-specified IMF-averaged nucleosynthetic yields, as opposed to stochastically sampling the IMF of massive stars for CCSNe. By default, \texttt{VICE} assumes a set of theoretical yields homogeneously applied to each element, which are computed for CCSNe from the \citet{LimongiChieffi2018} non-rotating massive star models\footnote{The default \texttt{VICE} yields do not reflect the most recent version of \texttt{VICE}'s yield integrator updated by \citet{Griffith2021}.} and the N1 explosion model for SNe Ia from \citet{Seitenzahl2013} given the number of SNe Ia events per solar mass of star formation from \citet{MaozMannucci2012}.

One-zone \texttt{VICE} models rely on the standard approximation that newly produced metals mix instantaneously throughout the star-forming gas reservoir, where the time evolution of the gas supply is specified by,
\begin{equation}
\label{eq:gce}
    \dot{M_g} = \dot{M_{\rm in}} -\dot{M_\star} - \dot{M_{\rm out}} + \dot{M_r},
\end{equation}
where $\dot{M_{\rm in}}$ is the rate of pristine gas infall, $\dot{M_\star}$ is the star formation rate (SFR), $\dot{M_{\rm out}}$ is the outflow rate, and $\dot{M_r}$ is the rate at which gas is recycled into the interstellar medium (ISM) from stellar envelopes. The mass recycling rate $\dot{M_r}$ is 
a continuous time-dependent function, which corresponds to a cumulative instantaneous return fraction of $r_{\rm inst} \sim 0.4$ over the star-forming lifetime of a galaxy ($\sim$1--10 Gyr) when assuming a \citet{Kroupa2001} IMF (JW20). The mass loading factor is defined as the ratio of the outflow to star formation rate, $\eta \equiv \dot{M_{\rm out}} / \dot{M_\star}$.  In addition, the SFR and the gas mass are related by the SFE timescale $\tau_\star = M_g /\dot{ M_\star}$, which describes the characteristic timescale of the gas mass to deplete due to star formation assuming that no additional gas is added to the ISM. The rate of CCSNe enrichment for an element $X$ depends on the SFR and the IMF-averaged fractional net yield from massive stars for that element, $y_X^{\rm CC}$. The rate of enrichment for SNe Ia depends both on the yield $y_X^{\rm Ia}$ and the SNe Ia delay time distribution (DTD), where by default \texttt{VICE} assumes a $t^{-1.1}$ power-law with a delay time of $t_D$ = 150 Myr (motivated by \citealt{MaozMannucci2012} and following \citealt{Weinberg2017}).

The majority of assumptions in GCE models produce second-order effects on the chemical abundance tracks, where only a few model parameters dictate first-order details in the [$\alpha$/Fe] vs.\ [Fe/H] plane: the SFE timescale $\tau_\star$, the mass-loading factor $\eta$, and the adopted nucleosynthetic yields \citep{Weinberg2017}. Moreover, the absolute scale of the yields is degenerate with the strength of the outflows (e.g., \citealt{Johnson2023,Sandford2024,Johnson2025b}), thus we set the yield scale in our GCE models by either allowing the yields to vary as free parameters when anchored to observationally constrained values for the Fe mass yield from CCSNe (\citealt{Rodriguez2023,Weinberg2024}, hereafter W24; see Section~\ref{sec:gce_scylla} for further detail), or assuming a fixed yield set for Mg and Fe from CCSNe and SNe Ia (Section~\ref{sec:gce_sfedriven}). We model the APOGEE chemical abundance distributions in the [Mg/Fe] vs.\ [Fe/H] plane, where we selected Mg as a representative $\alpha$-element given its status as a pure CCSNe element with negligible SNe Ia contribution (i.e., $y_{\rm Mg}^{\rm Ia}$ = 0).  All $\alpha$-elements in APOGEE, including Mg, follow qualitatively similar chemical abundance trends characterized by a dominant starburst (N20; H21).

In the absence of a sudden burst of star formation, the detailed form of the SFH has little impact on the shape of the chemical abundance track (\citealt{Weinberg2017}; JW20). However, this information is instead encoded in the density of stars in the chemical abundance plane. Thus, simultaneously fitting the stellar MDFs and $\alpha$-element abundance distribution functions (ADFs) is crucial for developing more accurate GCE models, as is correcting the stellar density functions for any observational bias introduced by the selection function. In contrast to previous work, which relied on $\chi^2$-minimization over a grid of precomputed GCE models to find the best-fit model track, we sample the posterior probability distribution of the chemical abundance tracks predicted by the GCE models given the 2D APOGEE chemical abundance distributions following Bayes' theorem. The log-likelihood function is given by \citep{Johnson2023},
\begin{equation}
\label{eq:loglike}
\log \mathcal{L} = \sum_i^N \log \left( \sum_j^K w_j \exp \left(- \frac{1}{2} \Delta_{ij} C_i^{-1} \Delta_{ij}^T \right) \right),
\end{equation}
where $i$ is the index for a given star, $N$ is the total number of stars, 
$j$ is the index for a given point on the model track $\mathcal{M}_j$, and $K$ = 1000 is the total number of points on the model track. The vector difference $\Delta_{ij}$ is the distance between the ([Fe/H], [Mg/Fe]) position of a star $i$ and point $j$ on the model track in the 2D chemical abundance plane, whereas $C_i$ is the covariance matrix of star $i$ (defined by $\delta$[Fe/H]$_i$ and $\delta$[Mg/Fe]$_i$) . 

The weight $w_j$ is applied to point $j$ on the model track, where the likelihood of an observed value of ([Fe/H]$_i$, [Mg/Fe]$_i$) along the model track is higher when the model SFR is enhanced, or when the APOGEE selection function is more complete in the region of chemical abundance space corresponding to the model point ([Fe/H]$_j$, [Mg/Fe]$_j$).  Therefore, the normalized weights are described as $w_j \propto \mathcal{S} (\mathcal{M}_j|\vec{\theta}) \times \dot{M_\star}(\mathcal{M}_j|\vec{\theta})$, where $\vec{\theta}$ corresponds to the parameters (e.g., $\eta$, $\tau_\star$) defining a given model track $\mathcal{M}_j$. We assigned a value of the selection function $\mathcal{S}$($\eta_{\rm XMC}$, $\xi_{\rm XMC}$) to each star $i$ based on its location in a given APOGEE field (Section~\ref{sec:apogee}). Then, we used the ([Fe/H]$_i$, [Mg/Fe]$_i$) positions of the stars in chemical abundance space to construct a 2D map of the average value of $\mathcal{S}$($\eta_{\rm XMC}$, $\xi_{\rm XMC}$) in bins of 0.1 dex in [Fe/H] and [Mg/Fe] for the L/SMC \added{(Figure~\ref{fig:sfunc})}. For each set of  model parameters $\vec{\theta}$ and model tracks $\mathcal{M}_j$, we compute $\mathcal{S}(\mathcal{M}_j|\vec{\theta})$ for each iteration of the sampling procedure by interpolating each point $j$ of the model track over the 2D map $\mathcal{S}$([Fe/H], [Mg/Fe]). 
The net effect of weighting by the APOGEE selection function is to increase the likelihood of observing a metal-poor versus metal-rich star \added{(Figure~\ref{fig:sfunc})}, where a higher fraction of the total underlying metal-poor stellar populations were observed given their preferential location in the less dense stellar outskirts of the MCs.

\added{When constructing predicted MDFs and ADFs for comparison to APOGEE data, the model tracks ([Fe/H]$_j$, [Mg/Fe]$_j$) should therefore generally be weighted by $w_j$, which includes the selection function $\mathcal{S}_j$. In regions of chemical abundance space where no data is observed, $w_j = 0$ in Equation~\ref{eq:loglike}, such that model points in this regime neither contribute to nor penalize the likelihood. In practice, the impact of $\mathcal{S}$ is minimal for the appearance of the SFH-constrained models in Section~\ref{sec:gce_scylla} as long as the starbursts occur in a region of chemical abundance space overlapping with the data (but see the left panels of Figure~\ref{fig:photsfh_fixedyield}), because the predicted stellar density is dominated by the input Scylla SFR $\dot{M}_\star$ and the total SF duration is fixed. 
In contrast, the total SF duration is a free parameter that is not well-constrained by the data
for the SFE-driven starburst models in Section~\ref{sec:gce_sfedriven}, where the SFR is specified by the model parameters $\vec{\theta}$. In this case, correcting by $\mathcal{S}$ can result in predicted chemical abundance tracks that move forward in time to high values of [Fe/H] beyond the data extent.
Weighting by $\mathcal{S}$ is therefore crucial for producing accurate predictions of the observed stellar density for the SFE-driven burst models, and is taken into account in representations of the model MDFs and ADFs in both Sections~\ref{sec:gce_scylla} and~\ref{sec:gce_sfedriven}.
}

To remain relatively agnostic regarding the model parameter values, we assumed weakly informative uniform priors on $\vec{\theta}$, such that the likelihood of the parameters $\mathcal{L}(\vec{\theta})$ is constant, and the posterior probability is approximately the likelihood (Equation~\ref{eq:loglike}). We summarize the adopted priors on the values of the free model parameters in Table~\ref{tab:gce_priors}, where we discuss the specific model parameters and priors in detail in Sections~\ref{sec:gce_scylla} and~\ref{sec:gce_sfedriven}.  For each GCE model, we use the Markov chain Monte Carlo ensemble sampler \texttt{EMCEE} \citep{Foreman-Mackey2013} to draw 10,000 samples from the posterior probability distribution, where we compute the ``best-fit'' model parameters from the 16$^{\rm th}$, 50$^{\rm th}$, and 84$^{\rm th}$ percentiles of the marginalized posterior probability distributions constructed from the latter 50\% of each chain.

\subsection{Models Constrained by Scylla SFHs}
\label{sec:gce_scylla}

\begin{figure}
    \centering
    \includegraphics[width=0.5\textwidth]{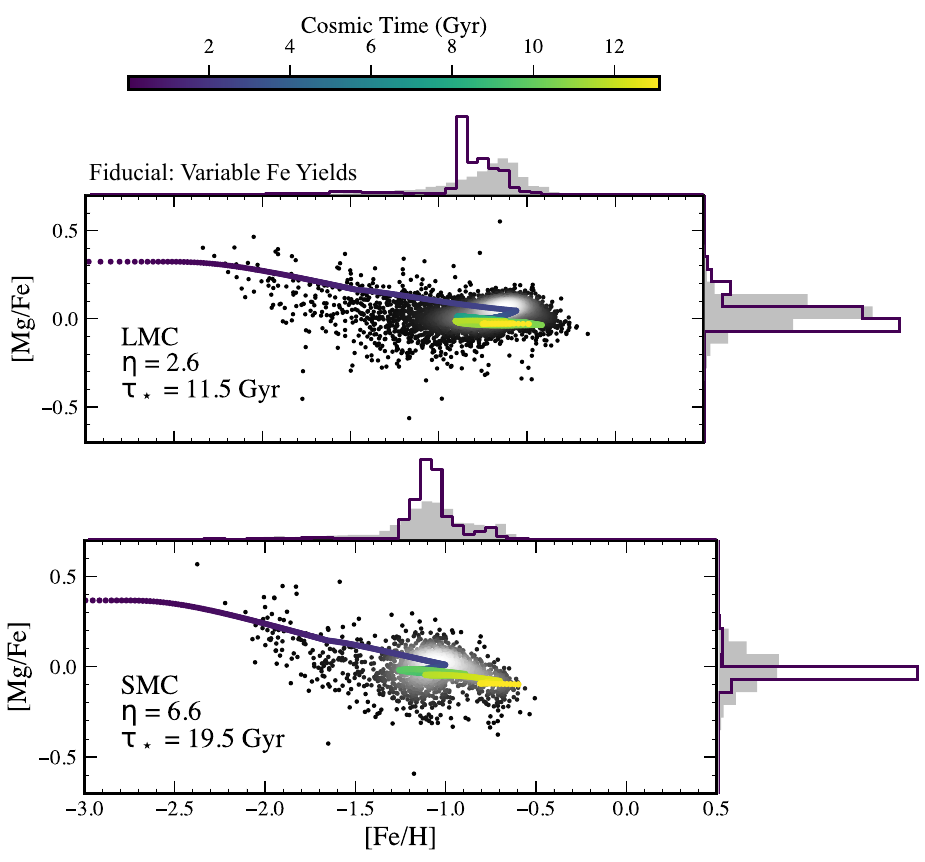}
    \caption{\textbf{Fiducial SFH-constrained models fitted to the MCs assuming variable Fe yields.}
    GCE models fitted to APOGEE chemical abundance distributions
    with Fe
    yields as free parameters (\yIa\ and \yCC, assuming the absolute yield scale specified in Table~\ref{tab:gce_priors}; \citealt{Rodriguez2023}; W24) in addition to the mass-loading factor \outflow\ and star formation efficiency timescale \sfe. This produces a better fit to the chemical abundance distributions (especially the SMC) compared to the fixed yield case (Figure~\ref{fig:photsfh_fixedyield}; Section~\ref{sec:yields}).
    The different yields (Table~\ref{tab:gce_pars}) for the LMC and SMC limit inter-galaxy comparisons for \outflow\ and \sfe, where the \yIa\ yields are significantly higher in the SMC than the LMC 
    (see Section~\ref{sec:evo}). However, both models show rapid enrichment to the equilibrium abundance followed by slower chemical evolution in a series of gas-driven starbursts (Section~\ref{sec:scylla_fiducial}).
    }
    \label{fig:photsfh}
\end{figure}

\begin{figure*}
    \centering
    \includegraphics[width=0.7\textwidth]{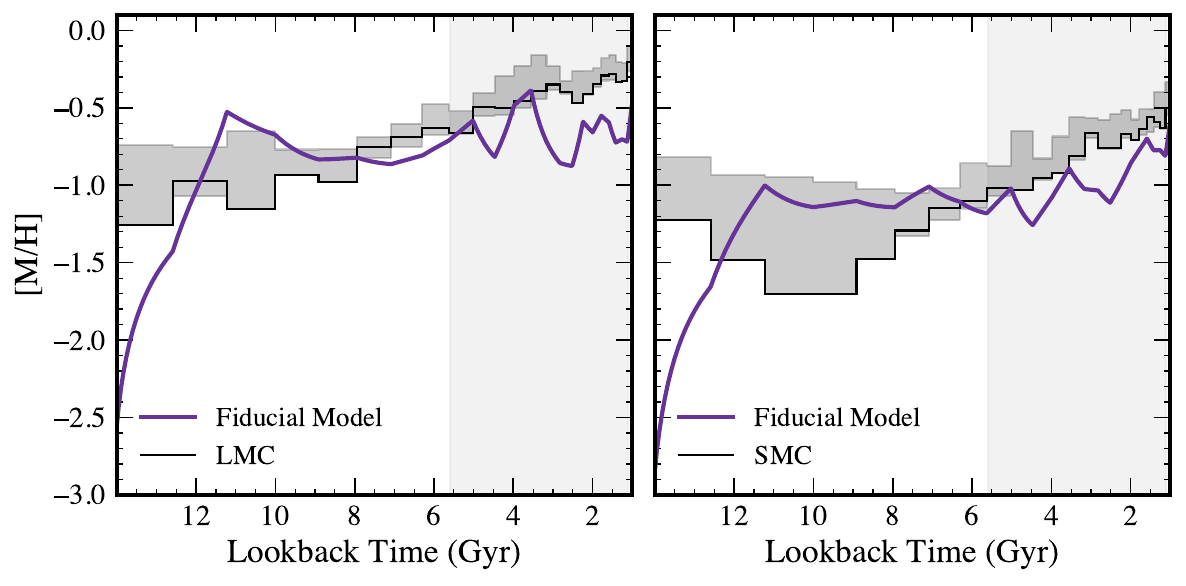}
    \caption{\textbf{AMRs from Scylla and from fiducial GCE models constrained by Scylla SFHs.} Best-fit global Scylla AMRs for the LMC (left panel) and SMC (right panel) as a function of lookback time (black lines; C24a,b) and their 1$\sigma$ uncertainties (shaded regions). The shaded grey vertical region in each panel represents \taurgb\ for each galaxy (Section~\ref{sec:fake}). We also show AMRs from fiducial SFH-constrained models (purple lines; Figure~\ref{fig:photsfh}) characterized by input Scylla SFRs 
    and Fe yields inferred from fitting APOGEE abundance distributions. 
    The shape of the model AMR is set by the input Scylla SFR and the accretion of pristine gas driving starbursts that result in non-monotonic behavior in metallicity with time (Section~\ref{sec:scylla_fiducial}). The metallicity difference between the observed and modeled AMRs for recent lookback times results from the known discrepancy between photometric and spectroscopic metallicities (Appendix~\ref{sec:appendix}).
    }
    \label{fig:amr}
\end{figure*}

To investigate the shape of chemical abundance tracks predicted directly from measured SFHs---given reasonable chemical evolution modeling assumptions---we constructed one-zone \texttt{VICE} models constrained by the global Scylla SFHs for the MCs (Figures~\ref{fig:sfr_lmc} and~\ref{fig:sfr_smc}). We fit these GCE models to the observed APOGEE chemical abundance distributions for all stars in the LMC and SMC (Section~\ref{sec:apogee}) using Equation~\ref{eq:loglike} to explore the implications of the best-fit model parameters for the evolutionary history of the MCs. We input the Scylla SFRs directly into \texttt{VICE} as a function of time in Gyr, where we omit the most recent 1 Gyr of evolution given that RGB stars should not have stellar ages $\lesssim$1 Gyr. Thus, the SF duration in this model is 13.1 Gyr starting from the beginning of cosmic time, where the number of time steps in the model is fixed to $K = 1000$.

By specifying $\dot{M_\star}$ as a function of time in \texttt{VICE}, Equation~\ref{eq:gce} implicitly solves for the infall rate of pristine gas $\dot{M_{\rm in}}$, such that starbursts in this model are assumed to be driven by gas accretion. Gas-driven starbursts, in which an amount of gas comparable to the current ISM mass of the galaxy is added to its reservoir on a shorter timescale than the depletion time of gas into stars, could result from the merger of a gas-rich system or a temporary enhancement in the accretion rate (JW20). \added{The number of starbursts in the \texttt{VICE} models are fixed by the number of significant SFR peaks in the observed SFHs (black lines in Figures~\ref{fig:sfr_lmc} and~\ref{fig:sfr_smc}) that are sufficient to drive a gas-driven starburst.}

\subsubsection{Implications of SFH-Constrained Models for Yields}
\label{sec:yields}

\added{The underlying nucleosynthetic yields are a well-known source of uncertainty in chemical evolution models (e.g., \citealt{Romano2010,Cote2017}). Theoretically derived stellar yields have been traditionally adopted in GCE models, including in the MCs (N20; H21), although they are significantly affected by poorly constrained processes in massive star evolution, such as stellar mass loss rates, rotational mixing,  convection, and direct collapse \citep{Ventura2013,Ertl2016,Frischknecht2016,Sukhbold2016,LimongiChieffi2018,Beasor2020,Griffith2021}. These uncertainties in theoretical yield prescriptions have thus motivated recent efforts to adopt empirically calibrated yields \citep{Johnson2023,Weinberg2024,Sandford2024}. In practice, stellar yields are often treated as nuisance parameters in GCE models, where model parameters such as those governing the SFH are of primary interest, and are either fixed to a standard set of yields (whether they are theoretically or empirically motivated; Section~\ref{sec:sfedriven_yield}), or tuned to reproduce the observed chemical abundance distributions. Here, we demonstrate that GCE models constrained by both observed SFHs and chemical abundances do not produce reasonable predictions with fixed sets of either theoretical or empirical yields, and must instead be inferred as part of the model (see also \citealt{Heiger2026}).
}

Initially, we fit only for the parameters \added{of the SFH-constrained model} that dictate much of the variation of the model track in chemical abundance space (\outflow\ and \sfe) and assumed a fixed set of yields. When adopting the \added{theoretically-motivated} default \texttt{VICE} yields (Table~\ref{tab:gce_priors}), the predicted chemical abundance tracks for the LMC and SMC (left panels of Figure~\ref{fig:photsfh_fixedyield}) show a steep decline in [Mg/Fe] vs.\ [Fe/H], where the ratio between \yCCMg\ and \yCC\ sets the height of the [Mg/Fe] plateau ([Mg/Fe]$_{\rm CC}$). This decline is partially driven by rapid enrichment to solar metallicity within $\sim$3 Gyr of evolution, which is caused by the large ratio between the default \yIa\ and \yCC\ yields\ (a factor of $\sim$10; Table~\ref{tab:gce_priors}). 
As a result, the ``best-fit'' \added{SFH-constrained} models for the LMC and SMC assuming the \added{theoretical} default yields do not match the APOGEE chemical abundance distributions \added{when constructing MDFs and ADFs that do not incorporate weighting by the APOGEE selection function}, where the models spend the majority of simulation time forming stars well outside the bulk of the chemical abundance distributions.

We thus considered an alternate yield set adopted from W24, which is anchored to an empirical measurement of the Fe mass yield from CCSNe \citep{Rodriguez2023}, assuming solar abundances from \citet{Mag2022} and \MgFeCC\ = 0.45 (Table~\ref{tab:gce_priors}). We calculated the associated values of \yCCMg\ and \yIa\ following Figure~5 of W24 (their Equations~10 and~22 assuming [Mg/Fe]$_{\rm eq}$ = 0 at equilibrium). The W24 yields result in significantly improved fits to the chemical abundance distributions (right panels of Figure~\ref{fig:photsfh_fixedyield}), especially in the case of the LMC. However, this yield set results in  models that are not able to simultaneously match the lower $\alpha$-enhancement of the SMC, as demonstrated by the high SFE ($\tau_\star^{-1}$) of the SMC compared to the LMC, in contrast to expectations for these systems (H21).

For our fiducial GCE models constrained by Scylla SFHs, we therefore assumed variable \added{empirical Fe} yields to better fit the chemical abundance distributions of both the LMC \textit{and} the SMC.
We set the absolute yield scale by fixing \yCCMg\ to the value from W24 and assuming $y_{\rm Mg}^{\rm Ia}$ = 0 (Table~\ref{tab:gce_priors}). We placed a uniform prior on \yCC\ over a narrow range of values (Table~\ref{tab:gce_priors}) to ensure that 0.3 $\leq$ \MgFeCC\ $\leq$ 0.6 in accordance with observational constraints on the [Mg/Fe] plateau in the Milky Way from APOGEE data \citep{Sit2025}, where \MgFeCC\ $\approx$ $\log_{10}$(\yCCMg/\yCC) $-$ $\log_{10}$($Z_{\rm Mg,\odot}$/$Z_{\rm Fe,\odot}$)  assuming the \texttt{VICE} solar abundances \citep{Asplund2009}. The free model parameters are therefore \outflow, \sfe, \yCC, and \yIa\ \added{for the fiducial SFH-constrained models.}

\subsubsection{Fiducial SFH-Constrained Models}
\label{sec:scylla_fiducial}

Figure~\ref{fig:photsfh} shows the \added{fiducial} best-fit model tracks in the [Mg/Fe] vs.\ [Fe/H] plane and best-fit MDFs and ADFs for the LMC and SMC, computed from the 50$^{\rm th}$ percentile values of the marginalized posterior probability distributions in Table~\ref{tab:gce_pars}. We also show the APOGEE chemical abundance distributions for comparison.

The different \added{best-fit} yields (\yCC\ and \yIa) for the LMC and SMC limit inter-galaxy comparisons between $\eta$ and $\tau_\star$, although the yield scales for each galaxy are sufficiently similar that we find that the outflows are stronger and the efficiency of star formation is lower for the SMC compared to the LMC, in agreement with previous work (H21). The fiducial GCE models, which provide the best match to the APOGEE data, imply that the LMC and SMC have consistent [Mg/Fe] plateaus (based on the values of \yCC; Table~\ref{tab:gce_pars}). However, we also find that $y_{\rm Fe}^{\rm Ia,SMC} > y_{\rm Fe}^{\rm Ia, LMC}$ by a factor of \factoryIa\ from the fiducial models,
implying that the SMC may have experienced more Fe enrichment by SNe Ia than the LMC (see discussion in Section~\ref{sec:evo}).\footnote{The larger relative SNe Ia Fe yield in the SMC is robust against the incorporation of the APOGEE selection function, where the net effect of the preferential weighting toward metal-poor stars is to increase (decrease) the inferred value of \yIa\ (\yCC) for both the LMC and SMC.}

A common feature of the GCE models constrained by the Scylla SFHs, regardless of the assumed yields \added{(Section~\ref{sec:yields})}, is that the predicted chemical tracks reach high metallcities within $\sim$3 Gyr.
Thus, Scylla SFHs, when combined with reasonable GCE assumptions, predict that both the LMC and SMC should have reached their respective equilibrium ISM abundances within this $\sim$3 Gyr timescale, where the composition of the ISM in the GCE models 
are relatively unaffected by changes in the gas mass \citep{FinlatorDave2008,Andrews2017,Weinberg2017,Johnson2024}. 

\added{After enriching to approximately the equilibrium ISM abundance, the predicted chemical tracks spend} the remainder of the time in the simulations within a narrower metallicity range forming stars in a series of gas-driven starbursts. The gas-driven bursts create loops in chemical abundance space as the ISM metallicity first decreases as a consequence of pristine gas infall, followed by an increase in both [Mg/Fe] and [Fe/H] triggered by CCSNe associated with the starburst, then a decrease in [Mg/Fe] from the delayed onset of SNe Ia from stars formed in the burst (JW20). \added{The modeled MDFs for the LMC and SMC show concentrations of stars formed at specific values of [Fe/H] as a result of these dilution events intrinsic to starbursts driven by pristine gas accretion. In particular, the modeled MDF for the SMC shows a secondary metal-rich peak (Figure~\ref{fig:photsfh}) in an attempt to match the younger, metal-rich population of stars in the SMC (e.g., \citealt{Povick2025}), which may originate from a recent starburst (see Section~\ref{sec:sfedriven_smc}).
}

The global Scylla AMRs for the MCs (Figure~\ref{fig:amr}) similarly suggest that the ISM of each galaxy had enriched to [Fe/H] $\gtrsim$ $-1.5$ within their first few Gyr of evolution. 
This is in qualitative agreement with MDFs derived from photometry in the inner regions of the MCs and based on large surveys of RGB stars with more complete spatial coverage than APOGEE, which show minimal contributions to the metal-poor tails of the distributions for [Fe/H] $\ll -1.5$ (e.g., \citealt{Grady2021,Massana2024}). The Scylla AMRs further indicate that subsequent enrichment in the MCs predominantly occurs within the last $\sim$4--6 Gyr. In the models, this enrichment period is characterized by the fundamental assumption that Scylla starbursts are driven by gas accretion, resulting in \added{two dominant dilution events, followed by a weaker third event, associated with enhanced SFRs in Scylla.\footnote{The number of gas-driven starbursts inferred by \texttt{VICE} from Scylla SFHs is not necessarily equivalent to observational definitions of distinct starbursts based on Scylla SFHs, which correspond to sustained periods of enhanced SFR (\citealt{Burhenne2025}; see also Section~\ref{sec:csfh}).} Thus, the modeled AMR is characterized by} non-monotonic behavior in metallicity as a function of time (see also Figures~\ref{fig:photsfh_fixedyield} and~\ref{fig:photsfh}). \added{
The observed AMRs do not show similarly strong non-monotonic behavior, suggesting either that the accretion of enriched, rather than pristine, gas is necessary to prevent significant dilution events in the model, or that models with starbursts driven by changes in SFE likely provide a more realistic description of the observations (Section~\ref{sec:discuss}).}

\subsection{SFE-Driven Starburst Models}
\label{sec:gce_sfedriven}

\begin{figure*}
    \centering
    \includegraphics[width=\textwidth]{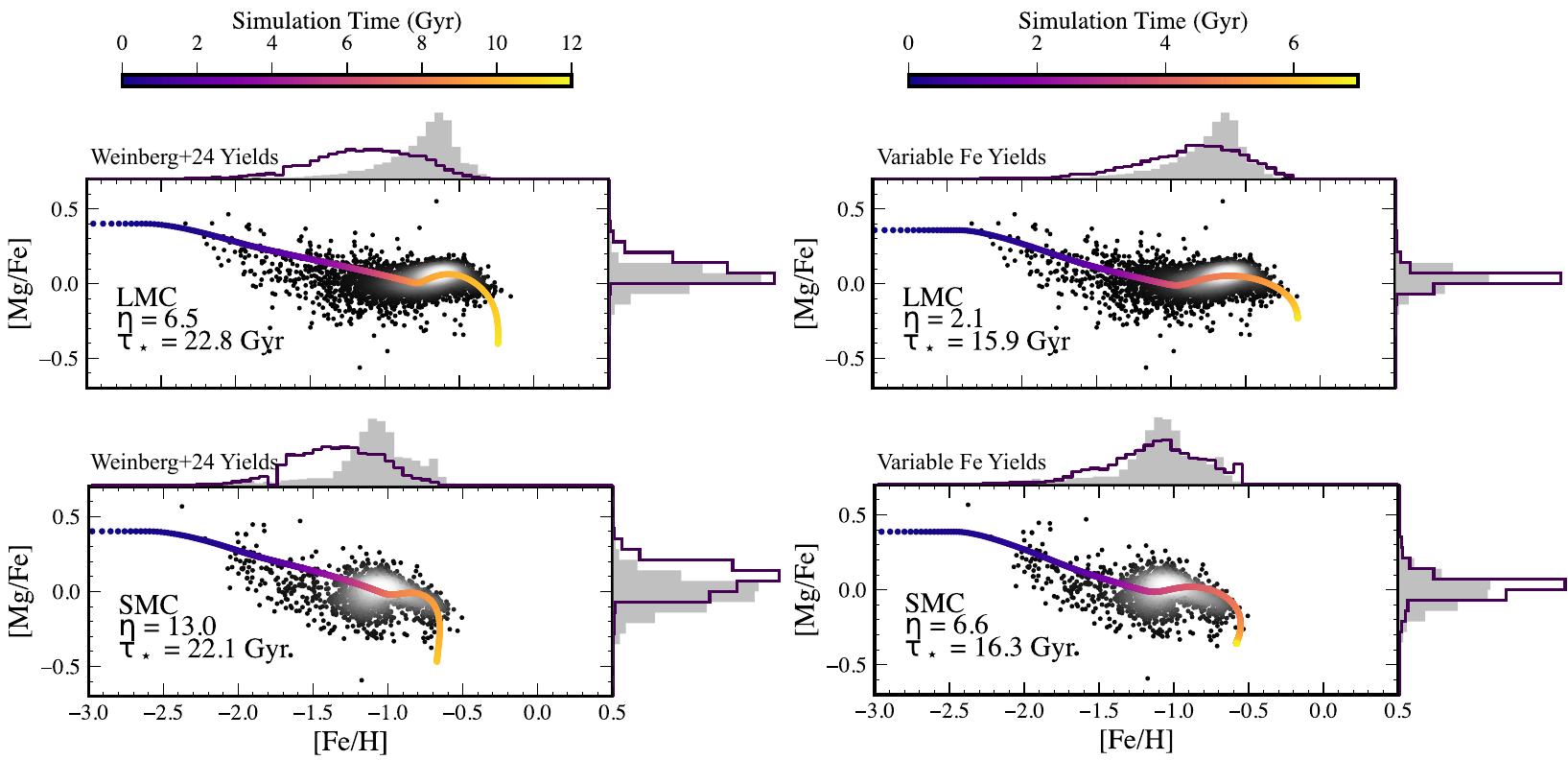}
    \caption{\textbf{SFE-driven starburst models fitted to the MCs for different yield assumptions.} One-zone \texttt{VICE} GCE models fitted to APOGEE chemical abundance distributions assuming a delayed-$\tau$ model for the infall rate and an analytical formulation for time-variable changes in SFE (Equation~\ref{eq:sfe}) that drive a sustained starburst (see H21; Section~\ref{sec:gce_sfedriven}). (Left panels) Models assuming the empirical W24 yields with \MgFeCC\ = 0.45 (Section~\ref{sec:yields}). (Right panels) Models where the Fe yields (\yIa\ and \yCC) are free parameters (adopting the absolute yield scale and priors in Table~\ref{tab:gce_priors}, where the best-fit Fe yields for these models are given in Table~\ref{tab:gce_pars}).  Although the W24 yields produce chemical abundance tracks that match the 2D chemical abundance distributions with long SF durations and low SFE,
    they overproduce the observed density of stars at the metal-poor end of the abundance distributions. Moreover, although fitting for the Fe yields produces models that better match the chemical abundance distributions than the W24 yield models, 
    inter-galaxy comparisons between the MCs are limited by the differing values of 
    the Fe yields. 
    However, each set of models demonstrates that trends such as a stronger, similarly timed starburst in the LMC 
    compared to the SMC are robust against the assumed yields (Section~\ref{sec:gce_sfedriven}; Table~\ref{tab:gce_pars}).
    }
    \label{fig:sfedriven_yield}
\end{figure*}

To evaluate whether we can recover starburst properties from the chemical abundance distributions of the MCs in broad agreement with those inferred directly from Scylla SFHs, we constructed \texttt{VICE} models where the SFE timescale \sfe\ is specified as a function of time. We fit these models to all stars in the LMC and SMC, as well as \textit{each} spatial subregion defined in Section~\ref{sec:region} (Section~\ref{sec:sfedriven_lmc},~\ref{sec:sfedriven_smc}). Analogously to \citet{Hendricks2014} and H21, we implement bursts of star formation as time-localized changes to the SFE timescale. These efficiency-driven starbursts result from a temporary increase in the SFE (decrease in \sfe) that depletes the gas reservoir more rapidly,  and could be caused by dynamical interactions either between the MCs or with the Milky Way. These interactions do not necessarily increase the gas supply, in contrast to the gas-driven starbursts modeled in
Section~\ref{sec:gce_scylla}. However, both efficiency- and gas-driven starbursts have similar effects on chemical abundance distributions, excepting the production of a population of $\alpha$-deficient stars in the former case (JW20). We therefore model the bursts in the MCs as being driven by changes in SFE and not by the accretion of gas, given their extended interaction histories and shallower gravitational potential wells, and for consistency with the literature. Similar to H21, the functional form of the SFE timescale is given by,
\begin{equation}
\label{eq:sfe}
    \frac{\tau_\star}{( 1 + (F_b - 1) \times \exp (-\frac{1}{2} ((t - \tau_b)/\sigma_b)^2 )}
\end{equation}
where \sfe\ is the value of the SFE timescale in the absence of a starburst, \Fb\ is the peak factor of increase in the SFE during the starburst (the strength of the burst), \taub\ is the simulation time at which the peak increase in SFE occurs (distinct from the onset of the burst, or the timing of the peak SFR), \sigmab\ is the scale factor for the burst duration, and $t$ is the simulation time in Gyr. We also include the total SF duration as a free parameter in the model, parameterized by the simulation time step $dt$ in Gyr. We fix the number of time steps 
to $K = 1000$ to enable comparisons of the statistical likelihood (Equation~\ref{eq:loglike}) despite differences in model run time.

For consistency with the literature and to enable direct comparisons between the MCs, we adopt the cosmologically-motivated delayed-$\tau$ model from H21 to parameterize the gas inflow history,
\begin{equation}
    \dot{M}_{\rm in} = \left(\frac{M_i}{\tau_i}\right) \left(\frac{t}{\tau_i}\right) e^{-t/\tau_i},
\end{equation}
where the assumed values for the inflow mass scale $M_i$ (the total mass accreted as the simulation time $t \rightarrow \infty$) and inflow timescale $\tau_i$ (the time at which inflow rate reaches its maximum) are presented in Table~\ref{tab:gce_priors}.

\begin{deluxetable*}{lcccccccc}
\tablecaption{Best-fit Galactic Chemical Evolution Model Parameters in the MCs\label{tab:gce_pars}}
\tablewidth{0pt}
\tablehead{
\colhead{Region} &
\colhead{$\eta$} &
\colhead{$\tau_\star$} &
\colhead{$10^{-4} \times$ \yIa} &
\colhead{$10^{-4} \times$ \yCCMg} &
\colhead{\Fb} &
\colhead{$\tau_b$} &
\colhead{$\sigma_b$} &
\colhead{$10^{3} \times dt$} \\
\colhead{} &
\colhead{} &
\colhead{(Gyr)} &
\colhead{} &
\colhead{} &
\colhead{} &
\colhead{(Gyr)} &
\colhead{(Gyr)} &
\colhead{(Gyr)}
}
\startdata
\multicolumn{9}{c}{Global Models Assuming Scylla SFR (Figure~\ref{fig:photsfh})} \\ \hline
\textbf{LMC} & $2.65^{+0.20}_{-0.10}$ & $11.47^{+0.49}_{-0.66}$ & $8.52^{+0.36}_{-0.15}$ & $5.93^{+0.05}_{-0.22}$ & \nodata & \nodata & \nodata & \nodata \\
\textbf{SMC} & $6.56^{+0.13}_{-0.23}$ & $19.50^{+0.52}_{-0.56}$ & $12.80^{+0.33}_{-0.19}$ & $5.14^{+0.11}_{-0.17}$ & \nodata & \nodata & \nodata & \nodata \\ \hline
\multicolumn{9}{c}{Global SFE-Driven Burst Models (Figures~\ref{fig:sfedriven_yield},~\ref{fig:sfedriven}, and~\ref{fig:gce_smc_hist})} \\ \hline   
LMC & $2.07^{+1.00}_{-0.75}$ & $15.87^{+1.86}_{-2.17}$ & $15.29^{+1.10}_{-1.31}$ & $5.25^{+0.24}_{-0.29}$ & $9.35^{+1.27}_{-1.29}$ & $5.17^{+0.25}_{-0.25}$ & $0.48^{+0.07}_{-0.07}$ & $6.70^{+0.67}_{-0.61}$ \\
SMC & $6.56^{+2.23}_{-3.39}$ & $16.29^{+2.16}_{-3.34}$ & $16.99^{+1.52}_{-1.46}$ & $4.91^{+0.52}_{-0.39}$ & $5.49^{+3.01}_{-2.64}$ & $4.25^{+1.00}_{-0.47}$ & $0.45^{+0.15}_{-0.24}$ & $6.82^{+0.89}_{-0.77}$ \\ \hline
\textbf{LMC} & $0.90^{+0.29}_{-0.23}$ & $7.97^{+0.60}_{-0.56}$ & \nodata & \nodata & $11.38^{+0.69}_{-0.55}$ & $2.88^{+0.11}_{-0.09}$ & $0.40^{+0.69}_{-0.55}$ & $3.62^{+0.11}_{-0.09}$ \\
\textbf{SMC} & $4.14^{+0.45}_{-0.50}$ & $15.01^{+0.60}_{-0.46}$ & \nodata & \nodata & $6.33^{+0.64}_{-0.50}$ & $2.73^{+0.05}_{-0.04}$ & $0.35^{+0.64}_{-0.50}$ & $3.66^{+0.05}_{-0.04}$ \\ \hline
SMC Degenerate & $4.00^{+0.59}_{-3.52}$ & $11.81^{+1.26}_{-5.51}$ & \nodata & \nodata & $5.57^{+0.81}_{-1.11}$ & $2.62^{+5.67}_{-0.08}$ & $0.34^{+0.81}_{-1.11}$ & $7.85^{+5.67}_{-0.08}$\\ \hline
\multicolumn{9}{c}{SFE-Driven Burst Models in the LMC (Figure~\ref{fig:sfedriven_subregion})} \\ \hline
LMC N & $0.74^{+0.33}_{-0.26}$ & $8.49^{+1.21}_{-1.02}$ & \nodata & \nodata & $10.63^{+1.05}_{-1.02}$ & $2.90^{+0.14}_{-0.14}$ & $0.42^{+1.05}_{-1.02}$ & $4.01^{+0.14}_{-0.14}$ \\
LMC S & $0.66^{+0.25}_{-0.22}$ & $9.36^{+1.08}_{-1.15}$ & \nodata & \nodata & $12.30^{+0.97}_{-1.09}$ & $3.09^{+0.15}_{-0.19}$ & $0.44^{+0.97}_{-1.09}$ & $3.90^{+0.15}_{-0.19}$ \\ \hline
LMC R1 & $0.68^{+0.24}_{-0.22}$ & $9.23^{+0.85}_{-0.59}$ & \nodata & \nodata & $11.67^{+0.83}_{-0.86}$ & $3.25^{+0.10}_{-0.08}$ & $0.42^{+0.83}_{-0.86}$ & $4.11^{+0.10}_{-0.08}$ \\
LMC R2 & $0.75^{+0.19}_{-0.20}$ & $7.91^{+0.91}_{-0.71}$ & \nodata & \nodata & $10.13^{+0.77}_{-0.74}$ & $2.87^{+0.09}_{-0.09}$ & $0.42^{+0.77}_{-0.74}$ & $3.46^{+0.09}_{-0.09}$ \\
LMC R3 & $0.90^{+0.56}_{-0.42}$ & $8.05^{+1.19}_{-0.94}$ & \nodata & \nodata & $11.68^{+1.00}_{-0.81}$ & $2.80^{+0.16}_{-0.14}$ & $0.39^{+1.00}_{-0.81}$ & $3.35^{+0.16}_{-0.14}$ \\ \hline
LMC NR1 & $0.92^{+0.28}_{-0.30}$ & $7.92^{+0.98}_{-0.86}$ & \nodata & \nodata & $10.02^{+0.96}_{-0.75}$ & $2.95^{+0.14}_{-0.15}$ & $0.39^{+0.96}_{-0.75}$ & $3.92^{+0.14}_{-0.15}$ \\
LMC NR2 & $0.91^{+0.32}_{-0.28}$ & $8.30^{+1.01}_{-0.96}$ & \nodata & \nodata & $11.46^{+1.19}_{-1.18}$ & $2.91^{+0.13}_{-0.18}$ & $0.40^{+1.19}_{-1.18}$ & $3.49^{+0.13}_{-0.18}$ \\
LMC NR3 & $0.62^{+0.35}_{-0.30}$ & $7.44^{+0.59}_{-0.53}$ & \nodata & \nodata & $10.52^{+1.01}_{-0.87}$ & $2.78^{+0.11}_{-0.13}$ & $0.39^{+1.01}_{-0.87}$ & $3.81^{+0.11}_{-0.13}$ \\ \hline
LMC SR1 & $0.70^{+0.45}_{-0.27}$ & $9.49^{+1.19}_{-1.06}$ & \nodata & \nodata & $11.67^{+1.01}_{-0.96}$ & $3.32^{+0.18}_{-0.26}$ & $0.47^{+1.01}_{-0.96}$ & $4.12^{+0.18}_{-0.26}$ \\
LMC SR2 & $1.22^{+0.25}_{-0.23}$ & $7.35^{+0.87}_{-0.73}$ & \nodata & \nodata & $11.06^{+0.82}_{-0.90}$ & $2.78^{+0.12}_{-0.13}$ & $0.40^{+0.82}_{-0.90}$ & $3.32^{+0.12}_{-0.13}$ \\
LMC SR3 & $0.75^{+0.50}_{-0.36}$ & $9.06^{+1.11}_{-1.10}$ & \nodata & \nodata & $10.90^{+1.26}_{-1.21}$ & $2.97^{+0.17}_{-0.15}$ & $0.37^{+1.26}_{-1.21}$ & $3.84^{+0.17}_{-0.15}$ \\ \hline
\multicolumn{9}{c}{SFE-Driven Burst Models in the SMC (Figures~\ref{fig:sfedriven_subregion},~\ref{fig:gce_smc_hist}, and~\ref{fig:sfedriven_smc}) } \\ \hline
SMC Body & $3.78^{+0.50}_{-0.58}$ & $9.39^{+3.57}_{-1.49}$ & \nodata & \nodata & $4.52^{+1.16}_{-0.81}$ & $2.21^{+0.19}_{-0.12}$ & $0.57^{+1.16}_{-0.81}$ & $3.45^{+0.19}_{-0.12}$ \\
SMC Wing & $3.11^{+1.08}_{-1.13}$ & $19.70^{+3.60}_{-2.87}$ & \nodata & \nodata & $7.21^{+1.67}_{-0.94}$ & $3.06^{+0.17}_{-0.13}$ & $0.44^{+1.67}_{-0.94}$ & $3.79^{+0.17}_{-0.13}$ \\ \hline
 SMC Primary & $4.31^{+0.43}_{-0.40}$ & $12.54^{+0.71}_{-0.98}$ & \nodata & \nodata & $5.73^{+0.63}_{-0.57}$ & $2.58^{+0.06}_{-0.06}$ & $0.36^{+0.63}_{-0.57}$ & $7.85^{+0.06}_{-0.06}$ \\
SMC Secondary & $0.65^{+0.90}_{-0.50}$ & $6.39^{+0.47}_{-0.57}$ & \nodata & \nodata & $4.97^{+1.51}_{-2.09}$ & $8.20^{+0.82}_{-1.53}$ & $0.23^{+1.51}_{-2.09}$ & $7.84^{+0.82}_{-1.53}$ \\ \hline
SMC Foreground & $4.64^{+0.94}_{-3.36}$ & $12.67^{+2.93}_{-7.51}$ & \nodata & \nodata & $8.81^{+2.34}_{-2.23}$ & $2.94^{+5.44}_{-0.16}$ & $0.53^{+2.34}_{-2.23}$ & $6.13^{+5.44}_{-0.16}$\\
SMC Background & $5.11^{+0.45}_{-1.17}$ & $14.91^{+0.86}_{-1.72}$ & \nodata & \nodata & $7.99^{+1.35}_{-1.19}$ & $2.86^{+0.24}_{-0.08}$ & $0.38^{+1.35}_{-1.19}$ & $5.53^{+0.24}_{-0.08}$\\ \hline
\enddata
\tablecomments{The best-fit parameters are calculated from the 16$^{\rm th}$, 50$^{\rm th}$, and 84$^{\rm th}$ percentiles of the marginalized posterior probability distributions. We list results for fiducial GCE models (bolded) assuming the Scylla SFR as input in which \yIa\ and 
\yCC\ are free parameters (Section~\ref{sec:gce_scylla}; Figure~\ref{fig:photsfh}). We also include parameters for fiducial global models in which temporal variations in the SFE drive a starburst (bolded) assuming fixed yields (Section~\ref{sec:gce_sfedriven}; Figure~\ref{fig:sfedriven}), and alternate versions of the global SFE-driven starburst models where the yields are free parameters (right panel of Figure~\ref{fig:sfedriven_yield}). For the SMC, we also provide a version of the global SFE-driven starburst model assuming default yields \textit{without} stringent priors on $\tau_b$ and $dt$, which results in a degenerate posterior probability distribution (Figure~\ref{fig:gce_smc_hist}).
The bottom rows list parameters for SFE-driven starburst models for different spatial subregions of the MCs (Section~\ref{sec:sfedriven_lmc},~\ref{sec:sfedriven_smc}; Figures~\ref{fig:sfedriven_subregion},~\ref{fig:gce_smc_hist}, and~\ref{fig:sfedriven_smc}), including models for two chemically distinct statistical populations in the SMC extracted from the global degenerate model for the SMC. 
The total star formation duration (parameterized by $dt$) and burst timing \taub\ are inversely related to the total Fe yield scale (dictated by \yIa\ + \yCC\ in our models), which results in low values of $dt$ when assuming the default \texttt{VICE} yields in the case of the fiducial SFE-driven burst models. The mass loading factor \outflow\ is also inversely related to \yIa. 
}
\end{deluxetable*}

Table~\ref{tab:gce_priors} lists the priors adopted for \Fb, \taub, \sigmab, and $dt$ when sampling the posterior probability distribution using Equation~\ref{eq:loglike}. We restrict \sigmab\ to a maximum value of 1 Gyr given that the longest expected starburst duration is $\sim$3 Gyr, or approximately 3\sigmab, based on the values measured by B25 from Scylla data in the MCs.  We also require that the total SF duration ($dt \times 10^3$) is at least 2 Gyr,  or approximately the ISM equilibrium timescale (Section~\ref{sec:gce_scylla}), but less than 5 Gyr, to ensure that a starburst occurs on a timescale such that the model track broadly matches the chemical abundance distributions.  We assume that the time of peak SFE in the simulation \taub\ occurs before the maximum SF duration of 5 Gyr. 

Without these restrictive priors on the timing parameters (\taub\ and $dt$), we recover degenerate solutions in the SMC characterized by a dominant population 
that experiences a starburst and a secondary population without a starburst (Section~\ref{sec:sfedriven_smc}). We therefore selected the priors for our fiducial SFE-driven burst models to trace the dominant solution in the posterior probability distributions for the SMC. In contrast, the LMC shows only a single solution in its posterior probability distributions even in the case of broad priors on the timing parameters ($\tau_b \sim \mathcal{U}(0,10)$, $dt \sim \mathcal{U}(2,10)$).

\added{As discussed in Section~\ref{sec:gce}, excluding corrections for the APOGEE selection function has a non-negligible impact on the SFE-driven burst models, although the resulting change in the inferred model parameters is typically marginal ($\lesssim$1.1$\sigma$). The most notable effects from excluding selection function corrections occur for more metal-rich, and less observationally complete, stellar populations (Figure~\ref{fig:sfunc}) located in the central regions of the MCs, such as the SMC body (Section~\ref{sec:sfedriven_smc}).

}

\subsubsection{Fiducial Yields for SFE-Driven Models}
\label{sec:sfedriven_yield}

\added{In contrast to the models with input Scylla SFHs (Section~\ref{sec:gce_scylla}), in which the nucleosynthetic yields can be empirically constrained by fits to the APOGEE chemical abundance distributions, the SFE-driven models are not anchored to observational SFH properties by definition. Instead, the SFE-driven models are designed to constrain the starburst properties across the MCs, such that the yields are not primary parameters of interest, as long as the adopted values are theoretically or empirically justifiable and provide reasonable fits to the abundance distributions for relative (not absolute) comparisons of starburst properties.
}

To facilitate comparisons between GCE models fitted to the LMC and SMC, or to different spatial subregions within each galaxy, we must \added{therefore} assume a fixed set of yields \added{across all SFE-driven models} 
because the SF duration (and thus the scaling of other timing parameters like \taub\ and \sigmab) depends on the rate of Fe enrichment \added{dictated by the yields} (\yCC\ + \yIa). \added{Moreover, the model parameters and predicted chemical tracks are significantly affected by the well-known degeneracy between outflows and the yield scale (Section~\ref{sec:gce}).}

\added{We adopted the theoretically-motivated set of default \texttt{VICE} yields for our fiducial SFE-driven models for greater ease of comparison to previous work in the MCs (N20; H21), which used theoretical CCSNe yields from \citet{ChieffiLimongi2004} and \citet{LimongiChieffi2006} and SNe Ia yields from \citet{Iwamoto1999} adopted in \texttt{flexCE} \citep{Andrews2017}. Although}
the SF durations in these models are short ($<$ 5 Gyr) \added{because the default yields} result in rapid Fe enrichment (Section~\ref{sec:gce_scylla}), \added{the resulting models} are able to produce a reasonable fit to both the shape of the median chemical abundance track in APOGEE \textit{and} the density of stars in the abundance plane, particularly for the metal-poor tail of the distribution. \added{The SFE-driven models assuming default yields are therefore sufficient for relative comparisons of the starburst properties between and within the MCs.}

This stands in contrast to \added{SFE-driven models} adopting the \added{empirical} yield set from W24 \added{assuming \MgFeCC\ = 0.45 as in Section~\ref{sec:yields}} (Figure~\ref{fig:sfedriven_yield}), which overproduce metal-poor stars in the predicted stellar density distributions \added{for both the LMC and SMC} despite seemingly good fits to the median chemical abundance tracks characterized by long SF durations ($\sim$10--12 Gyr) and low SFE (Figure~\ref{fig:sfedriven_yield}).\added{\footnote{The overproduction of metal-poor stars in the predicted MDFs of the SFE-driven burst models assuming W24 yields is caused by the slower chemical evolution of the model, as opposed to the weighting of the APOGEE selection function favoring metal-poor over metal-rich stars.}} \added{This demonstrates that empirically-motivated yields, even when anchored to measurements of the CCSNe Fe mass yield \citep{Rodriguez2023}, do not necessarily produce better fits to the observed chemical abundance distributions despite the longer SF durations of the associated models.}


\begin{figure}
    \centering
    \includegraphics[width=0.5\textwidth]{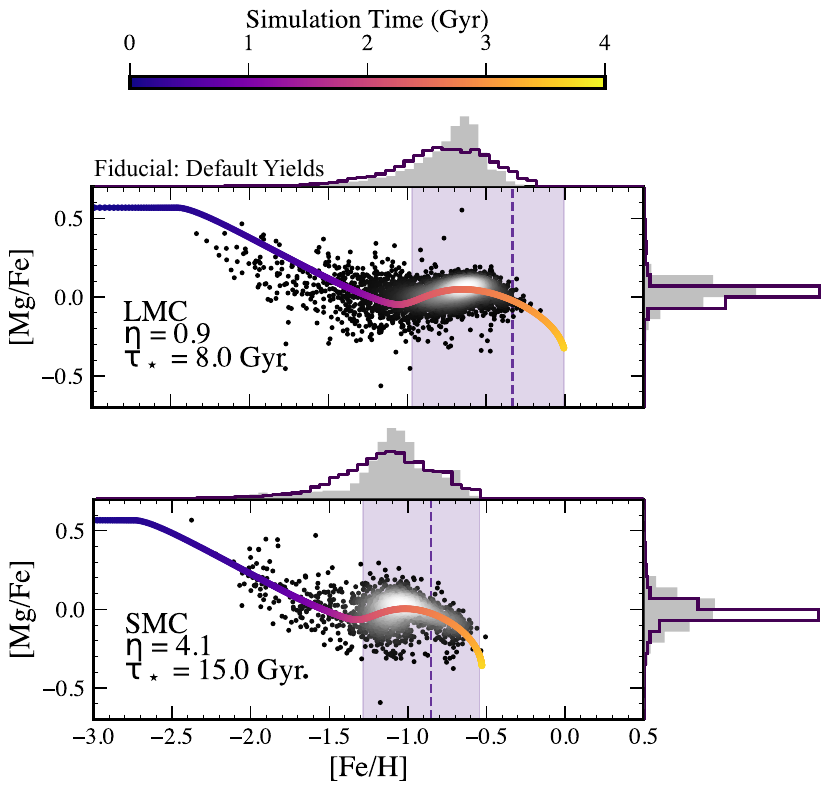}
    \caption{\textbf{Fiducial SFE-driven starburst models fitted to the MCs assuming default yields.} 
    GCE models fitted to APOGEE chemical abundance distributions assuming the theoretically-motivated
    \texttt{VICE} default yields (Section~\ref{sec:gce_sfedriven}). This produces a better match to the metal-poor tails of the MDFs than models assuming the empirical W24 yields (left panels of Figure~\ref{fig:sfedriven_yield}), albeit with shorter total SF duration, and enables direct comparison between the MCs in contrast to models where the Fe yields are free parameters (right panels of Figure~\ref{fig:sfedriven_yield}).
    The dashed vertical purple line (shaded purple vertical bands) represents the time of peak SFE \taub\ (burst duration 2.5\sigmab) converted to metallicity based on the modeled AMR.
    The SMC has stronger outflows and less efficient star formation than the LMC. The LMC also has a stronger starburst than the SMC, which may 
    also occur later than that of the SMC (Table~\ref{tab:gce_pars}).
    }
    \label{fig:sfedriven}
\end{figure}

\begin{figure*}
    \centering
    \includegraphics[width=\linewidth]{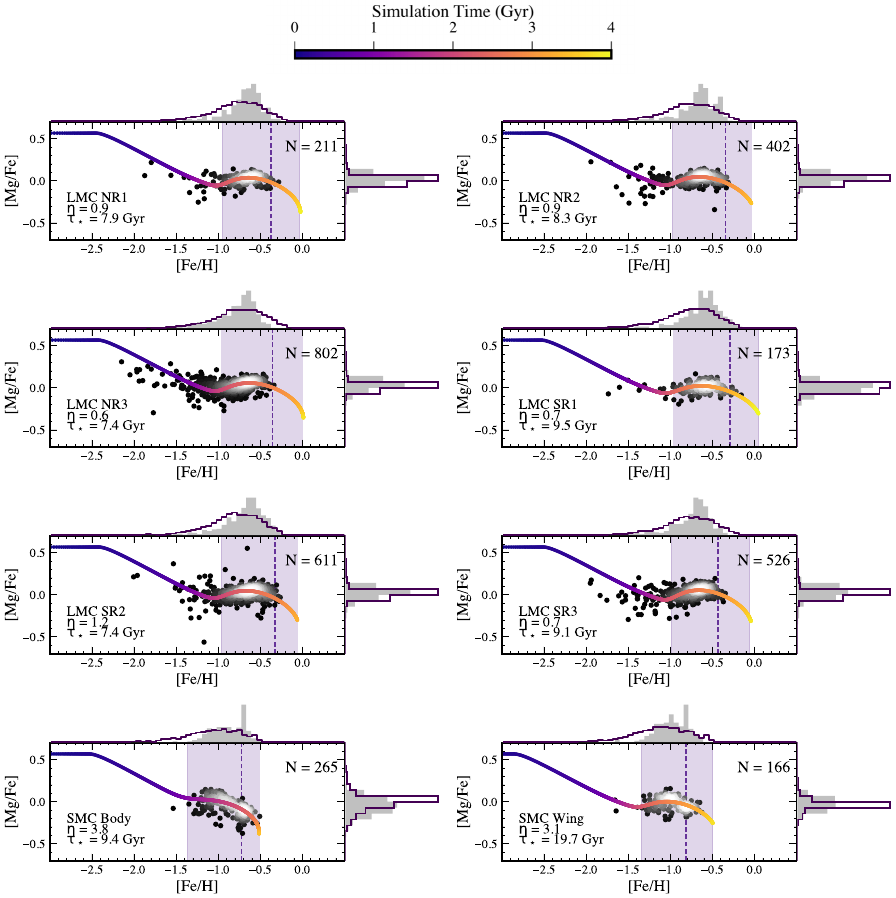}
    \caption{\textbf{Fiducial SFE-drvien starburst models fitted to spatial subregions in the MCs assuming default yields.} Similar to Figure~\ref{fig:sfedriven}, except for models fitted to chemical abundance distributions of LMC and SMC subregions (Section~\ref{sec:region}), including radial divisions in the northern and southern halves of the LMC disk, and the SMC body and wing (see Sections~\ref{sec:sfedriven_lmc},~\ref{sec:sfedriven_smc} for L/SMC models and Table~\ref{tab:gce_pars} for model parameters). 
    The models predict that the chemical evolution and starburst properties across the LMC are fairly uniform. 
    However, they suggest an earlier and weaker starburst at intermediate radii (R2), a later response in the LMC bar (SR1) and lower SFE in the region containing the SE disk (SR3), where the southern LMC disk shows more variation in its starburst properties than the north.
    In the SMC, the dominant stellar population in the wing has a lower SFE and a stronger and later starburst compared to same population in the body. 
    }
    \label{fig:sfedriven_subregion}
\end{figure*}

\added{To simultaneously reproduce the observed stellar density distributions and median chemical tracks of the MCs with empirical yields,} 
we \added{can instead} allow the Fe yields to vary as free parameters in the SFE-driven burst models 
(as in Section~\ref{sec:gce_scylla}, \added{adopting the absolute yield scale and priors in Table~\ref{tab:gce_priors}}). \added{Figure~\ref{fig:sfedriven_yield} shows that} we can achieve SF durations as long as $\sim$6--7 Gyr 
\added{while producing reasonable fits to the abundance distributions given the best-fit yields in Table~\ref{tab:gce_pars}}.  We present all best-fit parameters \added{for the variable Fe yield models, including \outflow\ and \sfe,} in Table~\ref{tab:gce_pars}. 
However, we do not adopt \added{the variable Fe yield models} as our fiducial \added{SFE-driven burst} models given the aforementioned limitations in placing the derived model parameters (including \outflow\ and \sfe) on the same \added{yield} scale for comparison \added{between and within the MCs}.\footnote{In addition, the lower number of data points in some spatial subregions, such as the SMC body and wing, makes it difficult to precisely constrain the parameter distributions in this 8-dimensional model (\outflow, \sfe, \yCC, \yIa, \Fb, \taub, \sigmab, $dt$).} \added{Moreover, the variable Fe yield and W24 models demonstrate that trends in the relative strength and timing of starbursts in the MCs inferred from the fiducial models with theoretically-motivated yields are robust against the underlying yield assumptions 
(see also Section~\ref{sec:sfedriven_global} and Section~\ref{sec:sfedriven_lmc}).}

\subsubsection{Global SFE-Driven Burst Models for the L/SMC}
\label{sec:sfedriven_global}

\added{Table~\ref{tab:gce_pars} presents best-fit parameters for the fiducial global SFE-driven burst models for the LMC and SMC (Section~\ref{sec:sfedriven_yield}). We also present a global SFE-driven burst model for the SMC in Table~\ref{tab:gce_pars} without restrictive priors on the timing parameters required to isolated the dominant stellar population in the SMC
(Section~\ref{sec:gce_sfedriven}) 
that provides evidence for a degenerate and distinct secondary population in the SMC \textit{without} a starburst
(see Section~\ref{sec:sfedriven_smc} for further discussion).}

In our fiducial global SFE-driven burst models (Table~\ref{tab:gce_pars}; Figure~\ref{fig:sfedriven})
, 
the LMC has weaker outflows and more efficient star formation than the dominant stellar population in the SMC, in 
agreement with H21 and models constrained by Scylla SFHs (Section~\ref{sec:gce_scylla}). These differences in the inferred model parameters (\outflow, \sfe) 
are consistent with expectations based on galaxy mass, where galaxies with deeper gravitational potential wells experience more efficient star formation and weaker mass-loading (e.g., \citealt{FinlatorDave2008,PeeplesShankar2011,Muratov2015}).

Moreover, the 
models indicate that the LMC has a stronger starburst at \sigmafb$\sigma$ confidence by a factor of $F =$ \factorfb\ and similar burst timing compared to the SMC, in terms of both \taub\ (\sigmataub) and \sigmab\ (\sigmasigmab), with the possibility that the peak SFE occurred later ($F =$ \factortaub) and that the burst was longer ($F$ = \factorsigmab) in the LMC. These trends in relative burst strength and timing are robust against the assumed yield scale, where increasing the SF duration only increases the statistical significance of the tentative difference in burst timing between the MCs (see Table~\ref{tab:gce_pars} and Figure~\ref{fig:sfedriven_yield}). 

Thus, our results suggest that the dominant starbursts in the LMC and SMC traced by APOGEE RGB stars, which have estimated mean stellar ages of \agergblmc\ Gyr and \agergbsmc\ Gyr respectively (Section~\ref{sec:fake}), occurred approximately at the same time. This is in qualitative agreement with the global Scylla SFRs, which demonstrate that the RGB stars trace simultaneous $\sim$3 Gyr ago bursts in the LMC and the main body of the SMC, which dominates its SF density (Figures~\ref{fig:sfr_lmc} and~\ref{fig:sfr_smc}; B25). Additionally, our models imply that the LMC experienced a stronger response to the event triggering its starburst than the SMC. This agrees with the factor of 1.5 difference in burst strength between the LMC and SMC independently found by H21 using \texttt{flexCE} models. However, the relative strength of the starbursts in the MCs predicted from GCE modeling is at odds with the SFH-based results of B25, who found that the SMC experienced a factor of 
$\sim$3 increase in its SFR during the $\sim$3 Gyr ago starburst compared to the 
$\sim$2$\times$ increase for the LMC during a similar time period (see also Section~\ref{sec:csfh}). Finally, we note that we are able to reasonably reproduce the chemical abundance distributions of the MCs with SF durations of only $\sim$4 Gyr (Table~\ref{tab:gce_pars}), which may result from a combination of short ($\sim$3 Gyr) ISM equilibrium timescales in the MCs (Section~\ref{sec:gce_scylla}) coupled with the fact that the RGB stars, although spanning a wide range of ages, mostly trace a single dominant starburst event occurring within the last few Gyr.

\subsubsection{SFE-Driven Starburst Models in the LMC}
\label{sec:sfedriven_lmc}

We repeated the modeling procedure detailed above (in the case of \textit{global} chemical abundance distributions) for spatial subregions in the LMC 
(Section~\ref{sec:region_lmc}). We present the best-fit model parameters in Table~\ref{tab:gce_pars} and a subset of model tracks, which correspond to radial zones separately analyzed in the northern and southern halves of the LMC disk, 
in comparison to the observed 2D chemical abundance distributions in Figure~\ref{fig:sfedriven_subregion}. 

In the LMC, we find that the chemical evolution and starburst characteristics are consistent within $\sim$1$\sigma$ when separating the galaxy into northern (N) and southern (S) halves of the disk. However, statistically detectable differences emerge when comparing across radial regions (R1, R2, and R3,  where R1 corresponds to the inner LMC, \rdisklmc\ $< 1/2$ \rinvlmc, and R3 corresponds to the outer LMC, \rdisklmc $>$ \rinvlmc). The onset of the dominant starburst occurs later in the central region containing the LMC bar (R1), where $\tau_{b, {\rm R1}} > \tau_{b,{\rm R2}}$ at 3$\sigma$ confidence ($\tau_{b, {\rm R1}} > \tau_{b,{\rm R3}}$ at 2.5$\sigma$). The strength of the starburst may also be weaker at intermediate radii (R2), where $F_{b, {\rm R2}} < F_{b, {\rm R1}} \sim F_{b, {\rm R3}}$ at 1.4$\sigma$), and the SFE may be lower at interior radii (R1; $\tau_{\star, {\rm R1}} > \tau_{\star, {\rm R2}}$ at 1.2$\sigma$). \added{The trends of a later starburst in R1 and a weaker starburst in R2 persist when instead assuming variable Fe yields (Section~\ref{sec:sfedriven_yield}), albeit at weaker statistical significance (1.1$\sigma$ and 1.3$\sigma$ respectively).}
A weaker and earlier starburst at intermediate radii (R2) is in good agreement with Scylla SFH results (Section~\ref{sec:csfh}).

Upon further subdividing the LMC disk into radial zones depending on their location in the northern or southern half of the disk, we find that the northern half of the LMC disk has consistent chemical evolution and starburst properties within 1$\sigma$ at all radii, whereas the southern LMC disk demonstrates more variation (in qualitative agreement with \citealt{Cohen2024a}). In the southern disk, the starburst is likely triggered last in the inner galaxy region (SR1) containing the LMC bar ($\tau_{b, {\rm SR1}} > \tau_{b, {\rm SR2}}$ at 1.9$\sigma$). The SFE may also be marginally lower in SR1 and the outer southern disk (SR3), which contains the SE disk, compared to intermediate radii (SR2; at 1.6$\sigma$ and 1.2$\sigma$, respectively).


\begin{figure*}
    \centering
    \includegraphics[width=\linewidth]{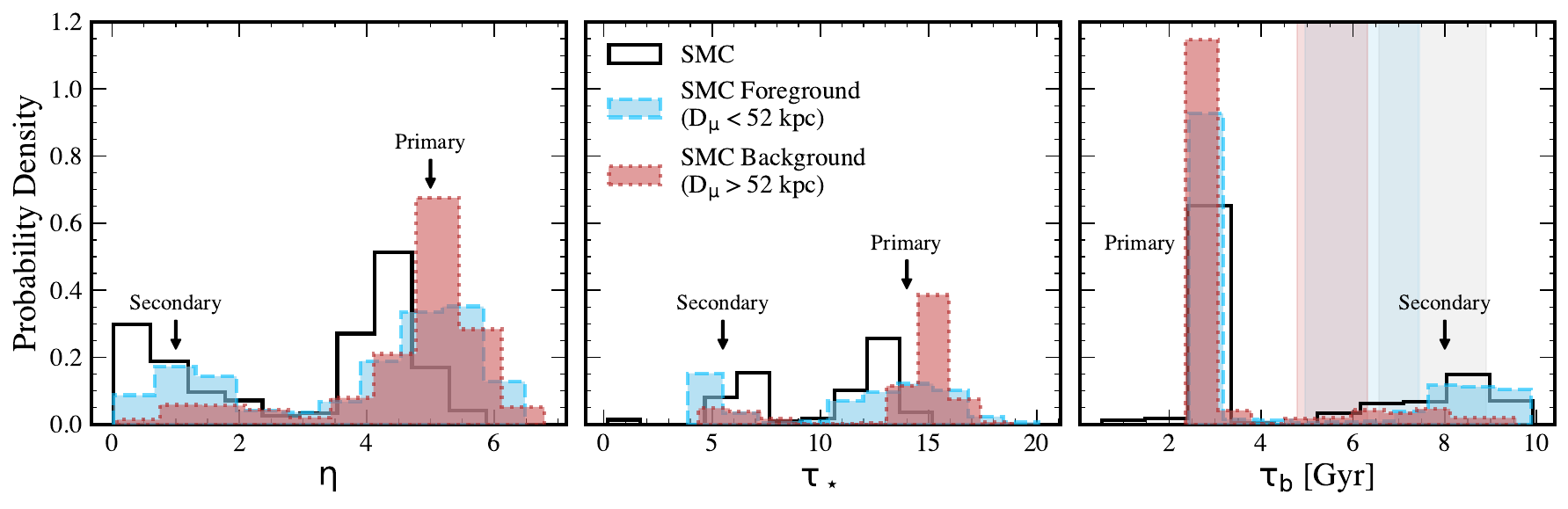}
    \caption{\textbf{Degenerate SFE-driven burst models assuming default yields for foreground and background samples along the line-of-sight to the SMC.} Marginalized posterior probability distributions 
    (Section~\ref{sec:gce})
    for the mass-loading factor $\eta$ (left), SFE timescale \sfe\ (middle), and timing of the peak SFE of the dominant starburst (right) for degenerate SFE-driven starburst models in the SMC. We show models for all stars in the SMC (black outlined histograms), a foreground SMC sample ($D_\mu$ $<$ 52 kpc; blue dashed histograms; \citealt{Almeida2024}; Section~\ref{sec:abund}), and background SMC sample ($D_\mu$ $>$ 52 kpc; red dotted histograms) \textit{without} stringent restrictions on the timing of the starburst or total SF duration (as in the fiducial models; Table~\ref{tab:gce_priors}). The degeneracy in the 
    distributions reveals two populations with distinct properties: (1)
    a dominant, or primary, SMC population, likely associated with the main body, that has stronger outflows, lower SFE, and experiences a starburst, and (2) a secondary population that is more pronounced in the foreground, and is characterized by weak outflows, higher SFE, and no clear starburst signature ($\tau_b \gtrsim dt$; shaded vertical regions).
    }
    \label{fig:gce_smc_hist}
\end{figure*}

When comparing locations in the northern and southern halves of the LMC disk at fixed radius, we find that the properties of the the inner galaxy are generally consistent, with marginal evidence in favor of a stronger and later starburst in SR1 than NR1 (at 1.2$\sigma$ and 1.3$\sigma$ confidence, respectively). Additionally, the outer southern disk (SR3) may have lower SFE compared to the outer northern disk (NR3; at 1.3$\sigma$ confidence).
Overall, the fairly uniform nature of starburst properties derived from chemical abundance distributions in the LMC is in qualitative agreement with SFH fits from Scylla (Section~\ref{sec:csfh}). Across all localized regions considered by B25, the LMC was characterized by a single dominant starburst $\sim$3 Gyr ago, with evidence that this burst 
may have been stronger in the LMC bar region.

\subsubsection{SFE-Driven Starburst Models in the SMC}
\label{sec:sfedriven_smc}

Similar to Section~\ref{sec:sfedriven_lmc}, we model the chemical evolution of spatial subregions in the SMC corresponding to its main body and wing (Section~\ref{sec:region_smc}), presenting the \added{fiducial} model parameters and associated tracks in Table~\ref{tab:gce_pars} and Figure~\ref{fig:sfedriven_subregion}. We also present a model for the \added{for 
secondary SMC population without a starburst}
(Table~\ref{tab:gce_pars} and Figure~\ref{fig:gce_smc_hist}) \added{that emerges from the degenerate global SMC burst model (Section~\ref{sec:sfedriven_global})}. 
\added{Here, we show that both the SMC body and wing have contributions from the dominant and secondary SMC populations, such that the properties of the secondary population are informative for understanding the behavior of starbursts in each spatial subregion of the SMC.}

Figure~\ref{fig:gce_smc_hist} shows marginalized posterior probability distributions for the most relevant model parameters ($\eta$, $\tau_\star$, $\tau_b$) \added{in the degenerate global SMC burst model, where the distributions reveal two peaks corresponding to the dominant (or primary) and secondary populations.} We extracted the model parameters for the two SMC solutions based on the criterion $\tau_b$ $<$ 4 Gyr for the primary population with higher probability density ($\tau_b$ $>$ 4 Gyr for the secondary population), 
which results in a clean separation of the full posterior probability distributions. 

The primary SMC population is characterized by significant outflows ($\eta \sim 4$), low SFE ($\tau_\star \sim 12$ Gyr$^{-1}$), 
and a starburst that occurs $\tau_b \sim 3$ Gyr into the model, well before the cessation of SF.\footnote{Although the total SF duration is longer in these alternate models, the chemical evolution significantly slows for $t \gtrsim 4$ Gyr, such that only a few Gyr of evolution is required to match the overall chemical abundance distribution as in the fiducial models. The longer SF durations preferred by these models may result from the secondary mode without a starburst, which requires $\tau_b \gtrsim (dt \times 10^3)$.} 
In comparison, the secondary population has negligible outflows ($\eta \sim 0.7$), high SFE ($\tau_\star \sim 6$ Gyr$^{-1}$), and \textit{no starburst} ($\tau_b \gtrsim dt \times 10^3$). 
The mass-loading factor (SFE timescale) of the secondary SMC population is inconsistent with the dominant population at the \sigmasecondarysmceta\ (\sigmasecondarysmcsfe) level (Table~\ref{tab:gce_pars}), but is consistent with the same model parameters for the SR2 subregion in the southern LMC disk within $\lesssim$1.1$\sigma$ (and marginally consistent with spatial subregions in the northern LMC disk within $\lesssim$2$\sigma$).

Tracing the primary stellar population in the SMC (Table~\ref{tab:gce_pars}), the fiducial SFE-driven burst models 
indicate that
the wing experienced a lower SFE compared to the same population in the main body (\sigmasfewingbody, $F = $ \factorsfewingbody), as well as a stronger starburst (\sigmafbwingbody, $F = $ \factorfbwingbody) with a later onset (\sigmataubwingbody, $F = $ \factortaubwingbody).\footnote{Excluding corrections for the APOGEE selection function in the SMC (Figure~\ref{fig:sfunc}) has a significant impact on the main body, which is relatively metal-rich and has a sparsely sampled metal-poor tail (Figure~\ref{fig:sfedriven_subregion}), owing the the lower observational completeness in the central regions of the SMC. Without selection function corrections, the SMC body has weaker SFE and stronger outflows, and larger burst strength and delayed (more metal-rich) burst timing. In this case, $F_b$ is consistent between the SMC body and wing.} In contrast, the Scylla SFHs suggest that the SMC wing 
did not experience a starburst similar to the main body, but rather a \added{near-}constant SFH over the last $\sim$3 Gyr (Figure~\ref{fig:sfr_smc}).
However, the \added{approximately} flat SFR of the SMC wing likely results from the superposition of multiple stellar populations along the line-of-sight (C24b), the presence of which may be related to our findings of two \added{distinct} statistical populations in the SMC \added{with and without a starburst.}
Indeed, B25 found evidence of a starburst occurring $\sim$3 Gyr ago in some localized regions contained within our wing field (their Wing-Bridge/North-Bridge and Wing-Bridge/Center regions), but not in others (their Wing-Bridge region). 

\begin{figure*}
    \centering
    \includegraphics[width=\linewidth]{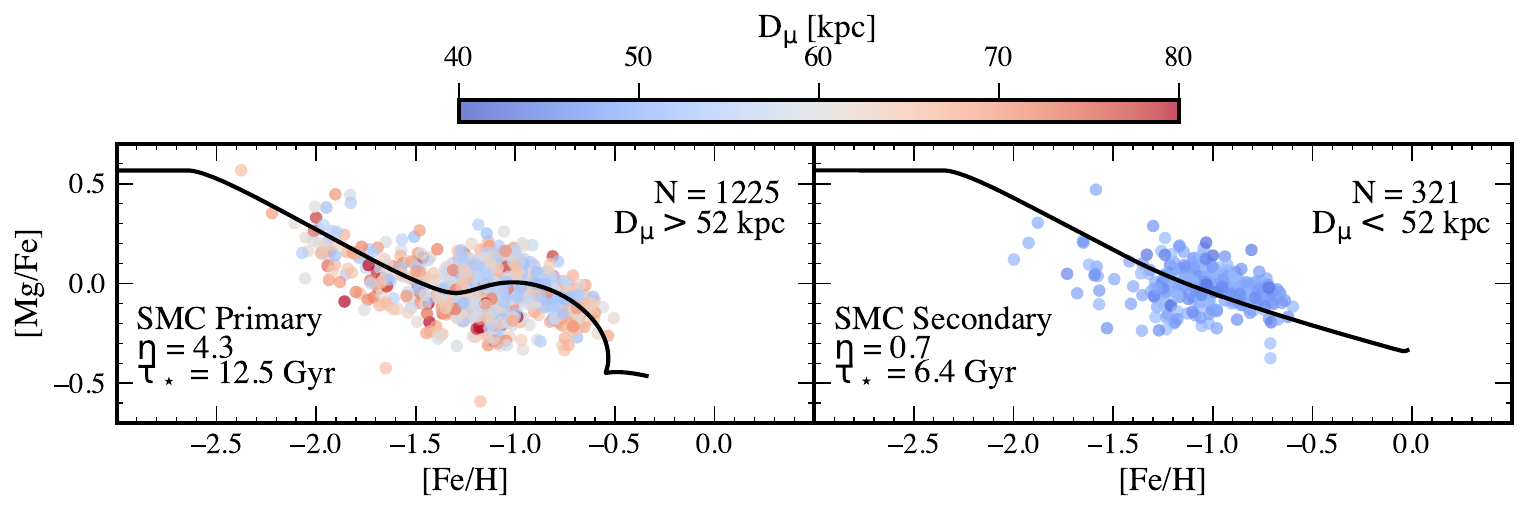}
    \caption{\textbf{Fiducial SFE-driven burst models assuming default yields for dominant (primary) and secondary populations in the SMC.}
    Similar to Figures~\ref{fig:abund_smc} and~\ref{fig:sfedriven_smc} for the SMC, except separated into models corresponding to the primary (left) and secondary (right) 
    statistical populations of the SMC (Section~\ref{sec:sfedriven_smc}, Table~\ref{tab:gce_pars}). We show the SMC background ($D_\mu$ $>$ 52 kpc) and foreground ($D_\mu$ $<$ 52 kpc) samples in the left and right panels respectively for reference, \textit{though these GCE models were not fitted to the background and foreground samples}. Although stars tracing the primary and secondary populations are present both in the background and foreground of the SMC (see also \citealt{Almeida2024}), the primary and secondary populations are roughly anchored to chemical features in the background and foreground populations respectively.  
    }
    \label{fig:sfedriven_smc}
\end{figure*}

\added{Compared to the primary SMC population,} the origin of the secondary SMC population 
is not immediately clear. The 
secondary population is recovered in both the SMC body and wing \added{when each subregion is modeled without restrictive priors on the timing parameters (Section~\ref{sec:gce_sfedriven})}, albeit with 
larger relative contribution in the wing.
Given that the wing contains more metal-rich stars located in the SMC foreground \added{than the main body} (Section~\ref{sec:abund}; Figure~\ref{fig:abund_smc}), we explored whether the well-known distance bimodality along the line-of-sight to the SMC could be related to the two chemically distinct statistical populations in the SMC. 

We separated the SMC into foreground and background samples delineated by $D_\mu = 52$ kpc (following \citealt{Almeida2024}; Section~\ref{sec:abund}) and modeled their chemical evolution assuming broad priors on the timing parameters (Section~\ref{sec:gce_sfedriven}). Table~\ref{tab:gce_pars} includes the best-fit model parameters for the SMC foreground and background, where the large uncertainties on the values reflect the degenerate solutions visible in Figure~\ref{fig:gce_smc_hist}.
The degenerate solutions recovered from the SMC foreground and background samples are consistent with the primary and secondary populations detected in the degenerate global SMC burst model (Figure~\ref{fig:gce_smc_hist}). The main distinction is that the probability density of the primary solution is higher (lower) in the background (foreground) population, and vice versa for the secondary solution.


Figure~\ref{fig:sfedriven_smc} shows 
models with and without a starburst corresponding to the primary (left panel) and secondary populations (right panel) respectively in the SMC (Table~\ref{tab:gce_pars}), compared to the background and foreground samples in the SMC. 
Although the correspondence between primary (secondary) and background (foreground) populations is approximate, the primary population is anchored to features in the chemical abundance distribution of the SMC background sample, such as the metal-poor tail ([Fe/H] $\lesssim$ $-1.5$) and the increase in [Mg/Fe] as a function of metallicity ([Fe/H] $\gtrsim$ $-1.5$). The SMC foreground sample lacks a pronounced metal-poor tail and has a less clear burst signature in the [Mg/Fe]--[Fe/H] plane, indicating that the secondary population likely represents a relatively metal-rich population that has more scatter in its chemical abundance distribution, or is more stochastically sampled owing to low number statistics, \added{such that a starburst cannot be clearly identified by the model. This secondary, more metal-rich SMC population may trace the younger, metal-rich group of stars observed in the SMC MDF (see Section~\ref{sec:scylla_fiducial}).}

In addition, when comparing the 
SMC foreground sample 
to 
the statistically similar LMC SR2 subregion, we find significant overlap in their chemical abundance distributions 
between $-1.5 \lesssim {\rm [Fe/H]} \lesssim -0.9$, preceding the 
burst signature in the LMC. We discuss potential origins for the secondary SMC population in the context of the 
LMC-SMC interaction in Section~\ref{sec:burst}.

\section{Discussion \& Conclusions}
\label{sec:discuss}

We have combined CMD-based star formation histories from the pure-parallel \textit{HST}-based Scylla survey 
(\citealt{Murray2024}; C24a,b) with chemical abundance measurements of RGB stars from APOGEE (N20; \citealt{Povick2024}) to systematically investigate the connection between bursts of star formation and chemical abundance distributions in the interacting MC system. Specifically, we have modeled the chemical evolution of the MCs using 
\texttt{VICE} (JW20) following a two-pronged approach to (1) predict 
abundance signatures in the [Mg/Fe]--[Fe/H] plane directly from SFHs measured via CMD fitting techniques given reasonable GCE assumptions (Section~\ref{sec:gce_scylla}) and (2) infer SFH properties from the 
abundance distributions assuming a 
dominant SFE-driven starburst across spatial subregions with both Scylla and APOGEE data in the MCs (Section~\ref{sec:gce_sfedriven}).  

To our knowledge, this work presents the first GCE models simultaneously constrained by 
measurements of input 
SFHs and chemical abundance distributions in the MCs. 
The predictions from this constrained modeling approach broadly agree with the chemical abundance patterns of the MCs, with the main limitations of the methodology being the requirement of starbursts driven by (pristine) gas infall (Figure~\ref{fig:amr}) and the stellar yields. When \textit{independently constrained} by CMD-derived galaxy SFHs, GCE models adopting theoretical yields fail to reproduce observed chemical abundance distributions (Figure~\ref{fig:photsfh_fixedyield}), whereas empirically calibrated prescriptions such as those from W24 or the yields inferred in this work (Figure~\ref{fig:photsfh}) substantially improve agreement with data. The inability of theoretical yields to reproduce 
abundances given independent SFH constraints is not surprising given the poorly understood processes underlying models of massive stellar evolution (e.g., \citealt{Ventura2013,Ertl2016,Frischknecht2016,Beasor2020,Griffith2021}).

\added{We similarly find that models in which the SFH is analytically specified through time-dependent changes in SFE provide reasonable descriptions of the data while recovering starburst properties (Section~\ref{sec:burst}), albeit regardless of the underlying assumption of theoretical or empirical yields (Section~\ref{sec:sfedriven_yield}). However, efficiency-driven starburst models may better reproduce the observed shape of the [Mg/Fe]--[Fe/H] sequence in the MCs than accretion-driven models, where the former predicts a population of $\alpha$-deficient stars formed from suppressed SF following the burst, in contrast to decreases in [Fe/H] from (pristine) gas infall \citep{JohnsonWeinberg2020}. Indeed, the smooth evolution of the chemical tracks and the decline in [Mg/Fe] at high metallicity predicted by SFE-driven burst models  (Figure~\ref{fig:sfedriven}) may better match APOGEE chemical abundance distributions than gas-driven starburst models with associated $\sim$0.1--0.3 dex decreases in [Fe/H] (Figure~\ref{fig:photsfh}). Thus, the enhanced accretion of gas is not necessary to reasonably model the abundance distributions of the MCs, although we cannot exclude the possibility of starbursts driven by the infall of enriched gas. Moreover, an efficiency-driven starburst scenario for the MCs is a more conservative interpretation given theoretical expectations of \textit{dynamically} triggered starbursts in interacting pairs of low-mass galaxies \citep{Martin2021,WilliamsonHugo2021}.
}

In addition, the  Bayesian framework utilized in this work represents a significant advance in modeling the chemical evolution of the MCs, where we have modeled the \textit{density} of stars corrected for the APOGEE selection function in the 2D chemical abundance plane in addition to median 2D chemical abundance trends. 
\added{This work demonstrates that the impact of incorporating the selection function is minor for stellar populations in the MCs with abundance distributions that are sufficiently metal-poor and/or have well-sampled metal-poor tails (Section~\ref{sec:gce}).}
We discuss the implications of our main findings  (Sections~\ref{sec:gce_scylla} and~\ref{sec:gce_sfedriven}) for the evolutionary history of the MCs in Section~\ref{sec:evo} and for the propagation of starbursts in interacting pairs of low-mass galaxies in Section~\ref{sec:burst}, respectively, \added{prior to summarizing future directions in Section~\ref{sec:future}.}

\subsection{Chemical Evolution of the MCs}
\label{sec:evo}

When constraining GCE models by Scylla SFRs and APOGEE chemical abundances (Section~\ref{sec:gce_scylla}), we find that the predicted chemical tracks for both the LMC and SMC evolve to the equilibrium ISM metallicity of [Fe/H] $\sim-1$ in $\sim$3 Gyr, followed by a series of gas-driven starbursts in which the systems gradually enrich to [Fe/H] $\sim-0.5$ (where the input SFR sets the predicted gas infall history). This rapid chemical enrichment is qualitatively consistent with $\sim$1 Gyr equilibrium timescales in the Milky Way \citep{Johnson2025a}, in which metal-enriched outflows balance the fast re-enrichment process following dilution events (e.g., \citealt{Dalcanton2007}). These rapid equilibrium timescales broadly agree with Scylla AMRs (Figure~\ref{fig:amr}), which indicate that the LMC and SMC enriched to [Fe/H] $\gtrsim-1.5$ within the first few Gyr of their evolution, followed by a subsequent enrichment rate of $\sim$0.05 dex per Gyr.

In addition, the constrained models in Section~\ref{sec:gce_scylla} suggest that a higher Fe yield from SNe Ia may be required in the SMC than the LMC ($y_{\rm Fe}^{\rm Ia,SMC} > y_{\rm Fe}^{\rm Ia, LMC}$, $F$ = \factoryIa; Table~\ref{tab:gce_pars}) to match the observed chemical abundance distributions. This could indicate the need for metallicity-dependent SNe Ia rates 
impacting IMF-integrated net nucleosynthetic yields in SMC-mass and lower mass galaxies.\footnote{Generally, the assumption that SNe Ia yields are independent of the mass and metallicity of the progenitor stars is reasonable for GCE modeling \citep{Andrews2017,Weinberg2017}. Moreover, these studies demonstrate that altering the minimum SNe Ia delay time minimally impacts the abundance distributions, where longer delay times result in small increases in the metallicity of the knee of the $\alpha$-element distribution.} \added{In potential support of this scenario, \citet{Hasselquist2024} found that multi-element chemical abundance distributions of the MCs can be accurately modeled with types of CCSNe and SNe Ia that are \textit{consistent} between the MCs, but possibly different from the MW. If the higher SNe Ia Fe yield in the SMC was instead caused by metallicity-dependent SNe Ia explosion mechanisms, such as sub-Chandrasekhar mass SNe Ia \citep{McWilliam2018,Kirby2019,Kobayashi2020,delosReyes2022,Heiger2026}, this would imply that the SMC and LMC experienced fundamentally different types of SNe Ia. Such metallicity-dependent SNe Ia explosions would produce different chemical signatures for Fe-peak elements in the SMC and LMC that have not yet been detected (\citealt{Hasselquist2024}; but see Section~\ref{sec:future}).
}

\added{Moreover,} recent SNe surveys have provided evidence that SNe Ia rates depend inversely on galaxy stellar mass \citep{Li2011,GraurMaoz2013,Brown2019,Wiseman2021} and therefore metallicity \citep{Kistler2013,Gandhi2022,Johnson2024}. These enhanced SNe Ia rates in low-mass galaxies ($M_\star \lesssim 3 \times 10^9 M_\odot$) may result from higher binary fractions \citep{Badenes2018,Moe2019} and possibly the formation of massive white dwarfs that more readily detonate \citep{Kistler2013} in low metallicity environments \citep{Johnson2024}. It is unlikely that metallicity-dependent SNe Ia rates originate from a higher relative fraction of low-mass stars forming in metal-poor systems, 
where evidence for IMF variations in Local Group dwarf galaxies, including the SMC, is tentative at most (\citealt{Geha2013,Gennaro2018a,Gennaro2018b,El-Badry2017,Filion2022,Filion2024}; R.~Cohen et al.\ 2026, in preparation; but see \citealt{Kalirai2013}).

\subsection{Starbursts in Interacting Pairs of Dwarf Galaxies}
\label{sec:burst}

By modeling the observed increase in $\alpha$-enhancement at high metallicity in the LMC and SMC (N20; H21) as resulting from a single sustained starburst in each galaxy driven by changes in SFE (Section~\ref{sec:gce_sfedriven}), we find that the dominant starbursts in the LMC and in the primary stellar population in the SMC (Section~\ref{sec:sfedriven_smc}) have statistically consistent onsets and durations, with tentative evidence for a slightly earlier burst onset in the SMC, in broad agreement with literature SFHs (\citealt{Weisz2013,Massana2022}; B25). From the age distribution predicted for APOGEE RGB stars from Scylla SFHs (Section~\ref{sec:fake}), we anticipate that these temporally coincident starbursts should trace mean ages of \agergblmc\ and \agergbsmc\ Gyr ago in the LMC and SMC.

In addition, we find that the strength of the starbursts \Fb, parameterized by the factor of decrease in the SFE timescale \sfe, are distinct between the MCs, in agreement with the independent GCE modeling results of H21. However, both this work and H21 find that the inferred relative burst strength in the LMC is \textit{stronger} than in the SMC by a factor of \factorfb\ (1.5 in H21), in contrast to Scylla survey findings demonstrating that the response to the $\sim$3 Gyr ago burst in the LMC is \textit{weaker} than the SMC (B25; Section~\ref{sec:sfh}). Moreover, we find that the magnitude of the enhancement in the peak SFR---not to be confused with the peak SFE (Table~\ref{tab:gce_pars})---during the starbursts inferred from our GCE models 
(factors of $\sim$6.5 and 4.5 respectively) is higher than that measured directly from the SFHs in both the LMC and SMC (
factors of $\sim$2 and 3; Section~\ref{sec:sfh}).

The reason for this discrepancy between GCE model predictions for burst strength in the LMC versus the SMC
compared to SFHs measured from CMD fitting is unclear. In the models, \Fb\ is determined from the increase in [$\alpha$/Fe] at high metallicity, where the SMC indeed has a flatter chemical abundance distribution than the LMC (Figures~\ref{fig:abund_lmc} and~\ref{fig:abund_smc}). Although this flatter chemical abundance trend may be driven by the foreground population in the SMC (which is more prominent in its eastern wing; Section~\ref{sec:sfedriven_smc}), the presence of a secondary population in the SMC is unlikely to suppress the apparent burst strength, which is driven by the dominant stellar population in the SMC main body.
This suggests that additional factors may be suppressing the apparent increase in the $\alpha$-enhancement of the SMC during its major starburst, which could be related to its complexity as a star-forming system (e.g., \citealt{Murray2024smc}).

The available observational evidence from other systems of interacing dwarf galaxies, although limited, suggests that the SMC should 
experience a stronger response than the LMC to the same dynamical trigger, as found from the measured SFHs.
For example, studies of individual pairs of interacting dwarf galaxies have found stronger responses in the secondary \citep{Privon2017,Paudel2020}, in agreement with theoretical expectations from \citet{Besla2012} that starbursts should preferentially occur in the lower-mass companion.
Although there exist systematic studies on the impact of dynamical interactions on star formation rates in isolated pairs of low-mass galaxies ($8 \lesssim \log M_\ast \lesssim 9.7$; \citealt{Stierwalt2015,Sun2020,Chauhan2025}; see also \citealt{Subramanian2024} for lower masses), 
these studies have focused on SF enhancements in dwarf galaxy pairs \textit{as a system}, as opposed to any differences between higher- and lower-mass companions within starbursting pairs of dwarf galaxies. 
However, these studies have found that SF enhancements of factors of $\sim$1--3 are common in interacting pairs of low-mass galaxies, including those near a massive host \citep{Stierwalt2015}, similar to findings from Scylla data in the MCs (B25).

\subsubsection{Internal Variations in Starburst Properties}

When separately modeling the chemical evolution of distinct spatial subregions in each dwarf galaxy (Section~\ref{sec:sfedriven_lmc},~\ref{sec:sfedriven_smc}), 
we find that the timing, duration, and strength of the dominant starburst in the LMC is fairly uniform across its disk, with slightly more variation 
in the south (c.f.\@ \citealt{Cohen2024a}), the most notable exception being 
a later burst onset in the region containing the LMC bar.
These predictions based on GCE models are in qualitative agreement with Scylla survey results on the spatial variations of starbursts in the MCs from B25. We find evidence for more substantial differences in starburst properties between the dominant SMC populations (Section~\ref{sec:sfedriven_smc}) located in the main body 
and the eastern wing. 
In detail, the SMC wing \added{likely} experienced a stronger burst (\added{at \sigmafbwingbody\ confidence}) with a later onset \added{(\sigmataubwingbody}) compared to the main body (Table~\ref{tab:gce_pars}), \added{where larger samples of metal-poor stars in the SMC will enable more precise constraints of its internal starburst properties.} The wing also has a lower efficiency of star formation compared to the SMC body, as does the outskirts of the southern LMC disk compared to the rest of the galaxy. Additionally, we find evidence for a distinct secondary stellar population in the SMC without a starburst that may be more chemically similar to regions in the southern LMC disk (Section~\ref{sec:sfedriven_smc}).

The spatial variation, or lack thereof, in the properties of the dominant starbursts in the MCs may have broader implications for how starbursts propagate in interacting pairs of low-mass galaxies. Specifically, \citet{Sun2020} found that interacting pairs of dwarf galaxies in SDSS ($8 \lesssim \log M_\ast \lesssim 10$) with similar mass companions (mass ratios 4:1 to 1:4) show SF enhancements that may be strongest in the galaxy center (within a half-light radius), although enhanced SF activity is present beyond this radius. This behavior for interaction-triggered SF enhancements over broad spatial scales in low-mass galaxies is similar to that found for massive galaxies ($\log M_\ast \gtrsim$ 9) in SDSS MaNGA following the first pericentric passage of the interacting pair \citep{Pan2019}. However, both enhancement and suppression of SF can occur in the outskirts of massive galaxy pairs depending on interaction  stage \citep{Pan2019,Thorp2019} and stellar mass ratio \citep{Steffen2021}, where the widespread significance of SF suppression in interacting galaxies remains unclear in the low-mass regime \citep{KadoFong2024,Huang2025}.

At face value, the uniformity of the burst strength in the LMC could therefore be consistent with this global picture of SF enhancement as a function of radius. Alternately, the presence of strong radial migration in the LMC \citep{Cohen2024a,Lucy2024,Lucy2025} that preferentially transports older stars from interior radii to the outer disk (e.g., \citealt{SellwoodBinney2002,Roskar2008,Minchev2012,Bird2013}) may drive the lack of spatial dependence in starburst properties. The more pronounced uniformity in the properties of the northern LMC disk (Section~\ref{sec:sfedriven_lmc}) may be driven by spiral structure in the LMC Arm setting the strength of radial migration (see C24a; \citealt{Roskar2012}).

However, the stronger burst in the dominant populations in the SMC wing compared to its central body may be at odds with the above scenario on global SF enhancement in interacting dwarf galaxies, although the spatial distribution of SF demonstrates significant variation on the level of individual star-forming dwarf galaxies (e.g., \citealt{McQuinn2012,Privon2017,Sacchi2018}). This work also suggests a scenario in which the main body of the SMC may have responded first to a single dynamical trigger (see discussion by B25), followed by simultaneous responses 
in the primary SMC population of the eastern wing and in the LMC. 

\subsubsection{Implications for the Origin of the SMC Wing}

Given that the SMC wing may have formed from material tidally stripped from the inner SMC 
and pulled toward the Magellanic Bridge in the direction of the LMC during their most recent interaction \citep{Oey2018,Zivick2018,Niederhofer2021,James2021,Omkumar2021,Dias2022,Almeida2024}, the more recent stellar structure may indeed have distinct starburst characteristics compared to the main body and the western halo of the SMC (\citealt{Rubele2018}; C24b). 
Nevertheless, the older RGB stars ($\gtrsim$1 Gyr ages) employed in this work 
as chemical tracers precede the $\lesssim$250 Myr ago interaction that formed the SMC wing in its present-day capacity \citep{Zivick2018,Choi2022,Rathore2025,Garver2026}, thus a stronger and later burst signature would either be 
driven by the metal-poor population (Section~\ref{sec:abund}) innate to the outskirts of the SMC, or alternately a superposition of this background population and the foreground stars that may have been stripped from the SMC main body. This work supports the former scenario, where the starburst signature in the eastern wing is driven by the dominant stellar population in the SMC (Section~\ref{sec:sfedriven_smc}).

Potentially complicating hypotheses for the origin of stellar populations in the SMC wing, B25 recently found that the SFH and AMR of their Wing/Bridge region is similar to the southwestern LMC, 
positing that the Wing/Bridge may contain stars stripped from the outer LMC, at face value in contrast to \citet{Almeida2024}. In 
support of an LMC origin for a subset of stars located along the line-of-sight to the SMC, we independently find statistical evidence for a secondary stellar population without a clear starburst signature that is chemically distinct from the dominant population in the SMC, but consistent with regions in the southern LMC disk (Section~\ref{sec:sfedriven_smc}). This population, which appears to be preferentially located in the SMC foreground, is found both in the eastern wing and main body of the SMC, although more prominent in the wing (see also \citealt{Almeida2024}). A possible LMC origin for a subset of stars in the SMC foreground may be consistent with the findings of \citet{Almeida2024} that \added{the foreground} stars are more metal-rich than the background SMC population (which resembles the western halo of the SMC), given the significant overlap in chemistry between the LMC and SMC (N20; H21), and that the foreground stars have distinct outflowing kinematics relative to the SMC main body. The secondary SMC population may therefore originate from stars tidally stripped from the southern LMC disk \textit{or} from stars formed in-situ 
from gas tidally stripped or ejected from the LMC. Future work is necessary to confirm whether such a scenario is consistent with dynamical models of the LMC-SMC interaction (e.g., \citealt{Besla2012,Choi2022,Rathore2025,Rathore2025smc,Garver2026}).

\subsection{Future Prospects}
\label{sec:future}

Future advances in chemical evolution modeling of the MCs that incorporate robust star-by-star distance constraints \citep{Oden2025} could aid in further disentangling the nature of the stellar populations along the line-of-sight to the SMC (e.g., \citealt{Almeida2024}). 
In addition, multizone models of chemical evolution \citep{SchonrichBinney2009,Minchev2013,Minchev2014,Johnson2021,Chen2023} could be applied in the LMC to explicitly account for the effects of radial migration \citep{Cohen2024a,Lucy2024,Lucy2025}. Hierarchical joint modeling frameworks could enable more detailed investigations of the underlying processes driving nucleosynthetic yields in the MCs, such as the possibility of metallicity-dependent SNe Ia rates in the SMC. \added{In particular, multi-element GCE modeling including Fe-peak elements with significant contributions from SNe Ia, such as Ni and Mn, 
may reveal whether larger IMF-averaged SNe Ia yields persist across elements in the SMC, thereby supporting an enhanced SNe Ia rate scenario.} 
Moreover, stellar age constraints could be directly incorporated into chemical evolution models (as in \citealt{Johnson2023}), particularly in the case of existing spectrophotometric ages in the LMC \citep{Povick2024}. Alternately, novel methods that link the chemical enrichment histories of galaxies directly to CMD-based measurements of their resolved SFHs \citep{Garling2025,Heiger2026} will likely produce improved observational constraints on the AMRs of the MCs. Finally, larger sample sizes of stellar chemical abundance measurements from SDSSV Magellanic Genesis \citep{Nidever2026} and the 4MOST 1001MC survey \citep{4most1001mc} will be transformative for understanding the chemical evolution of the MCs.

\begin{acknowledgments}
    The authors thank Karl Gordon for his core contributions to the Scylla survey, and Himansh Rathore for insightful discussions on modeling the LMC-SMC interaction.
    IE acknowledges financial support from programs HST GO-15891, GO-16235, and GO-16786, provided by NASA through a grant from the Space Telescope Science Institute, which is operated by the Association of Universities for Research in Astronomy, Inc., under NASA contract NAS 5-26555.
\end{acknowledgments}

\appendix

\section{Lifetime Star Formation and Metallicity Properties}
\label{sec:appendix}

\begin{figure}[h]
    \centering
    \includegraphics[width=0.7\linewidth]{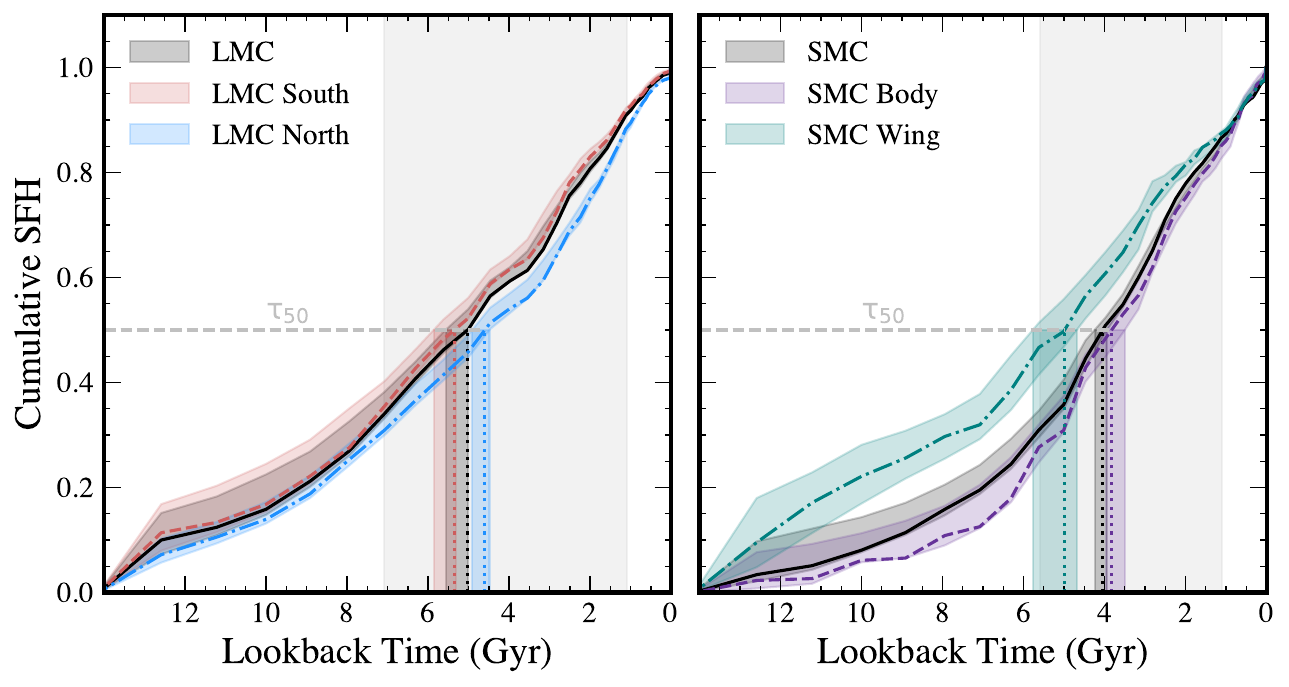}
     \caption{\textbf{Scylla CSFHs in the LMC and SMC.} 
     Examples of best-fit CSFHs (solid lines) and 1$\sigma$ uncertainties (shaded regions) for the LMC (right) and SMC (left). We also show an example of a lifetime SFH metric, $\tau_{50}$ (Table~\ref{tab:sfh_abund}), or the lookback time at which the galaxy had formed 50\% of its total mass ever formed in stars (vertical dotted lines), and its associated uncertainties computed from the uncertainties on the CSFH (shaded colored vertical regions). We show CSFHs for the northern and southern halves of the LMC disk (middle panel) and SMC body and wing (right panel), where these spatial subregions show the largest differences in mass assembly history within the LMC and SMC, respectively, over the age range defined by \taurgb\ (shaded grey vertical regions).}
    \label{fig:csfh}
\end{figure}

\begin{deluxetable*}{lccccccccc}
\tablecaption{Scylla SFH and APOGEE Chemical Abundance Characteristics by Spatial Subregion in the LMC and SMC\label{tab:sfh_abund}}
\tablewidth{0pt}
\tablehead{
\colhead{Region} &
\colhead{$\tau_{50}$} &
\colhead{$\tau_{90}$} &
\colhead{[M/H]${\rm RGB}$} &
\colhead{$\Delta$[M/H]${\rm RGB}$} &
\colhead{$M_{\rm tot}$} &
\colhead{$\langle$[Fe/H]$\rangle$} &
\colhead{$\sigma$([Fe/H])} &
\colhead{$\langle$[Mg/Fe]$\rangle$} &
\colhead{$\sigma$[Mg/Fe]} \\
\colhead{} &
\colhead{(Gyr)} &
\colhead{(Gyr)} &
\colhead{} &
\colhead{} &
\colhead{($10^6 M_\odot$)} &
\colhead{} &
\colhead{} &
\colhead{} &
\colhead{}
}
\startdata
\multicolumn{10}{c}{LMC} \\ \hline
LMC & $5.03^{+0.54}_{-0.01}$ & $1.19^{+0.03}_{-0.02}$ & $-0.35^{+0.11}_{-0.03}$ & $0.45^{+0.13}_{-0.07}$ & $10.79^{+0.63}_{-0.01}$ & $-0.75$ & $0.26$ & $0.03$ & $0.18$ \\
LMC N & $4.60^{+0.32}_{-0.11}$ & $0.96^{+0.03}_{-0.05}$ & $-0.42^{+0.08}_{-0.04}$ & $0.50^{+0.12}_{-0.16}$ & $3.10^{+0.12}_{-0.01}$ & $-0.75$ & $0.26$ & $0.03$ & $0.25$ \\
LMC S & $5.36^{+0.51}_{-0.05}$ & $1.28^{+0.03}_{-0.03}$ & $-0.33^{+0.14}_{-0.03}$ & $0.45^{+0.19}_{-0.06}$ & $8.04^{+0.56}_{-0.02}$ & $-0.71$ & $0.25$ & $0.02$ & $0.25$ \\
\hline
LMC R1 & $4.94^{+0.65}_{-0.01}$ & $1.19^{+0.05}_{-0.02}$ &  $-0.31^{+0.13}_{-0.04}$ & $0.44^{+0.16}_{-0.12}$ & $5.70^{+0.42}_{-0.01}$ & $-0.67$ & $0.24$ & $0.01$ & $0.18$ \\
LMC R2 & $5.25^{+0.47}_{-0.12}$ & $1.18^{+0.00}_{-0.05}$ & $-0.40^{+0.12}_{-0.04}$ & $0.46^{+0.19}_{-0.05}$ & $5.16^{+0.26}_{-0.01}$ & $-0.69$ & $0.24$ & $0.03$ & $0.18$ \\
LMC R3 & $5.63^{+0.07}_{-0.73}$ & $1.13^{+0.03}_{-0.17}$ & $-0.44^{+0.06}_{-0.13}$ & $0.57^{+0.22}_{-0.30}$ & $0.27^{+0.01}_{-0.01}$ & $-0.78$ & $0.26$ & $0.03$ & $0.18$ \\ \hline
LMC NR1 & $4.58^{+0.51}_{-0.14}$ & $1.06^{+0.07}_{-0.06}$ & $-0.37^{+0.11}_{-0.07}$ & $0.58^{+0.24}_{-0.18}$ & $1.83^{+0.06}_{-0.02}$ &
$-0.70$ & $0.24$ & $0.02$ & $0.25$ \\
LMC NR2 & $4.59^{+0.25}_{-0.16}$ & $0.85^{+0.04}_{-0.04}$ & $-0.47^{+0.11}_{-0.03}$ & $0.44^{+0.10}_{-0.25}$ & $1.71^{+0.07}_{-0.00}$ &
$-0.68$ & $0.24$ & $0.03$ & $0.25$ \\
LMC NR3 & $6.30^{+0.82}_{-1.33}$ & $1.20^{+0.04}_{-0.14}$ & $-0.45^{+0.14}_{-0.09}$ & $0.45^{+0.36}_{-0.23}$ & $0.09^{+0.00}_{-0.01}$ &
$-0.79$ & $0.26$ & $0.04$ & $0.25$ \\ \hline
LMC SR1 & $5.11^{+0.74}_{-0.02}$ & $1.24^{+0.06}_{-0.03}$ & $-0.30^{+0.15}_{-0.03}$ & $0.41^{+0.27}_{-0.10}$ & $4.87^{+0.42}_{-0.01}$ &
$-0.63$ & $0.25$ & $0.00$ & $0.25$ \\
LMC SR2 & $5.58^{+0.48}_{-0.16}$ & $1.33^{+0.03}_{-0.05}$ & $-0.36^{+0.14}_{-0.05}$ & $0.48^{+0.23}_{-0.06}$ & $4.36^{+0.25}_{-0.02}$ &
$-0.69$ & $0.24$ & $0.02$ & $0.25$ \\
LMC SR3 & $5.44^{+0.23}_{-0.69}$ & $1.03^{+0.13}_{-0.18}$ & $-0.43^{+0.09}_{-0.14}$ & $0.62^{+0.30}_{-0.33}$ & $0.23^{+0.01}_{-0.00}$ &
$-0.75$ & $0.25$ & $0.03$ & $0.25$ \\ \hline
\multicolumn{10}{c}{SMC} \\ \hline
SMC & $4.05^{+0.20}_{-0.11}$ & $0.80^{+0.00}_{-0.09}$ & $-0.67^{+0.10}_{-0.03}$ & $0.49^{+0.33}_{-0.05}$ & $9.04^{+0.40}_{-0.01}$ & $-0.91$ & $0.19$ & $-0.04$ & $0.18$ \\
SMC Body & $3.83^{+0.14}_{-0.33}$ & $0.70^{+0.08}_{-0.03}$ & $-0.73^{+0.19}_{-0.04}$ & $0.52^{+0.59}_{-0.08}$ & $2.16^{+0.10}_{-0.02}$ &  $-0.87$ & $0.19$ & $-0.05$ & $0.18$ \\
SMC Wing & $4.99^{+0.79}_{-0.15}$ & $0.80^{+0.09}_{-0.07}$ & $-0.46^{+0.16}_{-0.03}$ & $0.66^{+0.58}_{-0.13}$ & $0.41^{+0.02}_{-0.01}$ & $-0.97$ & $0.18$ & $-0.02$ & $0.18$ \\ \hline
\enddata
\tablecomments{The table column ``Region'' refers to the spatial subdivision of each galaxy based on its SFH characteristics (Section~\ref{sec:region}), where we include global regions that encompasses the entire Scylla footprint for each galaxy (Figures~\ref{fig:lmc} and ~\ref{fig:smc}).
        $\tau_{50}$ ($\tau_{90}$) is the lookback time at which the galaxy had formed 50\% (90\%) of its total mass ever formed in stars ($M_{\rm tot}$), which is scaled to a Kroupa IMF with mass limits 0.1–100 M$_\odot$ (e.g., \citealt{Telford2020}). The metallicity ([M/H]$_{\rm RGB}$) and 1$\sigma$ metallicity spread ($\Delta$[M/H]$_{\rm RGB}$) are predicted from the best-fit Scylla AMR (Section~\ref{sec:sfh}) at the mean RGB age for each galaxy (Section~\ref{sec:fake}) estimated from its best-fit Scylla SFH. The right columns are the mean chemical abundance ($\langle$[Fe/H]$\rangle$, $\langle$[Mg/Fe]$\rangle$) and 1$\sigma$ dispersion from APOGEE (Section~\ref{sec:abund}), computed via bootstrap resampling assuming Gaussian 
        measurement uncertainties. The precision on the chemical abundance statistics is $\leq$0.01 in each quantity.
}
\end{deluxetable*}

Here, we provide supplementary lifetime cumulative star formation history (CSFH) metrics, alongside additional quantities derived from Scylla SFHs such as the total stellar mass ever formed ($M_{\rm tot}$), for spatial subregions in the MCs (Table~\ref{tab:sfh_abund}). We also include statistics for global LMC and SMC SFHs based on the combined SFH results of all HST fields in the Scylla footprint for each galaxy. Figure~\ref{fig:csfh} shows CSFHs for a subset of subregions in the LMC and SMC, as well as an example calculation of lifetime CSFH metrics, such as $\tau_{50}$ and $\tau_{90}$, which correspond to the lookback times by which 50\% and 90\% of the total stellar mass had been respectively formed. We computed the uncertainties on $\tau_{50}$ and $\tau_{\rm 90}$ from the 1$\sigma$ uncertainties on the CSFHs, including both systematic and random uncertainty contributions.

We also computed a metric \mhrgb, the metallicity predicted by the best-fit Scylla AMR (Figure~\ref{fig:amr}) at \taurgb, for each subregion in each galaxy (Table~\ref{tab:sfh_abund}), to quantify the most relevant photometric metallicity for comparison to APOGEE data. We calculated the 1$\sigma$ uncertainties on \mhrgb\ from the uncertainties on the best-fit AMR at \taurgb. An additional metric, \dmhrgb, represents the photometric metallicity spread over the RGB age range defined by the 1$\sigma$ \added{range} on \taurgb, where the uncertainties on \dmhrgb\ are similarly computed from the uncertainties on the best-fit AMR (Figure~\ref{fig:amr}). 


In general, the CMD-derived SFHs predict systematically high mean metallicities (\mhrgb) compared to chemical abundance measurements (Table~\ref{tab:sfh_abund}; see also Figure~\ref{fig:amr}). This is despite the photometric depth of the observations, which constrain the well-known degeneracy between metallicity and age, and the broad agreement within the uncertainties between Scylla and independently constrained AMRs (Section~\ref{sec:sfh}). In order for the mean \feh\ from APOGEE to agree with the global LMC (SMC) AMR (Figure~\ref{fig:amr}), the RGB stars located in the Scylla footprint would need to have a stellar age of \agergblmcabund\ (\agergbsmcwingabund), in contrast to the relatively young median ages of all stellar populations probed by Scylla (Section~\ref{sec:fake}). Despite this offset in the metallicity normalization, the \textit{trends} in \feh\ with spatial location in the LMC broadly agree with those inferred from the Scylla SFHs (Table~\ref{tab:sfh_abund}).



\vspace{5mm}
\facilities{HST(ACS, WFC3, WFPC2, UVIS)}


\software{astropy \citep{astropy2013,astropy2018,astropy2022},
          MATCH \citep{Dolphin2002}, astrodrizzle \citep{Avila2015}, DOLPHOT \citep{Dolphin2000,Dolphin2016}, numpy, scipy, matplotlib, VICE \citep{JohnsonWeinberg2020}
          }






\end{document}